\documentclass[12pt]{article}
\usepackage{cite}
\usepackage[dvips]{graphicx}
\usepackage{amssymb}
\usepackage{rotating}
\usepackage{mathtools}
\usepackage{xcolor,soul}
\usepackage{booktabs}
\usepackage{slashed}
\usepackage[utf8x]{inputenc}

\def \tminus {\text{-}}

\def \neut {\tilde{\chi}^0}
\def \cha {\tilde{\chi}^\pm}
\def \glu {\tilde{g}}
\def \sq {\tilde{q}}
\def \stop {\tilde{t}}
\def \sbot {\tilde{b}}

\def \sl {\tilde{l}}

\newcommand\vecl[1]{{\bf{#1}}}          		 
\newcommand\vecs[1]{\boldsymbol{#1}}    		 

\newcommand\ovbb[0]{0\nu\beta\beta}			 

\newcommand{\re}[0]{\mrm{Re}}

\newcommand{\mrm}[1]{\mathrm{#1}}

\begin{document}

\vspace*{-1in}
\renewcommand{\thefootnote}{\fnsymbol{footnote}}

\vskip 5pt

\begin{center}
{\Large{\bf 
GUT Physics in the era of the LHC}}
\vskip 25pt

{
Djuna Croon$^1$,
Tom\'as E. Gonzalo$^{2,3}$\footnote{tomas.gonzalo@monash.edu},
Lukas Graf$^{4,5}$,
Nejc Ko\v snik$^{6,7}$,
Graham White$^1$
}

\vskip 10pt

{\it \small $^1$TRIUMF Theory Group, 4004 Wesbrook Mall, Vancouver, B.C. V6T2A3, Canada} \\
{\it \small $^2$Department of Physics, University of Oslo, N-0316 Oslo, Norway}\\
{\it \small $^3$ARC Centre of Excellence for Particle Physics at the Tera-scale,
 School of Physics and Astronomy, Monash University, Melbourne, Victoria 3800, Australia } \\
{\it \small $^4$Department of Physics and Astronomy, University College London, London WC1E 6BT, UK}\\
{\it \small $^5$Max-Planck-Institut f\"{u}r Kernphysik, Saupfercheckweg 1,
69117 Heidelberg, Germany}\\
{\it \small $^6$Department of Physics, University of Ljubljana, Jadranska 19, 1000 Ljubljana, Slovenia}\\
{\it \small $^7$Jo\v zef Stefan Institute, Jamova 39, P.\ O.\ Box 3000, 1001
  Ljubljana, Slovenia}

\medskip

\begin{abstract}

Grand Unified Theories (GUTs) are one of the most interesting high-energy completions of the Standard Model, because they provide a rich, powerful and elegant group-theoretical framework able to resolve a variety of problems remaining in our current understanding of particle physics. They usually act as motivators for many low energy BSM theories, such as left-right symmetric or supersymmetric models, and they serve to fill the gap between the experimentally reachable low energies and the physics in the ultraviolet. In recent years, however, they have fallen slightly from the spotlight, in favour of ``simplified'' models with more specific phenomenological predictions. The aim of this review is to summarize the state of the art on GUTs and argue for their importance in modern physics. Recent advances in experiments permit to test the predictions of GUTs at different energy scales. First, as GUTs can play a role in the inflationary dynamics of the early Universe, their imprints could be found in the CMB observations by the Planck satellite. Remarkably enough, GUTs could manifest themselves also in terrestrial tests; several planned experiments aim to probe the proton stability and to establish order of magnitude higher bounds on its lifetime. Moreover, the predictions of specific GUT models could be tested even at the LHC thanks to its high energy reach, via searches for exotic states or additional contributions to flavour anomalies. 

\end{abstract}

\end{center}

\noindent
{\small Keywords: Grand Unified Theories, Supersymmetry, Cosmology, Flavour}

\renewcommand{\thefootnote}{\arabic{footnote}}
\setcounter{footnote}{0}

\clearpage
\tableofcontents
\clearpage

\section{Introduction}
\label{sec:introduction}

The Standard Model (SM)~\cite{Glashow:1961tr,Weinberg:1967tq,Salam:1968rm} of particle physics is an incredible successful theory of subatomic physics. It describes the electroweak and strong interactions of fundamental particles with surprising accuracy up to the energy scales of modern day experiments. Further supported by the discovery of the Higgs boson~\cite{Aad:2012tfa,Chatrchyan:2012xdj}, it stands as one of the best evidences that symmetries and the mechanism of spontaneous symmetry breaking play a critical role on the Universe at the smallest scales~\cite{Higgs:1964pj,Englert:1964et,Guralnik:1964eu}.

In spite of its success at explaining with astonishing precision most of the observed phenomena, the SM cannot be the ultimate theory of particle physics . The Higgs quartic coupling in the SM becomes negative at scales $\gtrsim 10^{10}$ GeV, rendering the vacuum state of the theory unstable at high energies~\cite{Degrassi:2012ry}. New physics must then surface below or around that scale to stabilise the vacuum. Furthermore, there is a continuously increasing amount of observations that are in tension with the predictions of the SM. From the discovery of neutrino oscillations~\cite{Fukuda:1998mi,Ahmad:2001an} to the recent anomalies in the flavour sector~\cite{Aaij:2014ora, Aaij:2015yra, Aaij:2017vad}, these phenomena cannot be explained with the SM alone and contributions from beyond the SM (BSM) physics may be required to accommodate them.

\begin{figure}
\centering
\includegraphics[width=0.7\textwidth]{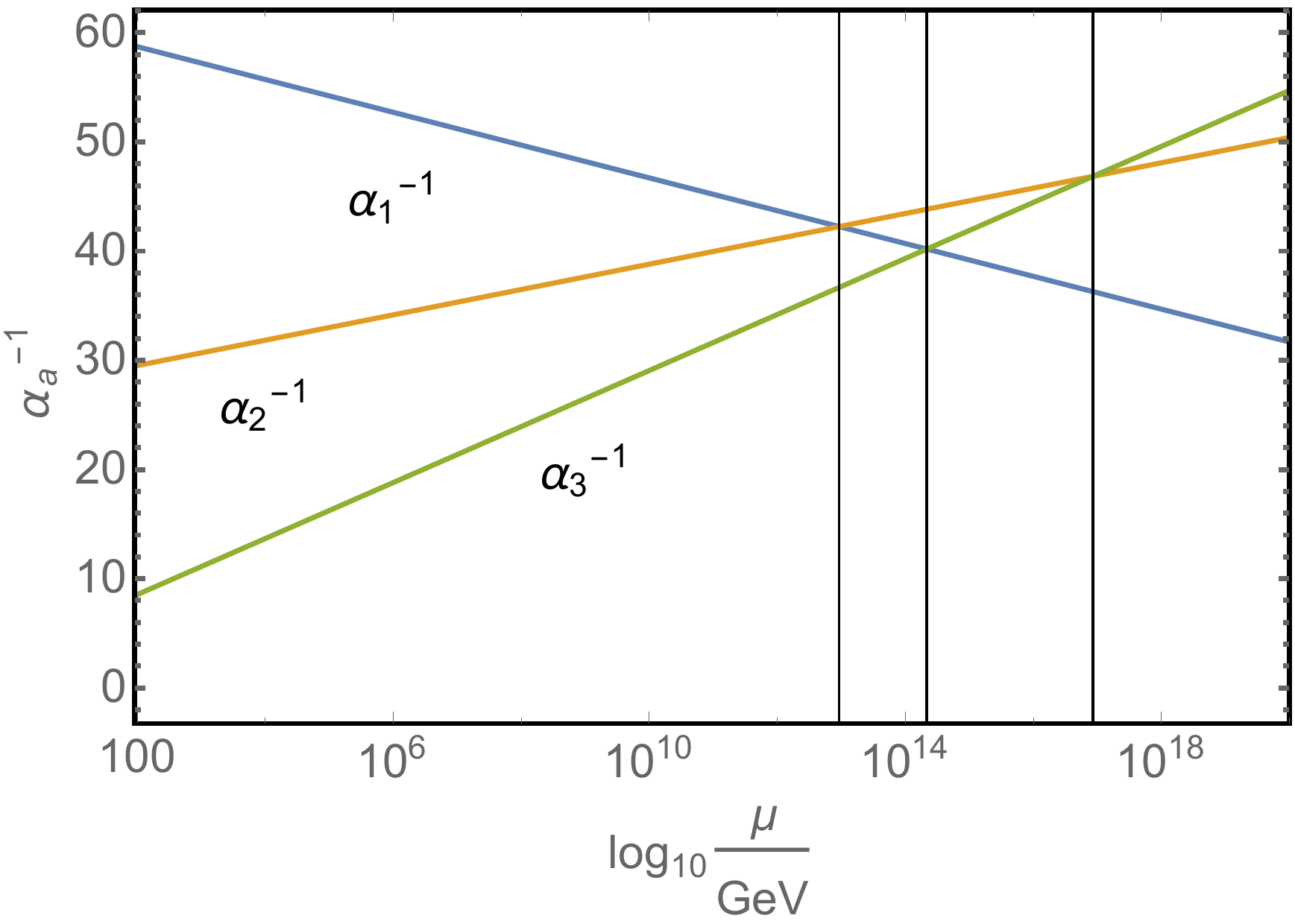}
\caption{\emph{Renormalization Group flow of the Standard Model gauge couplings}}
\label{fig:SMRGEs}
\end{figure}

Grand Unified Theories (GUTs)~\cite{Georgi:1974sy,Pati:1974yy,Mohapatra:1974hk,Fritzsch:1974nn,Georgi:1974my} are well motivated extensions of the SM that can address several of its outstanding issues. As the SM does for electromagnetism and weak nuclear decays, GUTs exploit the power of symmetries and group theory to unify the electroweak and strong interactions into a single force. As can be noticed in Figure~\ref{fig:SMRGEs}, the flow of the SM gauge couplings already hints at a possible unification at a high scale, thereby providing further motivation for GUTs as appealing BSM models. 

The predicted unification of forces provides an explanation for the \textit{ad hoc} nature of the SM charge assignments and their accidental anomaly cancellation~\cite{Langacker:1980js,GonzaloVelasco:2015hki}. Through the introduction of new fields and symmetries, GUTs can resolve many of the issues of the SM: they can provide an explanation for the lightness of neutrino masses, as well as introducing additional contributions that can accommodate some of the observed flavour anomalies. In addition, GUTs can also live alongside other BSM models, such as Supersymmetry (SUSY), with both theories complementing each other and on the whole becoming a better candidate for a successful BSM theory~\cite{Raby:2017ucc}.

Naturally living at high energies, it is expected that GUTs have some observable consequences for the cosmological evolution of the Universe, as they can play a role during the inflationary epoch and their phase transitions may be the source for matter-antimatter asymmetry or gravitational waves~\cite{Nath:2018rqn}. With or without SUSY, GUTs also make predictions that can be tested at low energy experiments such as particle colliders, which can search for new exotic states or deviations on flavour observables. Some of its high energy repercussions can also be probed with precision experiments, with nucleon decay measurements at the forefront. In short, Grand Unified Theories have observable effects in many fronts and can be probed by current and upcoming experiments in the near future.

Therefore, throughout this review we will focus on the description of GUT models and their observable consequences. We will introduce the basic concepts and summarise some of the modern research in GUTs. Out of all the possible observable probes of GUTs, we will focus on a subset of them. In the cosmological front we will outline the advances on inflationary GUTs, detail their observable gravitational wave signatures, from cosmic strings and phase transitions, and their relation with mechanisms for baryo and leptogenesis. The low energy front will cover collider searches for supersymmetry, leptoquarks and exotic states. Lastly, a number of precision tests of unification will be detailed, including nucleon decay, flavour observables and neutrinoless double beta decay.

As ultraviolet (UV) completions of the Standard Model that live at very high energies, GUTs are often closely related to theories of gravity, such as string theory. In fact, many unified theories arise as four-dimensional compactifications in some realization of superstring theory~\cite{Candelas:1985en,Witten:1985xc}. However, throughout this review we will not concern ourselves with these string theory realisations. For a review on embedding GUTs in the heterotic string and outcomes of string compactification for unified theories see~\cite{Raby:2017ucc}. 	

\section{Basics on GUT models}
\label{sec:basicsonguts}

Grand Unified Theories are extensions of the SM with larger symmetry groups. Strictly speaking, GUTs require that the unified group be a simple group, e.g. $SU(5)$, $SO(10)$ or $E_6$. However, here we use the term GUT more loosely, referring to any extension of the SM symmetries including product groups such as $SU(5)\times U(1)$ and $SU(4)\times SU(2)\times SU(2)$. Candidate groups for a realistic GUT model must satisfy two conditions: they must contain the SM group as a subgroup and they must have complex representations that reproduce the chiral structure of the SM. Although groups with pseudoreal representations have been studied as candidates for unified theories, $E_7$~\cite{Gursey:1976dn}, we will not consider them here.

\subsection{$SU(5)$}
\label{sec:su5}

The first appearance of a GUT in the literature dates back to 1974 when H. Georgi and S. Glashow proposed the unification of the SM gauge group into a simple group, $SU(5)$~\cite{Georgi:1974sy}. In their proposal all the left-handed fermions of a single generation fell into two representations of the group, $\mathbf{\bar{5}}$ and $\mathbf{10}$, in the following way
\begin{equation}
 \mathbf{\bar{5}} \leftrightarrow \left( \begin{matrix} d_1^c \\ d_2^c \\ d_3^c \\ e \\ -\nu \end{matrix} \right), \quad \mathbf{10} \leftrightarrow \left( \begin{matrix} 0 & u_3^c & -u_2^c & u_1 & d_1 \\ -u_3^c & 0 & u_1^c & u_2 & d_2 \\ u_2^c & -u_1^c & 0 & u_3 & d_3 \\ -u_1 & -u_2 & -u_3 & 0 & e^c \\ -d_1 & -d_2 & -d_3 & -e^c & 0 \end{matrix} \right).
 \label{SU5fields}
\end{equation}
and the gauge and Higgs sector of the theory were embedded into the $\mathbf{24}$ and $\mathbf{5}$ representations, respectively. In addition to the SM Higgs boson present in the representation $\mathbf{5}$, often a scalar $\mathbf{\bar{5}}$ representation is also present, which contains a second $SU(2)_L$ doublet, necessary for UV completions of two-Higgs doublet models (2HDM)~\cite{Branco:2011iw}.

The Georgi-Glashow (GG) model was the first attempt of a fully-unified model for particle physics, and it provided a neat explanation for the hypercharge quantisation in the SM. The traceless hypercharge generator can be written as~\cite{GonzaloVelasco:2015hki}
\begin{equation}
 Y= \alpha~ \text{diag}(-\tfrac{1}{3},-\tfrac{1}{3},-\tfrac{1}{3},\tfrac{1}{2},\tfrac{1}{2})
 \label{hypercharge}   
\end{equation}
which when acting upon the representations of $SU(5)$ results in the specific hypercharge assignments of the SM fields, i.e. for $\alpha = 1$, $Y(Q) = 1/6$, $Y(L) = -1/2$, $Y(u^c) = -2/3$, $Y(d^c) = 1/3$ and $Y(e^c) = 1$. In unified theories one often uses the ``GUT normalization'' of the hypercharge, which corresponds simply to a rescaling of the charges and gauge couplings as $Y_{GUT} = \sqrt{3/5}~Y$ and $g_1 = \sqrt{5/3}~g'$~\cite{Georgi:1974yf}.

Spontaneous symmetry breaking of $SU(5)$ happens when a scalar field in a non-trivial representation of the group acquires a vacuum expectation value (vev). The minimal representation of $SU(5)$ that can achieve this goal while keeping the SM phase unbroken is the $\mathbf{24}$~\cite{Georgi:1974sy,Buras:1977yy}. This vev provides a mass to the off-diagonal $SU(5)$ gauge bosons while the SM gauge bosons remain massless.

By virtue of the unification into a single gauge group, the GG model requires strict unification of the SM gauge couplings, which is hinted at but not really achieved in the SM~\cite{Georgi:1974yf,Buras:1977yy}, as can be seen in Fig.~\ref{fig:SMRGEs}, as well as that of Yukawa couplings for each of the two representations, a difficult task in its minimal version~\cite{Chanowitz:1977ye,Georgi:1979df}.

The minimal $SU(5)$ version suffers from other afflictions beyond the failed gauge and Yukawa unification mentioned above. One of these is the introduction of an artificial hierarchy, known as doublet-triplet splitting~\cite{Dimopoulos:1981zb,Masiero:1982fe}, in the components to the Higgs representation $\mathbf{5}$. The coloured components must be quite heavy to avoid rapid proton decay whereas the uncoloured components must be relatively light, for they correspond to the SM Higgs doublet, now know to have a mass of $m_h = 125.18$ GeV~\cite{Tanabashi:2018oca}. Solutions to this problem in several $SU(5)$ models have been proposed, such as the ``missing partner mechanism''~\cite{Masiero:1982fe,Grinstein:1982um} or the ``double missing partner mechanism''~\cite{Hisano:1994fn,Antusch:2014poa}. 

Another case where the minimal $SU(5)$ falls short is the lack of a mechanism for the generation of neutrino masses. Extended scalar sectors can generate neutrino masses in the type-I~\cite{Dorsner:2005ii} and type-III~\cite{Bajc:2006ia} seesaw mechanisms\footnote{See Sec.~\ref{sec:numasses} for details on neutrino mass generation through the seesaw mechanism.}, but the resulting theories are often non-renormalisable. Renormalisable $SU(5)$ models can also be constructed where the neutrino masses are generated via a mixture of type-I and type-III seesaw~\cite{Perez:2007rm} or the Zee mechanism~\cite{Zee:1980ai,Perez:2016qbo}.

Worst of all, however, is the fact that the vanilla $SU(5)$ model predicts rapid proton decay. The lifetime of the proton can be naively estimated as~\cite{Chanowitz:1977ye}
\begin{equation}
 \tau_p \sim \frac{M_X^4}{m_p^5},
\end{equation}
with $m_p$ the mass of the proton and $M_X$ the mass of the mediator field at the scale of unification. The apparent unification of gauge couplings happens at an energy scale $\mu \sim 10^{15}$ GeV, which gives a half life for the proton of the order of $10^{31}$ years, far below the experimental bound from the Super-Kamiokande experiment of $1.6 \times 10^{34}$ years~\cite{Miura:2016krn}. Particular choices of the Higgs sector of the $SU(5)$ model, however, avoid this issue, rendering non-minimal $SU(5)$ models viable candidates~\cite{Dorsner:2005fq, Dorsner:2006dj, Fornal:2018aqc}. Furthermore, $SU(5)$ models with vector-like fermions can be consistent with current limits and even predict an upper bound on the lifetime of proton decay~\cite{FileviezPerez:2018dyf}.

\begin{figure}
\centering
\includegraphics[width=0.7\textwidth]{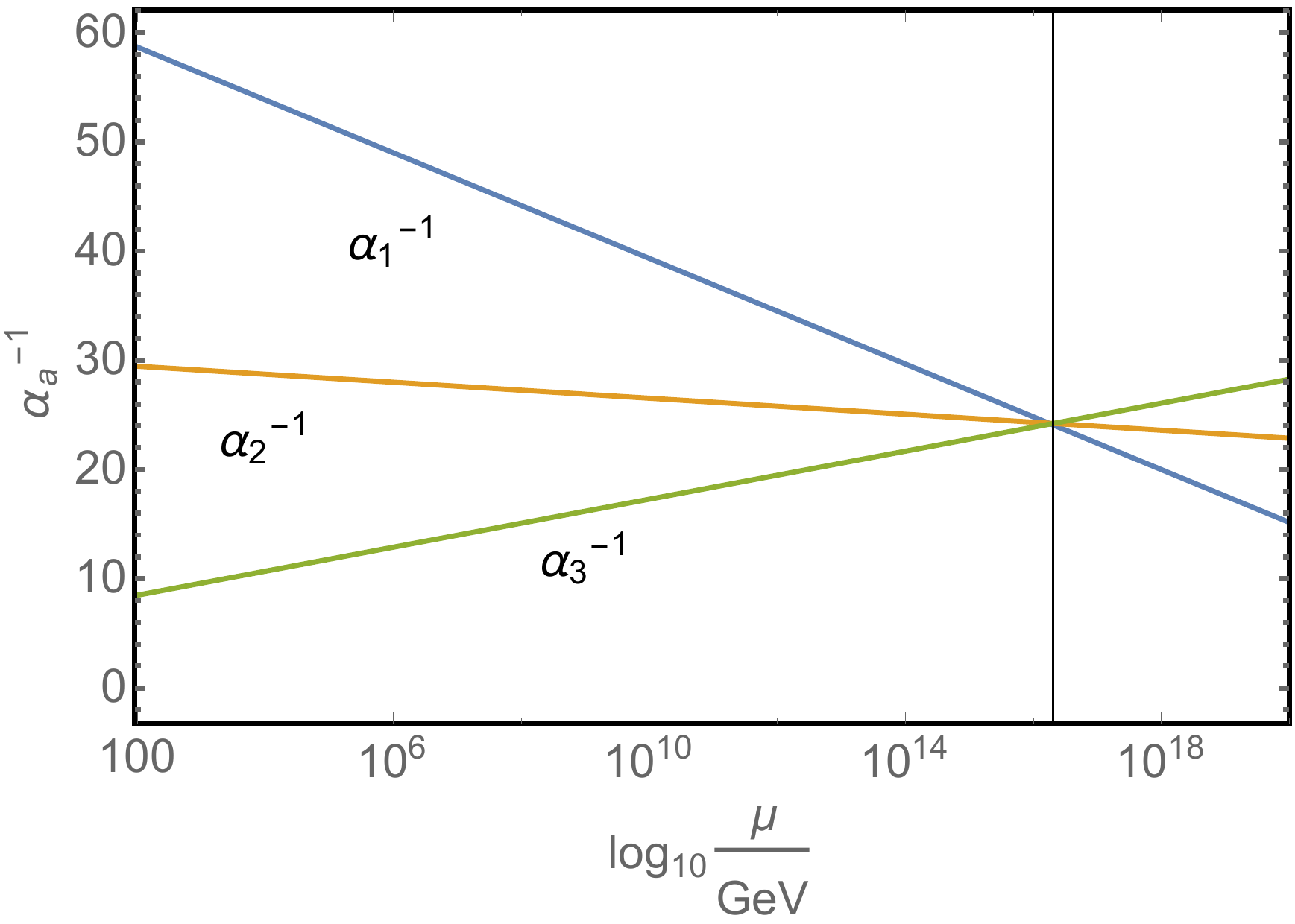}
\caption{\emph{Renormalization Group flow of the MSSM gauge couplings}}
\label{fig:MSSMRGEs}
\end{figure}

One of the fundamental issues with GUT models, which remains as a concern today, is the gauge hierarchy problem. The large hierarchy between the mass scale of unification and the electroweak scale poses a problem since it causes large loop corrections to the Higgs mass~\cite{Weinberg:1975gm}. Supersymmetry (SUSY) was proposed as a solution to this issue~\cite{Martin:1997ns} and even acted as a motivation for unified theories since some of its minimal realisations, such as the MSSM, predicted the unification of gauge couplings, as can be seen in Fig.~\ref{fig:MSSMRGEs}.

Supersymmetric GUTs are in fact rather popular and have in some cases been proven to be more successful at representing nature than regular GUTs~\cite{Nanopoulos:1982wk}. In SUSY $SU(5)$ theories the scale of unification is typically larger than in non-supersymmetric models\footnote{A detailed description of unification in SUSY $SU(5)$ can be found in \cite{Dorsner:2006ye}.}, around $\mu \sim 2\times 10^{16}$ GeV as can be seen in Fig.~\ref{fig:MSSMRGEs}. This has two advantageous consequences for these models: the larger mass scale for the gauge mediators imposes a further suppression on nucleon decay processes, consistent with experimental measurements~\cite{Nanopoulos:1982wk}; and pushes the unification scale beyond the scale of inflation, which helps to dilute the magnetic monopoles naturally present in the theory~\cite{tHooft:1974kcl}. Another issue in vanilla $SU(5)$ models that can be addressed in its supersymmetric version is the doublet-triplet splitting, where the Higgs doublets are made light via cancellations of the superpotential parameters ~\cite{Dimopoulos:1981zb, Witten:1981kv, Masiero:1982fe, Dimopoulos:1982af}.

\subsection{Flipped $SU(5)$}

An alternative solution to the issues of $SU(5)$ models, without supersymmetry, is what is now known as \textit{flipped} $SU(5)$~\cite{DeRujula:1980qc,Barr:1981qv}. The flipped version differs from regular $SU(5)$ in its gauge group, extended to $SU(5)\times U(1)$, and the manner in which the SM particle fields are embedded into representations of the group. In contrast to eq.~\eqref{SU5fields}, the matter representations in the flipped $SU(5)$ model are
\begin{equation}
 \mathbf{\bar{5}} \leftrightarrow \left( \begin{matrix} u_1^c \\ u_2^c \\ u_3^c \\ e \\ -\nu \end{matrix} \right), \quad \mathbf{10} \leftrightarrow \left( \begin{matrix} 0 & d_3^c & -d_2^c & u_1 & d_1 \\ -d_3^c & 0 & d_1^c & u_2 & d_2 \\ d_2^c & -d_1^c & 0 & u_3 & d_3 \\ -u_1 & -u_2 & -u_3 & 0 & \nu^c \\ -d_1 & -d_2 & -d_3 & -\nu^c & 0 \end{matrix} \right),
 \quad \mathbf{1} \leftrightarrow \Big( e^c \Big),
 \label{flippedSU5fields}
\end{equation}
where $\nu^c$ labels the right-handed neutrino field, whose presence provides a mechanism for neutrino mass generation, which was absent in vanilla $SU(5)$.

With these different embeddings of the SM fields, the hypercharge operator is no longer contained in $SU(5)$, as in eq.~\eqref{hypercharge}, but it is rather a combination of diagonal generators of both $SU(5)$ and $U(1)$. With standard normalisation the hypercharge operator can be written as~\cite{Derendinger:1983aj}.
\begin{equation}
    Y = -\frac{1}{5} T_{24} + \frac{1}{5} X,
\end{equation}
where $T_{24}$ is a diagonal generator of $SU(5)$ and $X$ the $U(1)$ charge.

Due to the extended gauge sector and modified unified conditions of flipped $SU(5)$, proton decay does not become an issue~\cite{Derendinger:1983aj}. In addition, in flipped $SU(5)$ magnetic monopoles cannot be created since the supergroup containing the charge operator is not a simple group~\cite{tHooft:1974kcl, Derendinger:1983aj}.

As was the case with regular $SU(5)$ models, flipped $SU(5)$ can be extended with the help of supersymmetry. The combination of solutions to the issues of the $SU(5)$ model that both SUSY and flipped $SU(5)$ offer makes SUSY flipped $SU(5)$ one of the most popular GUTs in the literature~\cite{Antoniadis:1987dx,Ellis:1988tx}, in spite of not realising a full unification of gauge couplings.

Flipped $SU(5)$ models are also well motivated from their UV completions, since they can easily be derived naturally from weakly-coupled string theory. As opposed to vanilla $SU(5)$, which undergoes symmetry breaking via a $\mathbf{24}$-dimensional representation, the flipped $SU(5)$ model does not require large dimensional representations, as it breaks via a $\mathbf{10}_1$, and it is therefore easier to obtain from a manifold compactification of string theory~\cite{Campbell:1987gz,Antoniadis:1987tv}.

\subsection{Pati-Salam and the left-right symmetry}
\label{sec:pslr}

Around the same time that the $SU(5)$ model was proposed, R. Pati and A. Salam suggested another extension of the SM~\cite{Pati:1974yy}. They proposed a rearrangement of the SM fields into a different group configuration, $SU(4)_c \times SU(2)_L \times SU(2)_R$. Though not really a fully unified theory, it provided a partial unification of leptons and quarks into a single colour group, $SU(4)_c$, while at the same time introducing another copy of $SU(2)$ for the right-handed sector of the theory. This model automatically contains a right-handed neutrino, embedded into a $SU(2)_R$ doublet with the right-handed charged lepton. Thus the SM fields fall into two representations of the group in the following way
\begin{equation}
\{\mathbf{4},\mathbf{2},\mathbf{1}\} \leftrightarrow \left(
\begin{array}{cccc}
u_1 & u_2 & u_3 & \nu \\
d_1 & d_2 & d_3 & e
\end{array}
\right),\quad
\{\mathbf{\overline{4}},\mathbf{1},\mathbf{2^*}\} \leftrightarrow \left(
\begin{array}{cccc}
d^c_1 & d^c_2 & d^c_3 & e^c \\
-u^c_1 & -u^c_2 & -u^c_3 & -\nu^c 
\end{array}
\right).
\label{psfermions2}
\end{equation}

One of the major successes of the Pati-Salam (PS) model was being the first appearance of a left-right symmetric model, with a right-handed sector $SU(2)_R$ and a heavy right-handed gauge boson $W_R$~\cite{Senjanovic:1975rk}. It was also the original proposal for the idea of quark-lepton complementarity. As an amalgamation of the two ideas, the PS group maximally contains the left-right symmetry group, $SU(3)_c \times SU(2)_L \times SU(2)_R \times U(1)_{B-L}$~\cite{Mohapatra:1974hk, Mohapatra:1974gc}, as well as the quark-lepton unified group, $SU(4)_c \times SU(2)_L \times U(1)_R$~\cite{Smirnov:1995jq,Perez:2013osa}.

Left-right symmetric models, \textit{\`a la} Pati-Salam or of the type $SU(3)_c \times SU(2)_L \times SU(2)_R \times U(1)_{B-L}$, are fairly popular because they naturally include a right-handed neutrino and can generate light neutrino masses via some type of seesaw mechanism~\cite{Mohapatra:1979ia,Mohapatra:1980yp}. Similar to PS, left-right symmetric (LR) models are not fully unified theories, yet they can be an intermediate step on the breaking chain of a PS model~\cite{Mohapatra:1980qe} or some other unified theory~\cite{Lindner:1996tf}. 

Symmetry breaking in the PS model can happen through a number of different paths, depending on the vev of the scalar fields in the theory. The most compelling paths preserve either the LR symmetry, with the LR group $SU(3)_c \times SU(2)_L \times SU(2)_R \times U(1)_{B-L}$ as an intermediate step, or quark-lepton unification, with $SU(4)_c \times SU(2)_L \times U(1)_R$ as a subgroup~\cite{Aulakh:2000sn}. Further symmetry breaking from the LR symmetry model happens when either a pair of $SU(2)$ doublets (one left-handed and one right-handed)~\cite{Senjanovic:1975rk,Senjanovic:1978ev}, or a pair of $SU(2)$ triplets (left and right-handed)~\cite{Mohapatra:1980qe} acquire a vev. In both PS and LR theories, the hypercharge operator is written as a linear combination of the diagonal $SU(2)_R$ generator and the $B-L$ charge ($U(1)_{B-L}$ generator embedded in $SU(4)_c$ in PS) as
\begin{equation}
 Y = T^3_R + \frac{1}{2}(B-L).
 \label{lrhypercharge}
\end{equation}

As opposed to the case of $SU(5)$ the proton is often stable in PS and LR models. This occurs because the gauge sector of the theory preserves $B$ and $L$ number independently and the only scalar fields that can mediate the transition are in antisymmetric representations, rarely seen in PS or LR models~\cite{Mohapatra:1980qe}. 

The addition of supersymmetry to PS and left-right symmetric models~\cite{Antoniadis:1988cm} is not as straightforward as with other GUT models. The simplest scenario with both SUSY and LR symmetry was shown to fail to achieve spontaneous symmetry breaking \cite{Kuchimanchi:1993jg}. In order to circumvent this issue one must either add extra fields, such as a parity-odd singlet\footnote{Although successful in achieving spontaneous symmetry breaking (SSB) in this SUSY LR model, the resulting vacuum state does not preserve the electromagnetic charge.}~\cite{Cvetic:1985zp} or an extra Higgs field~\cite{Aulakh:1997ba}, or alternatively supplement the Lagrangian with non-renormalizable operators~\cite{Aulakh:1998nn}. One of the main advantages of SUSY LR models, and the reason why so much effort is put on resolving the SSB issue, is that they naturally preserve $R$-parity, since $B-L$ is a gauge symmetry of the theory, which forbids the dangerous baryon and lepton number violating operators that appear in the MSSM, thereby making the lightest SUSY particle stable~\cite{Aulakh:1997fq}.

\subsection{$SO(10)$}
\label{sec:so10}

Although the GG and PS models seem quite distinct in their approach to unification, they have a common ancestor. Both $SU(5)\times U(1)$ and $SU(4)\times SU(2) \times SU(2)$ are maximal subgroups of another Lie group of larger dimension, $SO(10)$. This was first realised by H. Fritzsch and P. Minkowski~\cite{Fritzsch:1974nn}, and independently by H. Georgi~\cite{Georgi:1974my}, who proposed a model of unification with several intermediate steps. $SO(10)$ has since been the most popular choice as a unification group, since it provides a vast display of options for field configurations and symmetry breaking patterns.

One of the many key features of $SO(10)$ models is that they fully unify a generation of SM fermions into a single representation of the group. Thus the 16 Weyl fermions, including right-handed neutrinos, can be embedded into the fundamental $\mathbf{16}$ representation of $SO(10)$ as
\begin{equation}
\mathbf{16} = \{u_1^c,d_1^c,d_1,u_1,\nu^c,e^c,d_2,u_2,u_2^c,d_2^c,d_3,u_3,u_3^c,d_3^c,e,\nu\}_L.
\label{fermionsixteenplet}
\end{equation}

Due to the transformation properties of the $SO(10)$ group, the spinor representation $\mathbf{16}$ is a complex representation, thereby satisfying chirality as observed in the Standard Model. Additionally $SO(10)$ is a ``safe algebra''~\cite{Georgi:1972bb}, it does not suffer from anomalies as, for example, the $SU(5)$ case above, which makes model building in $SO(10)$ easier for it does not rely on some specific field configurations to cancel the gauge anomalies~\cite{Adler:1969gk}. 

Despite the large amount of $SO(10)$ models in the literature, a common feature is that the gauge fields are embedded in the adjoint representation of the group, that is $\mathbf{45}$,
\begin{align}
\notag \mathbf{45} &\to ~ \{\mathbf{8},\mathbf{1},~0\} \oplus \{\mathbf{1},\mathbf{3},~0\} \oplus \{\mathbf{1},\mathbf{1},~0\} \quad \quad \leftarrow \text{SM gauge bosons} \\ \notag
&\left.
\begin{array}{l}
\oplus ~ \{\mathbf{3},\mathbf{2},~\tfrac{1}{6}\} \oplus \{\mathbf{\overline{3}},\mathbf{2},\tminus\tfrac{1}{6}\} \oplus \{\mathbf{3},\mathbf{2},~\tfrac{1}{6}\}  \\
\oplus ~\{\mathbf{\overline{3}},\mathbf{2},\tminus\tfrac{1}{6}\}  \oplus \{\mathbf{\overline{3}},\mathbf{1},\tminus\tfrac{2}{3}\} \oplus \{\mathbf{3},\mathbf{1},~\tfrac{2}{3}\}  \\ 
\oplus ~ \{\mathbf{1},\mathbf{1},~1\}  \oplus \{\mathbf{1},\mathbf{1},\tminus1\}  \,\oplus \{\mathbf{1},\mathbf{1},~0\}. 
\end{array}  \right\} \; \leftarrow \text{leptoquarks},
\end{align}
which contains the SM gauge bosons as well as off-diagonal components which, as happened in $SU(5)$, can mediate quark-lepton transitions, known as leptoquarks. The Yukawa sector in $SO(10)$ models is often also quite recurrent because, at the renormalizable level, it can only be of the form~\cite{Mohapatra:1979nn}
\begin{equation}
\mathcal{L}_{Yuk} = \mathsf{Y}\cdot\mathbf{16}^T C_L C_{10} (\Gamma_i \Phi^i + \Gamma_{[i}\Gamma_{j}\Gamma_{k]} \Phi^{ijk} + \Gamma_{[i}\Gamma_{j}\Gamma_{k}\Gamma_{l}\Gamma_{m]}\Phi^{ijklm})\mathbf{16},
\end{equation}
where $\mathsf{Y}$ is the matrix of Yukawa couplings, $C_L$ and $C_{10}$ the charge conjugation matrices in the Poincar\'e and $SO(10)$ groups, $\Gamma_i$ the generators of $SO(10)$ in the spinor representation and $\Phi^i$, $\Phi^{ijk}$ and $\Phi^{ijklm}$ are scalar fields in the $\mathbf{10}$, $\mathbf{120}$ and $\mathbf{\overline{126}}$ representations, respectively. The SM Higgs field is, therefore, some linear combination of these fields and hence the SM fermion masses directly follow from the Yukawa matrix $\mathbf{Y}$ and the vacuum expectation values of the scalar fields. The particular choice of the scalar sector is typically guided by the principle of minimality. While the minimal regular (non-SUSY) $SO(10)$ model with SSB driven by the $\mathbf{45}$ and $\mathbf{126}$ Higgs representations has been revived and still represents a phenomenologically viable scenario~\cite{Bertolini:2010ng,Bertolini:2012im,Bertolini:2013vta,Kolesova:2014mfa,Graf:2016znk}, this is not the case of the minimal SUSY $SO(10)$ model~\cite{Bertolini:2006pe,Aulakh:2005mw}.

\begin{figure}[ht]
\centering
\includegraphics[width=0.8\textwidth]{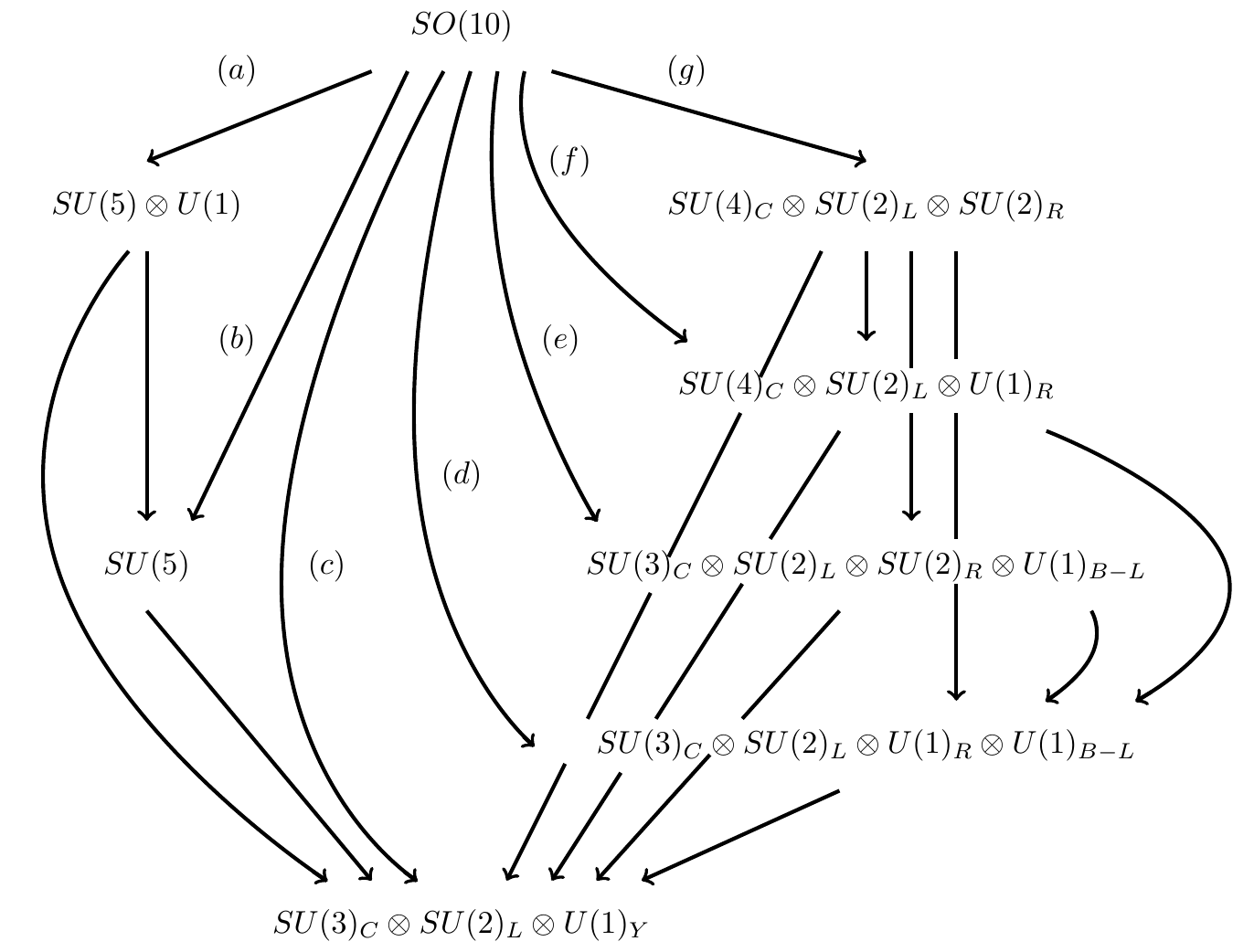}
\caption{\emph{Patterns of symmetry breaking from $SO(10)$ to the SM group~\cite{GonzaloVelasco:2015hki}.}}
\label{fig:SO10Breaking}
\end{figure}

Symmetry breaking in $SO(10)$ models can occur through one of many different paths. Since both $SU(5)\times U(1)$ and $SU(4)\times SU(2) \times SU(2)$ are subgroups of $SO(10)$, they can be an intermediate step on the symmetry breaking path towards the Standard Model, as can be any of their respective subgroup~\cite{Aulakh:2000sn,Aulakh:2002zr,Anastaze:1983zk}. Alternatively $SO(10)$ can be broken directly to the SM group, without intermediate steps~\cite{Clark:1982ai}. All the possible breaking paths from $SO(10)$ can be seen in Figure \ref{fig:SO10Breaking}. The particular symmetry breaking scenario that is realized in a $SO(10)$ model depends exclusively on the scalar sector of the theory and the configuration of the vacuum, and it constitutes one of the major differences among $SO(10)$ models in the literature.

Regular $SO(10)$ models may suffer from some of the same issues as regular $SU(5)$, namely rapid proton decay can occur with a low unification scale. The main solution to this problem, as it was with $SU(5)$, is the addition of supersymmetry. SUSY $SO(10)$~\cite{Clark:1982ai,Aulakh:1982sw} theories are rather popular and given the large number of degrees of freedom they possess, such as symmetry breaking pattern, field content, etc., they can easily avoid many of the traditional issues in unified theories.

Alike to the $SU(5)$ model, it is possible to construct alternative embeddings of the SM fermions into representations of the group via the addition of an Abelian group. In the \textit{flipped} $SO(10) \times U(1)$ model~\cite{Tamvakis:1987sd} the SM fermion content is not fully embedded into the $\mathbf{16}$ representation of the group, but rather into the direct product $\mathbf{16}_1 \oplus \mathbf{10}_{-2} \oplus \mathbf{1}_4$. This model loses its unified nature in favour of more degrees of freedom for the Yukawa and symmetry breaking sectors of the theory, which are no longer constrained by the statements above~\cite{Mohapatra:1979nn}.

\subsection{$E_6$}

The GUT models described so far have unification groups that are part of the infinite series $SU(n)$ or $SO(2n)$, which means that for each successful model with particular $n$ there is an infinite number of alternatives with order larger than $n$. For instance, the $SO(18)$ group has been studied as a candidate for gauge and family unification~\cite{Reig:2017nrz}. The exceptional algebras, however, are unique so they are more aesthetically appealing candidates as theories of unification~\cite{Langacker:1980js, Langacker:1998tc}. Among all exceptional algebras, only $E_6$ is large enough to contain the SM as a subgroup and admits complex representations.

The fermionic content in the $E_6$ theory is embedded in the fundamental $\mathbf{27}$-dimensional representation of the group, which contains the SM fermions plus exotic fields. The particular allocations of SM fermions into the $\mathbf{27}$ representation depends on the subgroup of $E_6$ that breaks into after SSB, which is typically either the trinification group, $SU(3)_c \times SU(3)_L \times SU(3)_R$~\cite{Gursey:1975ki,Achiman:1978vg,Shafi:1978gg} or $SO(10)\times U(1)$~\cite{Mohapatra:1986bd,Buccella:1987kc}. The decomposition of the fundamental $\mathbf{27}$ into these subgroups is
\begin{equation}
 \begin{array}{cll}
  \mathbf{27} & \to \{\mathbf{1},\mathbf{3},\mathbf{3}\} + \{\mathbf{3},\mathbf{3},\mathbf{1}\} + \{\mathbf{\bar{3}},\mathbf{1},\mathbf{\bar{3}}\}, & [SU(3)_c \times SU(3)_L \times SU(3)_R] \\
  \mathbf{27} & \to \mathbf{16}_1 + \mathbf{10}_{-2} + \mathbf{1}_4,  & [SO(10)\times U(1)]
\end{array}
\end{equation}

The field content in $E_6$ models is quite vast. There are 78 gauge bosons, of which only 45 survive at lower scales if $SO(10)$ is the breaking path, or even fewer in the case of $SU(3)\times SU(3)\times SU(3)$, just 24. The minimal scalar content needs at least a scalar field in the combination $\mathbf{\overline{27}} + \mathbf{\overline{351}} + \mathbf{\overline{351'}}$, which contains the SM Higgs, and a scalar responsible for SSB of $E_6$, which is dependent upon the pattern of symmetry breaking, e.g. a $\mathbf{78}$ for breaking to $SO(10)\times U(1)$.

One of the main motivations for $E_6$ as a unification group comes from superstring theory, where it was shown to emerge as a four-dimensional compactification of the heterotic $E_8\times E_8$ superstring theory~\cite{Candelas:1985en,Witten:1985xc}. In fact, the presence of compactified extra dimensions near the scale of $E_6$ breaking can trigger symmetry breaking of the $E_6$ group via the Hosotani mechanism~\cite{Hosotani:1983xw} straight into $SU(3)\times SU(2)\times U(1) \times U(1) \times U(1)$.

Most of the research in $E_6$ theories has been typically considered only within the scope of supersymmetry, which ties in with their motivation as low-energy limits of superstring theory where spacetime supersymmetry emerges naturally after compactification. Being a simple Lie group, $E_6$ benefits from the prediction of gauge coupling unification in supersymmetry, as did $SU(5)$ and $SO(10)$, which strengthens the motivation. In addition to the rich phenomenology of the MSSM, the $E_6$ model adds quite a few phenomenological predictions on its own, from exotic fermion states to new heavy gauge bosons~\cite{Hewett:1988xc,London:1986dk}.

\section{Selected topics in GUTs}
\label{sec:topics}

Model building in unified theories involves more than the selection of the group theory properties as introduced in Sec.~\ref{sec:basicsonguts}. There are a few outstanding issues that need to be addressed in order to construct a realistic model. Gauge coupling unification is typically one of the most pressing issues, which can often be resolved by intermediate steps in the breaking chain or by the addition of supersymmetry to the theory. In this section we describe the interplay between SUSY and GUTs, as well as other topics such as the generation of light neutrino masses and the unification of the Yukawa couplings.

\subsection{Supersymmetry and GUTs}
\label{sec:SUSYGUTs}

Supersymmetry is a very appealing theory on its own right. It is one of the most aesthetically pleasing extensions of the Standard Model and it has an extremely rich phenomenology that can be readily tested at colliders and other experiments. SUSY GUTs~\cite{Raby:2017ucc} are a conglomerate of the numerous advantages of unified theories and the predicting power of supersymmetry. One of the most attractive features of SUSY is that it can stabilise the electroweak scale against quantum corrections, the so called \textit{hierarchy problem}~\cite{Dimopoulos:1981yj,Witten:1981nf} and provides a mechanism for dynamic electroweak symmetry breaking~\cite{Ibanez:1982fr,Ellis:1982wr}.

In addition, if $R$-parity is conserved~\cite{Farrar:1978xj} the lightest supersymmetric particle (LSP) is stable. Therefore SUSY automatically predicts the existence of a Dark Matter candidate and can easily produce scenarios with the correct relic abundance~\cite{Ellis:1983ew,Jungman:1995df}. 

As previously mentioned, one of the major motivations for SUSY GUTs is that the minimal MSSM model predicts gauge coupling unification at some high scale $\sim 10^{16}$~\cite{Ellis:1990wk}. As was seen in Figure \ref{fig:MSSMRGEs}, just taking the one-loop RGE flow of the gauge couplings, the unification at the GUT scale is fairly successful. The RGEs for the gauge couplings at one-loop have an analytic solution of the form~\cite{Martin:1993zk}
\begin{equation}
 \alpha_i^{-1}(\mu) = \alpha_{GUT}^{-1} + \frac{b_i}{2\pi}\log\left(\frac{M_{GUT}}{\mu}\right)
\end{equation}
where $i=1,2,3$ labels the coupling of the $U(1)$, $SU(2)$ and $SU(3)$ subgroups of the SM gauge group, and $b_i$ are parameters that depend on the field content. For the MSSM these are $b_i=(33/5,1,-3)$. With a degenerate sparticle spectrum at 100 GeV, these one-loop RGEs unify at $M_{GUT} \sim 2.5 \times 10^{16}$ GeV with $\alpha_{GUT} \sim 0.0388$. 

\begin{figure}
\centering
\includegraphics[width=0.7\textwidth]{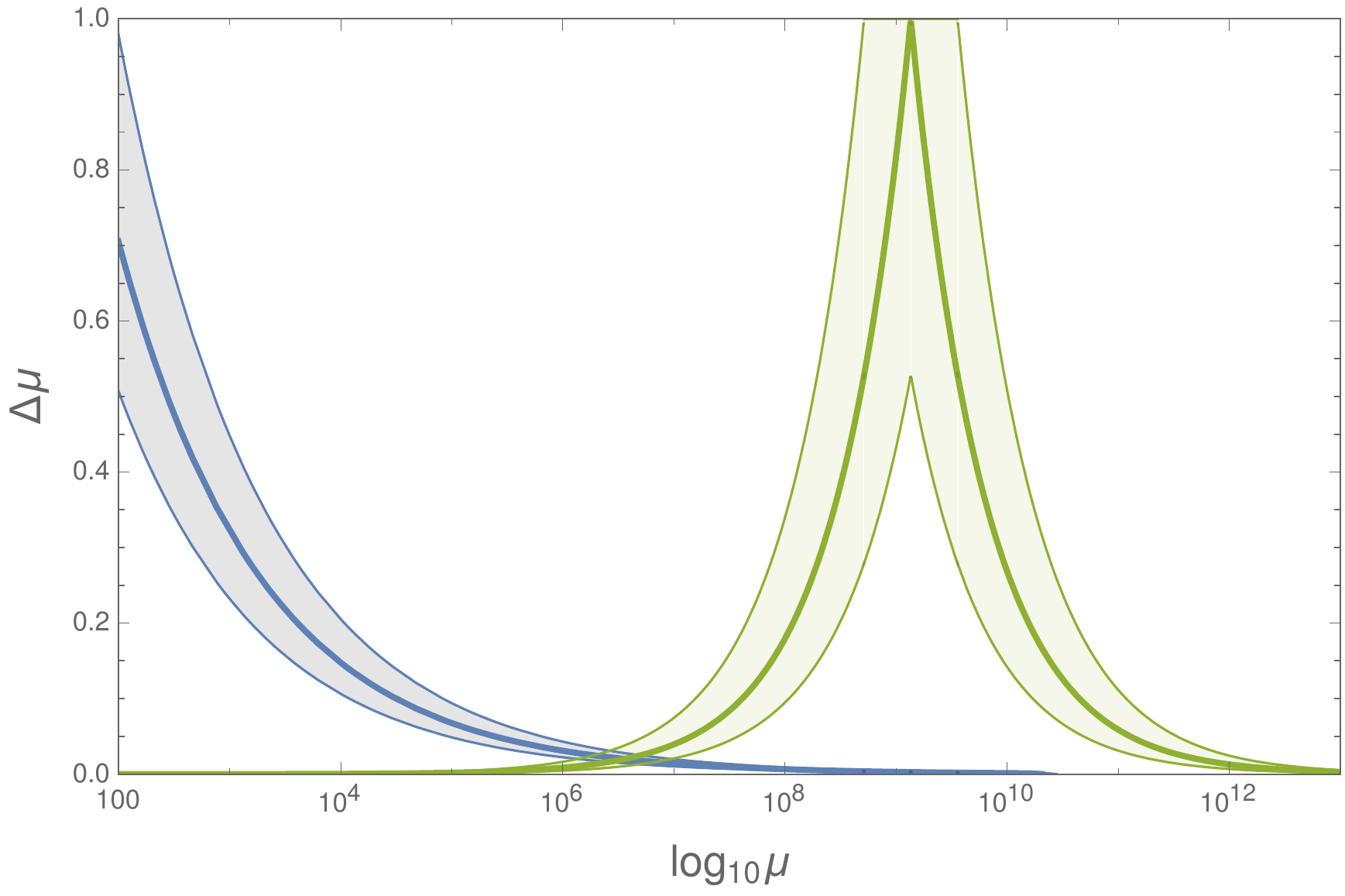}
\caption{\emph{Evolution of the unification measure with the SUSY scale for the MSSM with almost degenerate masses (blue) and scenario with lighter electroweakinos (green), calculated at one-loop. Shaded regions include threshold corrections over the solid lines.}}
\label{fig:SUSYunification}
\end{figure}

This picture, however, relies on a light and almost degenerate supersymmetric spectrum. For heavier or split spectra the situation changes drastically, often spoiling gauge unification altogether. A unification measure can be defined to assess how the unification of gauge couplings changes with the masses of the supersymmetric particles as
\begin{equation}
 \Delta\mu = \frac{\text{min}(\mu_{12},\mu_{23})}{\text{max}(\mu_{12},\mu_{23})},
\end{equation}
where $\mu_{ij}$ is the energy scale at which $\alpha_i^{-1}$ and $\alpha_j^{-1}$ unify. Figure \ref{fig:SUSYunification} shows how the unification measure varies with respect to the SUSY scale for an MSSM model with degenerate SUSY masses (blue). One can distinctly see that for larger sparticle masses, the unification of gauge couplings significantly worsens, from a 70\% unification for $M_{SUSY} \sim 100$ GeV to less than 30\% at $M_{SUSY} \sim 1$ TeV. Consequently, in addition to solving the little hierarchy problem without too much fine tuning, a light sparticle spectrum is clearly preferred to achieve gauge coupling unification. 

In spite of the appeal of SUSY GUTs, the combined effort of several collider experiments has not found any clear evidences of SUSY particles\footnote{See Section~\ref{sec:SUSYsearches} for more details on searches for Supersymmetry.}. Hence, minimal and light versions of the MSSM are in tension with experimental evidence and that makes achieving gauge coupling unification much harder. This tension relaxes slightly once the mass degeneracy condition is forgone. If the sparticle masses vary considerably across the spectrum, it is possible to evade experimental bounds for those masses more strongly constrained (e.g. gluinos, squarks, etc.) while keeping part of the spectrum light. Mass splittings across the supersymmetric spectrum can be taken into account by the contribution of threshold corrections at the SUSY scale, which are of the type~\cite{Hall:1980kf,Weinberg:1980wa}
\begin{equation}
 \lambda_i(M_{SUSY}) = \frac{1}{12\pi} \left( \sum_\phi S_i(\phi) \log\frac{m_\phi}{M_{SUSY}} + 8 \sum_\psi S_i(\psi) \log\frac{m_\psi}{M_{SUSY}}\right),
\end{equation}
with $\phi$ the scalar fields in the MSSM (sfermions), $m_\phi$ their masses and $S(\phi)$ their Dynkin indices; and $\psi$ the fermions in the MSSM (gauginos and Higgsinos), $m_\psi$ and $S(\psi)$ their masses and Dynkin indices. The shaded blue region in Figure \ref{fig:SUSYunification} corresponds to MSSM models with slightly non-degenerate masses. Although these models exhibit the same trend as before, decreasing the unification measure as $M_{SUSY}$ increases, some of these achieve a better gauge coupling unification than the degenerate case, with up to 60\% unification for $M_{SUSY} \sim 1$ TeV.

Supersymmetric models with large splittings in the particle spectrum can modify this picture significantly. A special case, known as Split Supersymmetry~\cite{Giudice:2004tc,ArkaniHamed:2004fb}, has all the sfermions decoupled at the GUT scale and only gauginos and Higgsinos remain light, protected by chiral symmetry. This model is very well motivated within the context of unified theory, because the decoupled fields form full multiplets of $SU(5)$, so the unification of the gauge couplings is not affected~\cite{Giudice:2004tc}. Hence, the behaviour of the unification measure for these models is identical to the semi-degenerate MSSM case from above (blue line and shaded region in Fig.~\ref{fig:SUSYunification}), but has the advantage of allowing a lighter spectrum since some of the strongest experimental constraints are on the squark masses, which are decoupled from the spectrum. These Split-SUSY models, however, predict the existence of a light gluino, which is unfortunately strongly constrained by experiments. Alternative versions of this model with light electroweakinos ($\sim 100$ GeV) and heavy gluinos ($\sim 5$ TeV) have been studied~\cite{Athron:2018vxy}, but these fail to provide successful gauge coupling unification for a light spectrum, slowly improving at larger scales, as can be noticed from the green solid line and shaded region (threshold corrections) in Figure \ref{fig:SUSYunification}. Split-SUSY and the light electroweakino model are just two extreme cases, the former requiring very light spectrum for successful unification and the latter a heavy spectrum. A number of models can be constructed with different spectra that have intermediate predictions for gauge coupling unification. In fact, with a precise analyses of threshold corrections, a number of supersymmetric models with large mass splittings have been shown to achieve exact unification, with a relatively light spectrum~\cite{Ellis:2015jwa}.

\subsection{Neutrino masses}
\label{sec:numasses}

The observed oscillations of neutrino flavours~\cite{Pontecorvo:1967fh, Fukuda:1998mi} require any successful extension of the SM to incorporate non-zero masses for at least two neutrino species. Effectively, these masses are generated by the 5-dimensional Weinberg operator
\begin{align} \label{eq:weinbergop}
\mathcal{O}_{W} = \frac{c_W}{\Lambda}LLHH,
\end{align}
where $c_W$ is the Wilson coefficient, $\Lambda$ denotes the operator's cut-off scale and $L$ and $H$ are the lepton and Higgs doublets, respectively. A typical UV-completion of this effective operator is some kind of seesaw mechanism~\cite{Mohapatra:1979ia, Schechter:1980gr}, which allows to satisfy elegantly the requirement of tiny neutrino mass size. Generally, these neutrino mass schemes assume a presence of new, heavy degrees of freedom, which are ideally motivated by other BSM physics. As has been shown, a number of different seesaw set-ups can be very naturally incorporated within the GUT framework. Provided that only a single type of new particle is added to the SM field content, there are three basic tree-level seesaw types \cite{Ma:1998dn}.

\paragraph{Seesaw Type I}
In the original and simplest seesaw mechanism of type I the right-handed neutrino singlets must be added to the model \cite{Minkowski:1977sc,Yanagida:1979as,Mohapatra:1979ia,GellMann:1980vs}. As the current experimental data require only two neutrinos to be massive, the minimal scenario must include two right-handed neutrino states. This extension then allows to write both Dirac and Majorana neutrino mass terms
\begin{align}\label{eq:lagSeesawI}
\mathcal{L}_{N} &= -\mathrm{y}^\nu \bar{L}^{\ell}\tilde{H}N^{I} - \frac{1}{2}[\mathrm{M}_M]_{IJ} {N^{I}}^TCN^{J} + \text{h.c.},
\end{align}
where $\mathrm{y}^{\nu}$ is the matrix of neutrino Yukawa couplings and $\mathrm{M}_M$ denotes the Majorana mass matrix. Hence, taking $\mathrm{m}_D = \mathrm{y}^{\nu} v$, with $v$ being the SM Higgs vev, the neutrino mass matrix can be written in the usual form
\begin{align}
\mathsf{M} = \begin{pmatrix}
0 & \mathrm{m}_D \\
\mathrm{m}_D^T & \mathrm{M}_M
\end{pmatrix}.
\end{align}
The block-diagonalisation of this matrix leads to the light mass of the oscillating neutrinos
\begin{align}
\mathrm{M}^{\mathrm{I}}_{\nu}=-\mathrm{m}_D \mathrm{M}_M^{-1} \mathrm{m}_D^T,
\end{align}
as the Majorana mass parameter can be chosen to be arbitrarily large. Considering the neutrino Yukawa couplings of order one and the Majorana mass around $10^{14}$ GeV, the desired neutrino mass sizes of order $m_{\nu} \approx 0.1$ eV are obtained. The type-I seesaw mechanism can be implemented in the GUT framework. Particularly, it arises very naturally in $SO(10)$ GUT, where the right-handed neutrino singlet can be accommodated together with all the other fermions in a single $\mathbf{16}_F$ spinor representation.

\paragraph{Seesaw Type II}
The second possibility to construct a seesaw mechanism is to assume a heavy scalar $SU(2)_L$-triplet $\Delta_L$,
\begin{align}\label{eq:Ltriplet}
\{\mathbf{1},\mathbf{3},2\} \equiv \Delta_L = \vecs{\Delta}_L \cdot \vecs{\tau} = \begin{pmatrix}
\frac{1}{\sqrt{2}}\Delta^+ & \Delta^{++} \\
\Delta^0 & -\frac{1}{\sqrt{2}}\Delta^+
\end{pmatrix},
\end{align}
which allows to write the following Lagrangian terms
\begin{align}
\mathcal{L}_{\Delta} = \left[\mathrm{y}_{\ell\ell'}^{\Delta}{L^{\ell}}^T C(i\tau^2)\Delta_L L^{\ell'} + \mu H^T(i\tau^2) \Delta_L^* H + \text{h.c.}\right] + M_{\Delta}^2 \mathrm{Tr}[\Delta\Delta^{\dagger}].
\end{align}
Diagonalisation of the type-II seesaw mass matrix~\cite{Magg:1980ut,Lazarides:1980nt,Mohapatra:1980yp} then generates neutrino mass
\begin{align}
\mathrm{M}_{\nu}^{\mathrm{II}} = \frac{\mu v^2}{M_{\Delta}^2} \mathrm{y}^{\Delta}
\end{align}
and for $M_{\Delta} \gg v$ the required suppression is obtained.

Also this seesaw can be responsible for neutrino mass generation in GUTs. For instance, in $SO(10)$ unification the left-handed scalar triplet is contained by the $\mathbf{126}$ Higgs field, which is usually considered to be present in the scalar sector of the theory. It has been shown that type-II seesaw can be the dominant neutrino mass scheme within both SUSY~\cite{Goh:2004fy} and non-SUSY~\cite{Mohapatra:2011yn} $SO(10)$ GUTs.

\paragraph{Seesaw Type III}
The third option for a UV-completion of the Weinberg operator in Eq.~\eqref{eq:weinbergop} is to introduce new fermionic $SU(2)_L$ triplets $\mathbf{T}_F^{I}$~\cite{Foot:1988aq}\footnote{Similarly to the right-handed neutrino singlets, only two triplets are necessary, although three (one per flavour) are considered here.}. Their interaction with the SM content is analogous to the type I seesaw, namely,
\begin{align}
\mathcal{L}_{T_F} = \mathrm{y}^{T_F}_{\ell J}{L^{\ell}}^T C(i\tau^2)(\vecl{T}_F^{J}\cdot\vecs{\tau})H + \mathrm{M}^{T_F}_{IJ} (\vecl{T}_F^{I})^T C \vecl{T}_F^{J} + \text{h.c.}.
\end{align}
The neutrino mass matrix for type III seesaw then reads
\begin{align}
\mathrm{M}_{\nu}^{\mathrm{III}} = (\mathrm{y}^{T_F})^T v^2 [\mathrm{M}^{T_F}]^{-1}\mathrm{y}^{T_F}
\end{align}
and for $\mathrm{M}^{T_F} \gg \mathrm{y}^{T_F}v$ the smallness of neutrino masses is ensured.

The incorporation of the type-III seesaw mechanism into GUTs has been studied in literature~\cite{Ma:1998dn,Bajc:2006ia,Perez:2007iw}. When implemented within $SU(5)$ models, type-III seesaw comes automatically in hand with the type-I seesaw, as both fields responsible for these mechanisms share the same adjoint representation $\mathbf{24}_F$.

\paragraph{Inverse Seesaw}
At low energies the light neutrino masses can be generated at tree level via the so called inverse seesaw mechanism. This string theory motivated\cite{Witten:1985xc} scheme can be constructed when a non-minimal lepton content of a given theory is assumed. Namely, extra singlet leptons must be added to the model, which is in general allowed for any gauge theory \cite{Schechter:1980gr}. The minimalistic extension of the SM particle content leading to inverse seesaw requires a pair of left-handed two-component lepton singlets $N^{c}$ and $S$ \cite{Mohapatra:1986bd}. Taking three generations of these new singlet fields, one can write the $9\times 9$ mass matrix of the neutral leptons in the basis $\{\nu^{\ell}_L,{N^{I}}^c,S^{A}\}$ (with $A=a,b,c$) as
\begin{align}\label{eq:seesawIS}
\mathsf{M}^{\text{IS}} = \begin{pmatrix}
0 & \mathrm{m}_D & 0 \\
\mathrm{m}_D^T & 0 & \mathrm{M} \\
0 & \mathrm{M}^T & \mu
\end{pmatrix},
\end{align}
where M and $\mu$ are the mass matrices corresponding to the $SU(2)_L$ singlets, while $\mathrm{m}_D$ is the Dirac neutrino mass matrix as usual. As predicted by some string models, the Majorana mass entries corresponding to $\nu_L$ and $N$ are zero. Thus, the only Majorana mass parameter is the matrix $\mu$, which corresponds to the extra singlet $S$. This entry is then responsible for lepton number violation. If $\mu$ is set to be zero, the $B-L$ symmetry is restored, the matrix $\mathsf{M}^{\text{IS}}$ degenerates and the three oscillating neutrinos become massless.

On the other hand, for non-vanishing $\mu$ such that $\mu \ll \mathrm{m}_D \ll \mathrm{M}$ the resulting mass matrix of the light neutrino eigenstates reads
\begin{align}
\mathrm{M}^{\text{IS}}_{\nu} = \mathrm{m}_D \mathrm{M}^{-1}\mu [\mathrm{M}^T]^{-1}\mathrm{m}_D^T.
\end{align}
The main difference from the standard seesaw scenarios is that in the present case neutrinos become light for $\mu \rightarrow 0$, not for large values of Majorana mass parameter. This is also the reason why one talks about `inverse' seesaw. As vanishing $\mu$ enhances the symmetry of the theory, the assumption of its small value can be considered to be natural~\cite{GonzalezGarcia:1988rw,tHooft:1979rat}.

\paragraph{Linear Seesaw}
A particularly interesting realisation of the inverse seesaw mechanism can be constructed within the $SO(10)$ GUTs framework with broken D-parity \cite{Malinsky:2005bi}. The so called linear seesaw mechanism consists in extending the minimal fermionic content of the $SO(10)$ model, contained by three copies of the $\mathbf{16}_F$ representation, by three gauge singlets $S^{A}$. The original version of this scheme was designed within the supersymmetric $SO(10)$ framework; however, it can be constructed also in non-supersymmetric scenarios. The mass matrix for the neutral fermions in the basis $\{\nu^{\ell}_L,{N^{I}}^c,S^{A}\}$ has the following form
\begin{align}\label{eq:seesawLS}
\mathsf{M}^{\text{LS}} = \begin{pmatrix}
0 & \mathrm{m}_{D} &  \mathrm{m}_L \\
\mathrm{m}_{D}^T & 0 & \mathrm{M} \\
\mathrm{m}_L^T & \mathrm{M}^T & 0
\end{pmatrix}.
\end{align}
Here, $\mathrm{m}_D$ denotes the Dirac neutrino mass, $\mathrm{M}$ is the heaviest Dirac neutrino mass term mixing $N$-$S$ and $\mathrm{m}_L$ stands for the small term mixing $\nu$-$S$, which breaks the $(B-L)$ symmetry. The light neutrino masses are then given by the expression
\begin{align}
\mathrm{M}_{\nu}^{\mathrm{LS}} \simeq \mathrm{m}_D^T\mathrm{M}^{-1}\mathrm{m}_L + (\mathrm{M}^{-1}\mathrm{m}_L)^T\mathrm{m}_D,
\end{align}
which depends linearly on $\mathrm{m}_D$ (and therefore also on corresponding Yukawa couplings). In the present scenario it is the large parameter $M$ given by the unification scale what ensures the smallness of neutrino masses. Hence, the lightness of neutrinos is independent of the $(B-L)$ symmetry breaking scale, which can consequently lie at low, experimentally accessible energies without spoiling the desired size of neutrino masses or the unification.

\paragraph{Other Neutrino Mass Models}
Despite the success of seesaw mechanisms, one can think of a number of alternative neutrino mass generation schemes. From the phenomenological point of view, these can be even more interesting, as they often predict (unlike the three usual seesaws) a low-energy origin of neutrino masses. The light neutrino masses are obtained using a small lepton-number-violating parameter (similarly as in the inverse seesaw), or they can be suppressed by loops and small Yukawa couplings. While the former option can be realised e.g. within supersymmetric models with R-parity breaking \cite{Hirsch:2004he}, the latter possibility refers to the models of neutrino mass generation via calculable radiative corrections (i.e. the Zee mechanism)~\cite{Zee:1980ai,Babu:1988ki}. A two-loop mechanism generating neutrino masses within a minimal $SO(10)$ GUT was identified by Witten~\cite{Witten:1979nr} and the same scheme can be constructed also in the flipped $SU(5)$ context~\cite{Leontaris:1991mq,Rodriguez:2013rma,Harries:2018tld}.

\subsection{Yukawa coupling unification and fermion masses}
\label{sec:yukawaunification}

In fully unified theories, such as $SU(5)$ and $SO(10)$, the gauge couplings must unify at some high energy scale. This is typically achieved automatically in SUSY GUTs, as mentioned above, due to the RGE flow of the MSSM gauge couplings (c.f. Fig.\ref{fig:MSSMRGEs}), but it can also be achieved through the addition of new scalar representations~\cite{Giveon:1991zm} or with a multi-step symmetry breaking pattern~\cite{Aulakh:1982sw}.

\begin{figure}
\centering
\includegraphics[width=0.45\textwidth]{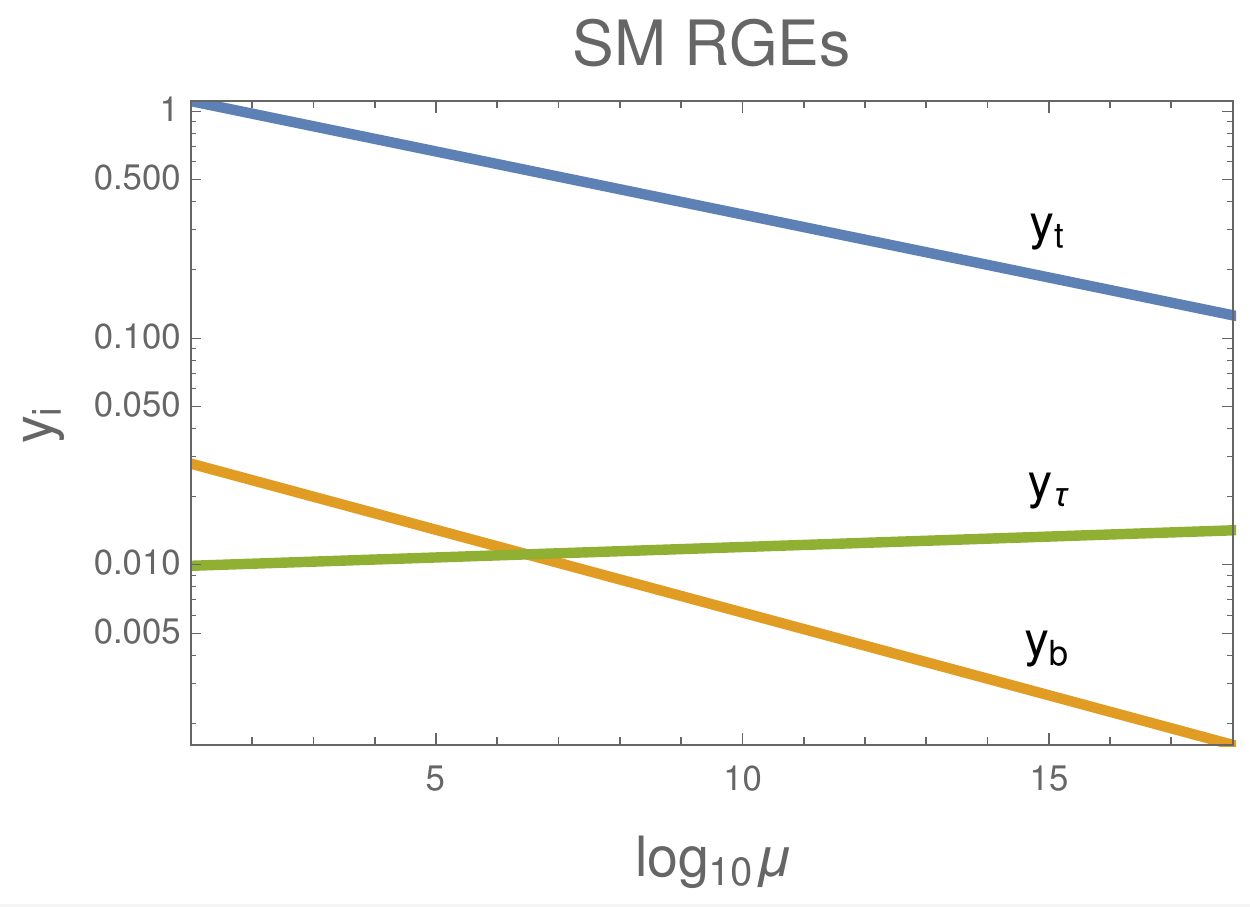}
\includegraphics[width=0.45\textwidth]{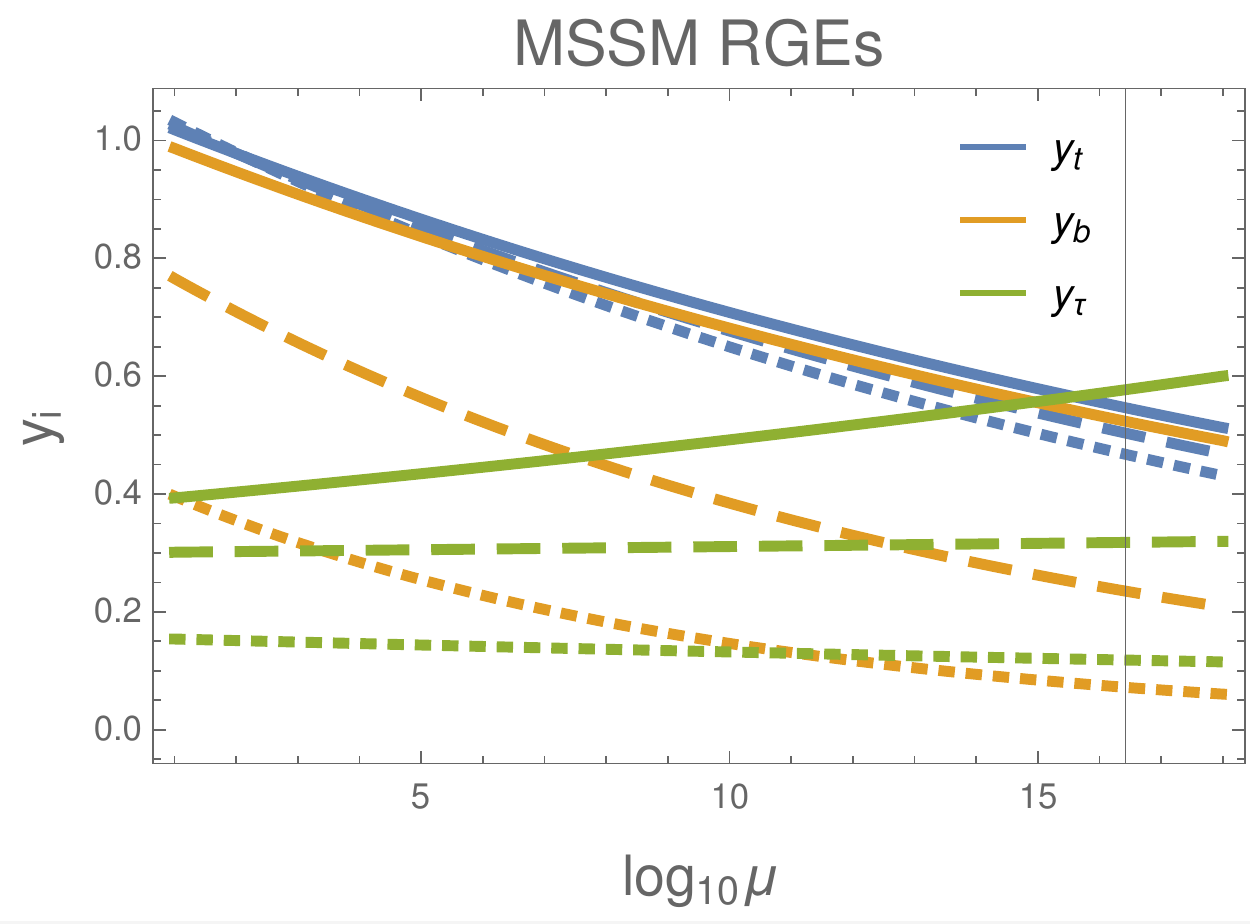}
\caption{\emph{One loop renormalisation group flow of the SM (left) and MSSM (right) Yukawa couplings, with $m_0 = 2$ TeV, $m_{1/2} = 3$ TeV, $A_0 = 0$ and $\tan\beta = 40$ (solid), $\tan\beta = 30$ (dashed) and $\tan\beta = 15$ (dotted).}}
\label{fig:YukawaRGEs}
\end{figure}

Along gauge coupling unification, $SU(5)$ and $SO(10)$ models also require the unification of the Yukawa couplings. The largest hierarchy on the fermion masses happens in the third generation where $m_t/m_b \sim 40$ and $m_b/m_\tau \sim 2.3$, hence Yukawa unification in GUTs is always studied within the third generation only. In $SU(5)$ the charged leptons live in the same representation as the down-type quarks, $\mathbf{\bar{5}}$, and as such it is expected that at the GUT scale $y_b = y_\tau$, whereas in $SO(10)$ all SM fermions are embedded into the same 16-dimensional representation, so the unification condition becomes $y_t = y_b = y_\tau$.

Although a natural prediction of GUTs, Yukawa unification is not easily achieved in the vanilla $SU(5)$ and $SO(10)$ models~\cite{Chanowitz:1977ye,Buras:1977yy,Georgi:1979df}. As can be seen in the left-side plot of Figure \ref{fig:YukawaRGEs}, the Yukawa couplings in the SM are far from unification. In spite of this, a few successful attempts to solve the unification of $y_b$ and $y_\tau$ in $SU(5)$ inspired models have been performed, either by including large scalar $SU(5)$ representations to the field content~\cite{Georgi:1979df,Giveon:1991zm,Antusch:2013rxa} or by adding Planck scale suppressed interaction of the Higgs field to the SM fermions~\cite{Ellis:1979fg}.

In SUSY GUTs, however, Yukawa unification can often be achieved in some regions of the full supersymmetric parameter space. As can be seen in the right-side picture in Figure \ref{fig:YukawaRGEs}, the Yukawa couplings in the MSSM tend to run towards convergence at high scales, and they can be seen to almost unify for large values of $\tan\beta$~\cite{Kelley:1991aj,Ananthanarayan:1991xp,Antusch:2008tf,Antusch:2009gu}. This occurs because the third generation fermion masses depend on $\tan\beta$ in the following way~\cite{Babu:1998er}
\begin{equation}
 m_t = \frac{v}{\sqrt{2}} y_t \sin\beta,\quad m_b = \frac{v}{\sqrt{2}} y_b \cos\beta, \quad m_\tau = \frac{v}{\sqrt{2}} y_\tau \cos\beta,
\end{equation}
which can realise the hierarchy $m_t \gg m_b,m_\tau$ even in $SO(10)$ or $E_6$ models where one expects $y_t \sim y_b \sim y_\tau$. These solutions with large $\tan\beta$ can spoil radiative EW symmetry breaking in unified models, since $B\mu \simeq \frac{M_A^2}{\tan\beta} \simeq 0$ implies that $m_{H_d}^2 - m_{H_u}^2 > m_Z^2$~\cite{Olechowski:1994gm}, contrary to the unified picture where $m_{H_u}^2 = m_{H_d}^2$. This issue can often be resolved either by \textit{ad hoc} splittings of the Higgs soft masses at the GUT scale, or by considering the effect of $D$-terms in the boundary conditions at the GUT scale~\cite{Blazek:2002ta}, which naturally imposes a splitting of $m_{H_d}^2 - m_{H_u}^2 = 4 m_D^2$.

\begin{figure}[t]
 \centering
 \includegraphics[width=0.8\textwidth]{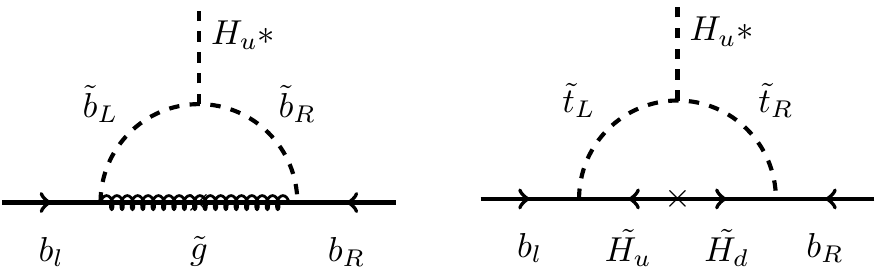}
 \caption{\emph{One-loop radiative corrections to $m_b$.}}
 \label{fig:mbcorrections}
\end{figure}

In addition to satisfying $m_t \gg m_b,m_\tau$ for unified Yukawa couplings, one can lift the hierarchy between $m_b$ and $m_\tau$ with the inclusion of radiative corrections on the $b$ mass. At one loop the $b$ quark couples to $H_u$ via a gluino or Higgsino loop~\cite{Babu:1998er}, as can be seen in Figure~\ref{fig:mbcorrections}, which adds a correction to $m_b$ of the type~\cite{Hall:1993gn}
\begin{equation}
 \delta m_b \simeq \frac{v}{\sqrt{2}} y_b \frac{\sin\beta}{16\pi^2}\left(\frac{8}{3}g^2_3 \frac{\mu m_{\tilde{g}}}{m_{\tilde{b}^2}} + y_t^2 \frac{\mu A_t}{m_{\tilde{t}^2}}\right).
 \label{mbcorrections}
\end{equation}

Though similar corrections appear for $m_t$ and $m_\tau$, they are negligible compared to $\delta m_b$. The correction on $m_t$ is not proportional to $\tan\beta$, which is required to be large to satisfy $t-b-\tau$ unification. Further, $\delta m_\tau$ does not have a gluino loop and the Higgsino contribution is inversely proportional to $m_{\tilde{\nu_t}}$ which is typically much larger than $m_{\tilde{t}}$, and therefore the contribution is small. These radiative corrections on $m_b$ are proportional to $\tan\beta$ and therefore can be significant, up to $50\%$~\cite{Blazek:2002ta}, which can spoil the hierarchy $m_t \gg m_b$. There are regions of the SUSY parameter space, however, where it is possible to reduce $\delta m_b$ while keeping $\tan\beta$ large~\cite{Hall:1993gn,King:2000vp,Blazek:2002ta}, thereby successfully predicting $t-b-\tau$ unification, even factoring in LHC searches~\cite{Baer:2012cp}.

In SUSY $SU(5)$ models the more straightforward boundary condition $y_b = y_\tau$ is imposed. It was found that, in addition to the large $\tan\beta$ scenarios from above, $b-\tau$ unification can also be achieved in a region of parameter space with low $\tan\beta$~\cite{Langacker:1993xb,Langacker:1994bc}. However, a sufficiently low $\tan\beta$ might struggle to lift sufficiently the tree level Higgs mass to the observed value, and hence there remains some tension between unified $b-\tau$ models of low $\tan\beta$ and the observed Higgs mass~\cite{Baer:2012by}.

A number of other mechanisms have been proposed to satisfy the Yukawa unification conditions. Intermediate breaking steps, such as the Pati-Salam group, can modify the Yukawas RGEs in a favourable manner achieving quasi-unification~\cite{Gomez:2002tj,Gogoladze:2010fu}. Non-canonical seesaw mechanisms in neutrino models require $b-\tau$ unification to match the observed neutrino mixings~\cite{Bajc:2002iw}. Or the inclusion of certain higher dimensional operators can successfully yield Yukawa unification~\cite{Anderson:1993fe}.

Beyond the unification of the Yukawa couplings for each of the families of SM fermions, the mass hierarchies among the different families remains an open question. Although GUTs by themselves do not make predictions on the nature of this hierarchy, they often include a fair amount of parameters and mixing matrices that are unconstrained and can fit the fermion masses. Additionally GUTs are often extended with family symmetries, continuous or discrete, which can, with a smaller set of parameters, accurately predict the fermion mass hierarchies, as well as their mixings and CP phases encoded in the CKM and UPMNS matrices~\cite{Altarelli:2010gt,Ishimori:2010au,Grimus:2011fk,King:2012in,deAnda:2018yfp}. We will not discuss family symmetries any further since they fall beyond the scope of this work.

\section{Modern day GUTs}
\label{sec:models}

Since their first appearance in the late 70s, a large number of GUT models have been proposed. These vary according to the symmetry group employed, the symmetry breaking mechanism and field content among others. Some of them were driven by the experimental results of the time and other by new theoretical insights. In this section we attempt to outline a small, non-exhaustive, subset of GUT models, aiming to explore those with strong phenomenological consequences and some that have been in the spotlight in recent years. We thus focus on left-right symmetric models, SUSY $SO(10)$, trinification models and $\mathrm{E}_6$SSM.

\subsection{Left-right symmetric models}
\label{sec:lr}

One of the minimal extensions of the SM is the earlier mentioned left-right symmetric model \cite{Mohapatra:1974hk, Pati:1974yy, Mohapatra:1974gc, Senjanovic:1975rk, Mohapatra:1980yp}. Despite not being real GUTs, LR models can very conveniently play the role of an intermediate symmetry restored between the electroweak scale and the GUT scale~\cite{Lindner:1996tf,Arbelaez:2013nga,Deppisch:2017xhv}. The LR framework has attracted a lot of attention particularly in connection with the LHC~\cite{Senjanovic:2010nq,Nemevsek:2011hz,Han:2012vk,Das:2012ii,Bambhaniya:2013wza,Chen:2013fna,Dev:2013oxa,Deppisch:2014qpa,Deppisch:2014zta,Deppisch:2015cua,Dev:2015pga,Mitra:2016kov,Helo:2018rll}, as it typically predicts new physics at energies that can be probed by the collider searches.

The fermionic particle content of LR models is given by a straightforward LR symmetric extension of the SM content, i.e. the right-handed doublets are introduced
\begin{align}
L_R^{\ell} = \begin{pmatrix}
N^{\ell} \\
\ell_R
\end{pmatrix}
&\leftrightarrow
\begin{pmatrix}
\nu_L^{\ell} \\
\ell_L
\end{pmatrix} = L^{\ell},
\\
Q^{i}_R = \begin{pmatrix}
u^{i}_R \\
d^{i}_R
\end{pmatrix}
&\leftrightarrow
\begin{pmatrix}
u_L^i \\
d_L^i
\end{pmatrix} = Q^{i}.
\end{align}
As a result, right-handed neutrinos are naturally included making the left-handed neutrinos acquire mass in the LR models, which is a highly desirable feature of a BSM model. The presence of the right-handed neutrino partners is also essential for cancellation of the $B-L$ gauge anomaly.

The Higgs sector of LR symmetric theories can vary. The minimal scenarios mostly include a scalar bi-doublet\footnote{Here, the representations are labelled the usual way in the order $\{SU(3)_C, SU(2)_L, SU(2)_R, U(1)_{B-L}\}$.}
\begin{align}
\Phi \equiv \{\vecl{1},\vecl{2},\vecl{2},0\} = \begin{pmatrix}
\phi^0_1 & \phi^+_2 \\
\phi^-_1 & \phi^0_2
\end{pmatrix},
\end{align}
containing the SM Higgs, which subsequently gives masses to quarks and leptons. The corresponding vev reads
\begin{align}
\langle \Phi \rangle = \begin{pmatrix}
v_{\Phi 1} & 0 \\
0 & v_{\Phi 2}
\end{pmatrix},
\end{align}
where $v\equiv \sqrt{v_1^2 + v_2^2}$ and it mixes the left-handed and right-handed gauge bosons as described below.

Besides the bi-doublet, typically a pair of scalar triplets 
\begin{align}
\Delta_L \equiv \{\mathbf{1},\mathbf{3},\mathbf{1},-2\},\quad \Delta_R \equiv \{\mathbf{1},\mathbf{1},\mathbf{3},-2\},
\end{align}
or doublets 
\begin{align} \label{eq:LRdoublets}
\chi_L \equiv \{\mathbf{1},\mathbf{2},\mathbf{1},-1\},\quad \chi_R \equiv \{\mathbf{1},\mathbf{1},\mathbf{2},-1\},
\end{align}
must be added to the Higgs sector in order to break the LR gauge group to the SM.
In fact, the right-handed scalar is enough to do so, but inclusion of the left-handed triplet (or doublet) preserves the LR symmetry (so called ``manifest LR symmetry''), i.e. the $SU(2)_L$ and $SU(2)_R$ gauge couplings are equal: $g_L=g_R$.

If no additional fermions besides the SM fermionic content are considered, at least two bi-doublets must be present in the scalar sector to account for the correct SM flavour physics~\cite{Arbelaez:2013nga}. In a model with a single bi-doublet the Yukawa Lagrangian implies that the up-quark mass matrix is proportional to the down-quark mass matrix (independently of the vev structure); thus, the CKM matrix becomes trivial $V_{\text{CKM}}=\mathtt{1}$.

Consequently, the LR symmetry breaking takes place in two steps. At first, the neutral component of right-handed scalar triplet (or doublet) gets the vev $v_R$ and breaks the LR gauge group to the SM gauge group. Subsequently, the bi-doublet acquiring its vev breaks the SM gauge group to $SU(3)_C\otimes U(1)_Q$. Based on the observations it can be assumed that $v_R \gg v_{\Phi 1}, v_{\Phi 2}$.

Depending on the scalar content of a particular LR model, different ways of light neutrino mass generation can be employed. Having right-handed neutrino singlets means that type-I seesaw is always the option. In general, the neutrino mass matrix can take the form
\begin{align}\label{eq:LRmassmatrix}
\mathsf{M}_{\nu} = \begin{pmatrix}
\mathrm{M}_{M,L} & \mathrm{m}_D \\
\mathrm{m}_D^T & \mathrm{M}_{M,R}
\end{pmatrix},
\end{align}
where $\mathrm{m}_D$ denotes the Dirac mass matrix, while $\mathrm{M}_{M,L}$ and $\mathrm{M}_{M,R}$ are the Majorana mass matrices corresponding to the left-handed and right-handed neutrinos, respectively.

The Yukawa couplings in LR models include the scalar bi-doublet,
\begin{align}
\mathcal{L}^{\Phi}_{\text{Yukawa}} = \mathrm{y}_{\ell\ell'}^{\Phi} {L^{\ell}}^T C \Phi L_R^{\ell'} + \tilde{\mathrm{y}}_{\ell\ell'}^{\Phi} {L^{\ell}}^T C \tilde{\Phi} L_R^{\ell'} + \text{h.c.},
\end{align}
where $\tilde{\Phi} = \sigma^2 \Phi^* \sigma^2$. Then the Dirac neutrino mass matrix and the mass matrix of charged leptons are in the broken phase given by
\begin{align}
\mathrm{m}_D &= \mathrm{y}^{\Phi}v_{\Phi 1} + \tilde{\mathrm{y}}^{\Phi}v_{\Phi 2}, \\
\mathrm{m}_{\ell} &= \mathrm{y}^{\Phi}v_{\Phi 2} + \tilde{\mathrm{y}}^{\Phi}v_{\Phi 1}.
\end{align}

In case that the right-handed scalar triplet $\Delta_R$ is responsible for the LR symmetry breaking, one can write also the Yukawa couplings for the right-handed lepton doublet in the form
\begin{align}
\mathcal{L}^{\Delta_R}_{\text{Yukawa}} = \frac{1}{2} \mathrm{y}^{\Delta_R}_{\ell\ell'}(L_{R}^{\ell})^T C(i\tau^2)\Delta_R L_R^{\ell'} + \text{h.c.},
\end{align}
where $\Delta_R = \vecs{\Delta}_R\cdot\vecs{\tau}$. After the triplet acquires its vev
\begin{align}
\langle \Delta_{R} \rangle = \begin{pmatrix}
0 & 0 \\
v_R & 0
\end{pmatrix},
\end{align}
the LR symmetry is broken and the right-handed neutrino receives Majorana mass $\mathrm{M}_{M,R}=\mathrm{y}^{\Delta_R}v_R \gg v$, which allows for type-I seesaw mechanism.

When the Higgs sector contains also the left-handed scalar triplet $\Delta_L$ with vev
\begin{align}
\langle\Delta_L\rangle = \begin{pmatrix}
0 & 0 \\
v_L & 0
\end{pmatrix},
\end{align}
it generates the left-handed Majorana mass matrix $\mathrm{M}_{M,L} = \mathrm{y}^{\Delta_L}v_L$ switching on type-II seesaw mechanism.

In principle, the type-I and type-II seesaws can be combined giving the ``full'' seesaw matrix \eqref{eq:LRmassmatrix}. The resulting light neutrino mass matrix reads
\begin{align}\label{eq:seesawI+II}
\mathrm{M}_{\nu}^{\text{I+II}} = \mathrm{M}_{M,L} - \mathrm{m}_D[\mathrm{M}_{M,R}]^{-1}\mathrm{m}_D^T.
\end{align}
Specifically, if $v_{\Phi 2}=0$ is assumed for simplicity, then the formula~\eqref{eq:seesawI+II} can be rewritten in terms of the parameters of the LR models as
\begin{align}
\mathrm{M}_{\nu}^{\text{LR}} = \mathrm{y}^{\Delta_L}v_L - \frac{v^2_{\Phi 1}}{v_R}\mathrm{y}^{\Phi}[\mathrm{y}^{\Delta_R}]^{-1}\mathrm{y}^{\Phi T}.
\end{align}
Hence, if the hierarchy $v_R \gg v_{\Phi 1} \gg v_L$ is satisfied, the neutrino masses become small.

In models with the LR symmetry breaking driven by the right-handed doublet $\chi_R$ instead of the triplet $\Delta_R$ the light neutrino masses can be obtained employing the inverse~\cite{Mohapatra:1986bd,Mohapatra:1986aw,Dev:2009aw,Brdar:2018sbk} and/or linear~\cite{Akhmedov:1995ip,Akhmedov:1995vm,Deppisch:2015cua} seesaw mechanisms, provided that a singlet fermion $\{\mathbf{1},\mathbf{1},\mathbf{1},0\}$ is added to the model particle content. Alternatively, it is also possible to construct the type-III seesaw mechanism, if a left-handed or right-handed fermionic triplet is present within the LR model~\cite{FileviezPerez:2008sr,Duerr:2013opa}. Lastly, neutrino mass generation in LR models via the Zee mechanism can be achieved with the addition of a charged scalar boson $\{\mathbf{1},\mathbf{1},\mathbf{1},2\}$~\cite{FileviezPerez:2016erl}.

\subsection{SUSY $SO(10)$ models}

Supersymmetric $SO(10)$ models are rather appealing GUTs, for they combine together the advantages of SUSY, Pati-Salam and $SU(5)$ models. As briefly outlined in Section~\ref{sec:so10}, $SO(10)$ models unify all fermions of a generation in the SM into a single representation, of dimension $\mathbf{16}$. This decomposes into the maximal subgroups as
\begin{equation}
 \begin{array}{cll}
  \mathbf{16} & \to \{\mathbf{4},\mathbf{2},\mathbf{1}\} + \{\mathbf{\bar{4}},\mathbf{1},\mathbf{2}\}, & [SU(4)_c \times SU(2)_L \times SU(2)_R], \\
  \mathbf{16} & \to \mathbf{10}_{-1} + \mathbf{\bar{5}}_{3} + \mathbf{1}_{-5},  & [SU(5)\times U(1)].
\end{array}
\end{equation}

As mentioned before, at the renormalizable level the Yukawa sector of $SO(10)$ includes the Higgs representations $\mathbf{10}$, $\mathbf{120}$ and $\mathbf{\overline{126}}$, which are promoted to superfields in SUSY $SO(10)$. Hence, the superpotential of the Yukawa sector is
\begin{equation}
 W_Y = \mathbf{16}^T \left( \mathsf{Y}_{10} \mathbf{10} + \mathsf{Y}_{120} \mathbf{120} + \mathsf{Y}_{126} \mathbf{\overline{126}} \right) \mathbf{16}.
\end{equation}
where $\mathsf{Y}_i$ are matrices of Yukawa couplings in family space. One of the most remarkable features of SUSY $SO(10)$ is that, starting from a Yukawa unified scenario, $\mathsf{Y}_{10}$ and $\mathsf{Y}_{126}$ are sufficient to reproduce the full mass spectrum of SM fermions, along with the measured values of mixings in the quark and neutrino sector~\cite{Babu:1992ia, Albright:2000dk,Fukuyama:2002ch, Goh:2003hf, Goh:2003sy, Bertolini:2004eq, Matsuda:2000zp,Matsuda:2001bg,Babu:2005ia, Dueck:2013gca,Deppisch:2018flu}. The mass matrices of SM fermions $\mathsf{M}_i$ can be written as~\cite{Bertolini:2006pe}
\begin{align}
 \mathsf{M}_d &= v_{10}^d \mathsf{Y}_{10} + v_{126}^d \mathsf{Y}_{126} \notag \\
 \mathsf{M}_u &= v_{10}^u \mathsf{Y}_{10} + v_{126}^u \mathsf{Y}_{126} \notag \\
 \mathsf{M}_l &= v_{10}^d \mathsf{Y}_{10} - 3 v_{126}^d \mathsf{Y}_{126} \notag \\
 \mathsf{M}_D &= v_{10}^u \mathsf{Y}_{10} - 3 v_{126}^u \mathsf{Y}_{126} \notag \\
 \mathsf{M}_L &= v_{L} \mathsf{Y}_{126} \notag \\
 \mathsf{M}_R &= v_{R} \mathsf{Y}_{126}
\end{align}
where $\mathsf{M}_D$, $\mathsf{M}_L$ and $\mathsf{M}_R$ are the Dirac and Majorana masses in types I and II seesaw (c.f. Sec.~\ref{sec:numasses}), and the $v$'s are the various vacuum expectation values of $\mathbf{10}$, $\mathbf{\overline{126}}$ and the left and right-handed $SU(2)$ triplets.

The minimal SUSY $SO(10)$ model therefore contains the Higgs superfields $\mathbf{10}$ and $\mathbf{\overline{126}}$, responsible for fermion masses, and a pair of representations $\mathbf{126}$ and $\mathbf{210}$ which trigger the symmetry breaking of $SO(10)$~\cite{Aulakh:2003kg,Aulakh:2004hm,Bajc:2004xe}. Although quite appealing due to its minimal set of model parameters, this model does not achieve the right level of gauge coupling unification and suffers from rapid proton decay~\cite{Bajc:2005qe,Aulakh:2005mw}. 

Many solutions have been implemented to resurrect minimal $SO(10)$ models. The spectrum of soft masses in the low energy MSSM strongly affects the outcome of gauge coupling unification, as was seen in Sec.~\ref{sec:SUSYGUTs}, hence modifications on the hierarchy of soft masses can help towards solving the issues with SUSY $SO(10)$ models~\cite{Bajc:2008dc,DeRomeri:2011ie,Deppisch:2014aga,Poh:2015wta,Ellis:2018khn}. Additionally, extended scalar sectors, either containing a $\mathbf{120}$~\cite{Dutta:2004zh,Aulakh:2008sn} or a $\mathbf{54}$~\cite{Fukuyama:2004ps} representation, can increase the unification scale through strong threshold effects, thereby alleviating the constraint of nucleon decay. Recently it has been shown that a combination of new Higgs representations with a modified spectrum of soft masses can accommodate gauge coupling unification and nucleon decay constraints, while still being able to predict a suitable spectrum of fermion masses~\cite{Babu:2018tfi}.

As with many GUT models, SUSY $SO(10)$ makes predictions that can be tested in a number of different fronts. Collider searches at the LHC~\cite{BhupalDev:2010he,Anandakrishnan:2013nca} as well as dark matter searches~\cite{Baer:2008jn} can discover the predicted light, TeV-scale, states. Precision tests such as nucleon decays~\cite{Goh:2003nv, Mohapatra:2018biy}, lepton flavour violation~\cite{Masiero:2002jn} and flavour observables~\cite{Dermisek:2005sw} can probe the validity of the models at high scales. For more details on probing SUSY $SO(10)$ and GUTs in general see Sections~\ref{sec:cosmo}-\ref{sec:precision}.

\subsection{Trinification}

As a maximal subgroup of $E_6$, the trinification gauge group $SU(3)_c \times SU(3)_L \times SU(3)_R$ is an alternative approach to SUSY $SO(10)$ on the road to $E_6$ unification. The matter content of trinification models per generation typically looks like~\cite{Babu:1985gi}
\begin{equation}
 \{\mathbf{1},\mathbf{3},\mathbf{\bar{3}}\} = \left(\begin{matrix}h_{11} & h_{22} & e \\ h_{21} & h_{22} & \nu \\ e^c & \nu^c & \phi  \end{matrix}\right), \quad
 \{\mathbf{3},\mathbf{\bar{3}},\mathbf{1}\} = \left(\begin{matrix}u & d & D\end{matrix}\right), \quad
 \{\mathbf{\bar{3}},\mathbf{3},\mathbf{1}\} = \left(\begin{matrix}u^c & d^c & D^c\end{matrix}\right)
\end{equation}
where $h_{ij}$ are the components of two Higgs doublets, $\phi$ a SM singlet field and $D$ and $D^c$ colour-triplets. An additional $Z_3$ symmetry is often considered to make the gauge couplings unify at the GUT scale, $g_c = g_L = g_R$.

Given the presence of exotic fields in the matter multiplets, trinification models struggle to trigger spontaneous symmetry breaking without making the matter content impossibly heavy. Additional Higgs multiplets~\cite{Sayre:2006ma,Stech:2012zr,Willenbrock:2003ca,Wang:1992hu} and/or non-renormalizable operators~\cite{Cauet:2010ng,Dvali:1994wj,Carone:2005ha,Nath:1988xn} are usually introduced to alleviate this issue. Unfortunately these models tend to produce tension with current limits on proton decay~\cite{Maekawa:2002qv} and collider searches~\cite{Hetzel:2015bla,Stech:2014tla}.

The fundamental challenge behind the issues of trinification is the complicated vacuum structures and the large number of parameters they depend on. Solutions to this problem involve the addition of family symmetries which reduce the number of parameters and facilitate the study of the symmetry breaking vacuum~\cite{Camargo-Molina:2016bwm}, further simplified by embedding the theory into larger dimensional groups such as $E_8$~\cite{Camargo-Molina:2016yqm,Camargo-Molina:2017kxd}.

\subsection{$\mathrm{E}_6$SSM}

The Exceptional Supersymmetric Standard Model (ESSM or $\mathrm{E}_6$SSM)~\cite{King:2005jy,King:2005my} is an extension of the MSSM motivated as a low energy effective theory from a $E_6$ unified GUT model at high scales. At low scales it has the gauge group $SU(3)_c \times SU(2)_L \times U(1)_Y \times U(1)_N$, where the additional $U(1)_N$ factor is leftover from the symmetry breaking of $E_6$. All the superfields in the theory are contained in three copies of the $\mathbf{27}$ representation of $E_6$, which decompose under the $SU(5)\times U(1)_N$ subgroup as~\cite{Athron:2010zz}
\begin{equation}
 \mathbf{27}^i \to \mathbf{10}^i_1 + \mathbf{\bar{5}}_2^i + \mathbf{\bar{5}}_{-3}^i + \mathbf{5}^i_{-2} + \mathbf{1}_5^i + \mathbf{1}_0^i,
\end{equation}
where $\mathbf{10}_1^i$ and $\mathbf{\bar{5}}_2^i$ are the matter multiplets for all three generations, with the standard embeddings of matter fields in $SU(5)$ (c.f. Sec.~\ref{sec:su5}). The superfields $\mathbf{\bar{5}}_{-3}^i$ and $\mathbf{5}_{-2}^i$ contain the two Higgs doublets of the MSSM, $H_u$ and $H_d$, plus two copies of pairs of exotic doublets, $H_u^{1,2}$ and $H_d^{1,2}$ and three copies of exotic triplets $D_i$ and $\bar{D}_i$. Lastly, the singlets $\mathbf{1}_5^i$ and $\mathbf{1}_0^i$ correspond to exotic singlet fields $S_i$, responsible for $U(1)_N$ breaking at low scales, and right-handed neutrino fields $N_i$, respectively.

Anomaly cancellation of the $U(1)_N$ factor in the $\mathrm{E}_6$SSM model is guaranteed so long as the only decoupled state is the singlet neutrino field. $N_i$ can be as large as necessary to provide light neutrino masses through type I seesaw mechanism and generate the baryon assymmetry of the Universe via leptogenesis~\cite{King:2008qb,Nevzorov:2017gir}. The remaning fields of the $\mathbf{27}^i$ multiplets charged under $U(1)_N$ remain at energies below the breaking of $U(1)_N$ and hence anomalies cancel. Light coloured states have dangerous consequences, however, for they can mediate baryon and lepton number violating interactions leading to rapid proton decay. In order to avoid that, the original $\mathrm{E}_6$SSM model postulates the existence of an approximate $Z_2$ symmetry that forbids those interactions. An exact $Z_2$ symmetry can also be considered~\cite{Nevzorov:2012hs,Nevzorov:2013ixa}, but in such a case additional exotic states must be introduced to ensure that the exotic quarks are not stable.

Gauge coupling unification in the $\mathrm{E}_6$SSM model requires the addition of incomplete multiplets of $E_6$ at low scales, since full multiplets do not modify the unification properties of the RGE flow. A pair of fields $H$ and $\bar{H}$ in conjugate representations are added, to ensure no anomalies are reintroduced. Alternatively, a Pati-Salam intermediate step has been postulated that achieves gauge coupling unification without the need of additional superfields. This ``minimal'' $\mathrm{E}_6$SSM model, however, predicts unification at the Planck scale so quantum gravity corrections may play a role and affect the outcome of unification~\cite{Howl:2007hq}. 

The $\mathrm{E}_6$SSM has a rather rich phenomenology since most of the predicted exotic states live at low energies. The constrained $\mathrm{E}_6$SSM (c$\mathrm{E}_6$SSM) is a version of the full $\mathrm{E}_6$SSM that exploits the properties of unification of $E_6$ and assumes universal scalar and gaugino soft masses at the GUT scale~\cite{Athron:2009bs,Athron:2009ue}. Predictions of the c$\mathrm{E}_6$SSM include contributions to the Higgs mass and rare decays~\cite{Athron:2012sq, Athron:2014pua} together with light exotic states, such as the $Z_N'$ associated with the $U(1)_N$ broken symmetry, and the colour triplet fermions $D$ and $\bar{D}$, all of which can be probed at the LHC~\cite{Athron:2011wu}. Lastly, as in the MSSM the lightest supersymmetric particle is stable, so it is a valid candidate for dark matter. In contrast with the regular neutralino dark matter in the MSSM, the dark matter candidate in the $\mathrm{E}_6$SSM is a mixture of binos, winos and higgsinos, as well as the inert singlinos and higgsinos in $H_{u,d}^i$ and $S_i$~\cite{Hall:2009aj,Hall:2011zq,Hall:2010ix,Athron:2015vxg,Athron:2016gor}.

\section{Cosmology and the early Universe}
\label{sec:cosmo}

\subsection{Inflation and GUTs}
\label{sec:inflation}
Cosmic inflation plays an important role in theories of Grand Unification, as it is needed to dilute relics such as magnetic monopoles, which are produced ubiquitously in GUT models.\footnote{However, there are exceptions, most notably the flipped $SU(5)$ SUSY GUT theories.} The requirement to dilute these relics therefore determines the scale of inflation in specific models \cite{Rocher:2004et}. 
Moreover, since generic inflation models are associated with a scale $\Lambda_{\rm inflation}\sim10^{16}$ GeV, it becomes attractive to associate the inflaton with a GUT scalar. 

To agree with observations, inflation models need to predict a large number of observable e-folds ($N = \int_{t_0}^{t_e} H dt \gtrsim 60$), as well as small spectrum density of fluctuations $\delta\rho/\rho \sim 10^{-5}$. 
For an effectively single field model, this can be illustrated by the tension between the Lyth bound (a measure of the field excursion necessary to solve the problems inflation was invented to solve) given in terms of the number of e-foldings $N$ \cite{Lyth:1996im}, 
\begin{equation} \Delta \phi \sim \left(\frac{r}{0.002} \right)^{1/2} \left(\frac{N}{60} \right) M_p \end{equation} 
and the amplitude of the Cosmic Microwave Background (CMB) anisotropies, which implies \cite{Ade:2015lrj,Akrami:2018odb},
\begin{equation}
     \Lambda_\mathit{inf}^4 = \left(2.2 \times 10^{16} \text{ GeV}\right)^4 \left( \frac{r}{0.2} \right). \end{equation} 
It is clear, then, that inflation requires a very flat scalar potential. Hence, it is attractive to consider inflation models in which the potential is dynamically generated \cite{Coleman:1973jx,Croon:2014dma,Croon:2015fza,Croon:2015naa}.

A successful example of such a model was realised as early as 1983 \cite{Shafi:1983bd}. This paper considered a potential of a Coleman-Weinberg form \cite{Coleman:1973jx}, 
\begin{equation}\label{eq:CWinflation}
    V(\phi) = A \phi^4 \left( \log \frac{\phi^2}{v_\phi^2} - \frac{1}{4} \right) + C
\end{equation}
Such a potential can only be made compatible with CMB constraints if $A$ is very small (presently, $A\lesssim 10^{-14}$ \cite{Barenboim:2013wra}). Therefore, the potential in Eq.~\ref{eq:CWinflation} cannot be due to loops of $SU(5)$ gauge bosons. 
Instead, \cite{Shafi:1983bd} considered a model in which the inflaton transformed as a singlet of $SU(5)$, couples weakly to the adjoint and fundamental Higgs fields, and therefore obtains a vacuum expectation value when $SU(5)$ breaks into the SM. 
The original CW-GUT inflation model \cite{Shafi:1983bd} predicts primordial gravitational waves with tensor-to-scalar index $0.02\leq r \leq 0.1$ \cite{Okada:2014lxa}. Although this is currently not in tension with the CMB-constraints \cite{Ade:2015lrj,Akrami:2018odb}, several modifications have been proposed which predict smaller $r$ \cite{Cerioni:2009kn,Panotopoulos:2014hwa,Barenboim:2013wra,Karam:2018mft}.

An alternative class of GUT inflation models are based on no-scale supergravity.
It was realised in 2013 \cite{Ellis:2013xoa} that particular realisations of no-scale supergravity (SUGRA) models of inflation can be equivalent to the Starobinsky model \cite{Starobinsky:1980te}, in which inflation is realised from a non-minimal Einstein-Hilbert action $S = \frac{1}{2}\int dx \sqrt{-g}(R+R^2/6 M^2)$. The correspondence can be seen by a conformal transformation, such that the model is equivalent to canonical gravity plus a scalar field \cite{Whitt:1984pd,Ellis:2013xoa}. The scalar potential then becomes
\begin{equation}
    V(\phi) = \frac{3}{4} M^2 \left(1 - e^{-\sqrt{2/3} \phi} \right)^2.
\end{equation}
Starobinsky-like models are attractive candidates for inflation models, as they make viable predictions for inflationary observables without the need to introduce a large set of finely tuned parameters.
Examples of no-scale SUGRA inflation models include sneutrino inflation, which can be consistently implemented in flipped $SU(5)$ SUSY GUTs \cite{Croon:2013ana, Ellis:2013nka, Ellis:2014xda, Ellis:2016ipm, Gonzalo:2016gey}.
Other no-scale GUT inflation models identify the inflaton with the Higgs boson, and circumvent the problems associated with conventional Higgs inflation \cite{Ellis:2014dxa,Ellis:2016spb}.

A phenomenological approach was taken by \cite{Hertzberg:2014sza}. Here it was assumed that inflation is driven by the vacuum energy associated with unification. It was shown that several examples of large-field ($\Delta \phi \sim M_p$) models of inflation give predictions consistent with the CMB-constraints \cite{Ade:2015lrj,Akrami:2018odb}.

GUT-inflation has also been studied in combination with other cosmological scenarios, most importantly with non-thermal leptogenesis and the seesaw mechanism for neutrino masses \cite{Rehman:2008qs,Boucenna:2014uma,SravanKumar:2018tgk}.\footnote{We expand more on the topic of the baryon assymmetry in subsection \ref{sec:baryonassymmetry}.} Models of sneutrino inflation are well suited for this purpose \cite{Croon:2013ana, Ellis:2013nka, Ellis:2014xda,Ellis:2016ipm, Gonzalo:2016gey}.

\subsection{Cosmological constraints on cosmic strings}
\label{sec:gwcosmicstrings}
Cosmic strings are generic cosmological predictions of many GUT theories \cite{Kibble:1976sj,Kibble:1980mv,Jeannerot:2003qv}. 
Field theories with broken gauge symmetries may have a vacuum state that is not unique, such that different points in physical space may have distinct (but degenerate) vacuum configurations. By continuity of the field, the interpolating field values must be taken on in between these points, which gives rise to an energetic object called a topological defect, or (in the one-dimensional case) a string. 

The simplest description of cosmic strings after their formation approximates the fundamental \emph{Nambu-Goto} strings.
Nambu-Goto strings are characterized by the dimensionless string tension $G \mu$, where $\mu$ is the mass per unit length and $G$ is Newton's constant. Strings produced at the GUT scale typically have a mass per unit length of the order of $\mu \sim 10^{21} \,\rm{kg}\, \rm{m}^{-1}$ and a thickness of $10^{−24} \, \rm{m}$, such that the tension is $G \mu \sim 10^{-6}$ \cite{Achucarro:2008fn}. For comparison, strings formed around the electroweak scale are expected to have much smaller tensions, $G \mu \sim 10^{-34}$. 
As the expansion of the Universe stretches strings, while the string tension stays constant and in the absence of a decay mechanism, $\rho_{\rm strings}$ would grow with the scale factor and eventually dominate the energy density of the Universe. Cosmic strings can decay into gravitational or scalar radiation, however. In the presence of such a decay channel an attractor scaling solution is reached, in which the strings maintain a constant fraction of the energy budget.  

Cosmic strings could be detected through gravitational lensing and anisotropies in the CMB \cite{Vilenkin:1981iu,Brandenberger:1985tx}. Cosmic strings imprint on the CMB as line-like discontinuities, caused by a boost of photons towards the observer as a string moves across the line of sight \cite{Gott:1984ef,Kaiser:1984iv}. Planck data constrains the Nambu-Goto string tension $G\mu<1.8 \times 10^{−7}$ \cite{Ade:2015xua}, the non-local string tension $G\mu<10.6 \times 10^{−7}$ \cite{Ade:2015xua} and the Abelian Higgs string model $G\mu<2.0 \times 10^{−7}$ \cite{Lizarraga:2016onn}.

If the strings decay gravitationally, such radiation can be observed in Gravitational Wave (GW) experiments \cite{Vilenkin:1981bx,Vachaspati:1984gt,Caldwell:1991jj,Siemens:2006yp,Sanidas:2012tf}. Strong gravitational radiation bursts may be produced by cusps  \cite{Damour:2000wa,Damour:2001bk,Damour:2004kw,Siemens:2006vk,Olum:1998ag}:
the LIGO/VIRGO collaboration reported an experimental upper limit of $G\mu < 10^{-8}$ in some regions of the parameter space, in which gravitational backreaction determines the size of the loops \cite{Aasi:2013vna}.
Pulsar Timing Arrays (PTAs) potentially give more stringent bounds, as they can already probe the stochastic GW background; depending on the model, $G \mu < \mathcal{O}(10^{-12}-10^{-11})$ \cite{Sanidas:2012ee,Abbott:2017mem,Blanco-Pillado:2017rnf}.
However, the relative importance of the gravitational decay channel has been the source of some disagreement in the literature. 
Simulations of Nambu-Goto strings \cite{Lorenz:2010sm,Ringeval:2005kr,Ringeval:2017eww,Blanco-Pillado:2013qja,Blanco-Pillado:2017oxo,Martins:2005es} and full field-theoretic simulations of the Abelian Higgs model \cite{Hindmarsh:2008dw,Hindmarsh:2014rka,Hindmarsh:2017qff} differ in the typical scale on which the strings form loops. Loops of the order of the string width $r_s$ can radiate heavy particles (as the natural mass of coupled particles is $m\sim r_s^{-1}$); loops of the typical inter-string spacing $\xi$ are expected to decay gravitationally \cite{Brandenberger:1986vj}. Recent field-theory simulations \cite{Hindmarsh:2017qff} suggest a mechanism to transport energy from large to small scales, which is not currently understood. Numerical results~\cite{Hindmarsh:2016dha,Hindmarsh:2016lhy,Lopez-Eiguren:2017ucu} also indicate that the simplest analytical models are due an update, when non-Abelian gauge groups are considered. 
Furthermore, different groups simulating Nambu-Goto strings differ in the distribution of the loop size. Simulations in which the gravitational radiation back-reacts on the string have smoother features, which hinders the formation of smaller loops \cite{Lorenz:2010sm,Ringeval:2005kr,Ringeval:2017eww}. In these simulations, the PTA constraints on the stochastic background and LIGO/VIRGO constraints on burst are stronger \cite{Abbott:2017mem}.

The shape of the fractional energy density $\Omega_{GW} \equiv f/\rho_c (d\rho_{\text{GW}}/df)$ power spectrum 
from cosmic strings is expected to be nearly scale-invariant, with an amplitude defined by the characteristic string tension $G \mu$, such that if it has a large enough amplitude, the signal would be seen in frequency windows of different experiments \cite{Figueroa:2012kw}. This distinguishes the power spectrum from other sources. In particular, an observation of GW at pulsar timing arrays, if coming from cosmic strings, 
will draw attention to interferometer searches for this source.

\subsection{Gravitational waves from phase transitions}
\label{sec:gwphasetransitions}

\begin{figure}
    \centering
    \includegraphics[width=0.5\textwidth]{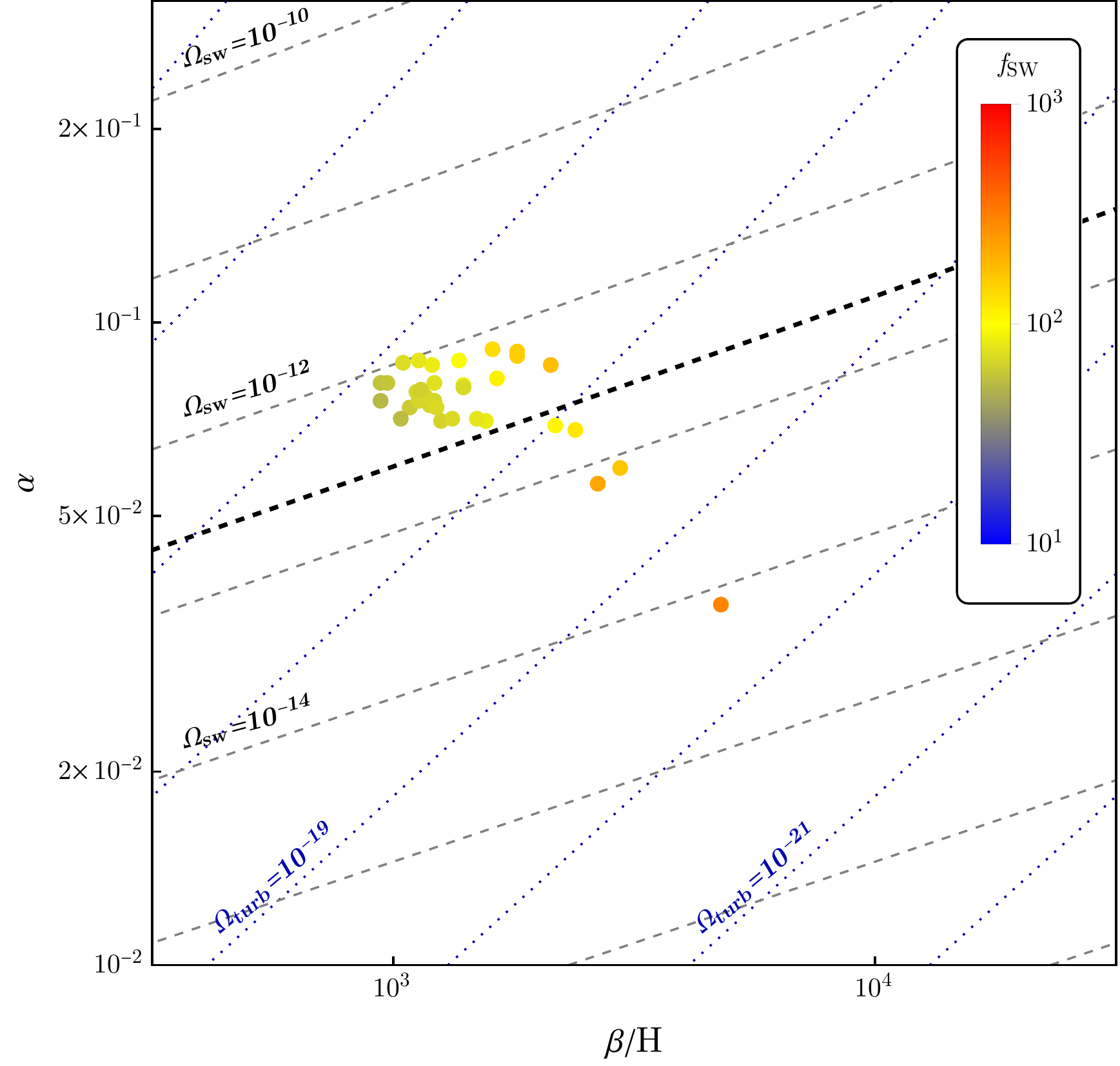}
    \caption{\emph{Plane of thermal parameters with contours of constant sound wave (in black) and turbulence (in blue) peak amplitude.  The thicker line shows the peak sensitivity of the Einstein telescope~\cite{Punturo:2010zz}. Points denote thermal parameters produced by a Pati-Salam phase transition with a Pati-Salam scale of $M_{\rm PS} = 10^5$ GeV. The peak frequency for the sound wave spectrum is indicated by the colour scaling. Figure taken from \cite{Croon:2018kqn}}.}
    \label{fig:PSPT}
\end{figure}

\begin{figure}
    \centering
    \includegraphics[width=0.48\textwidth]{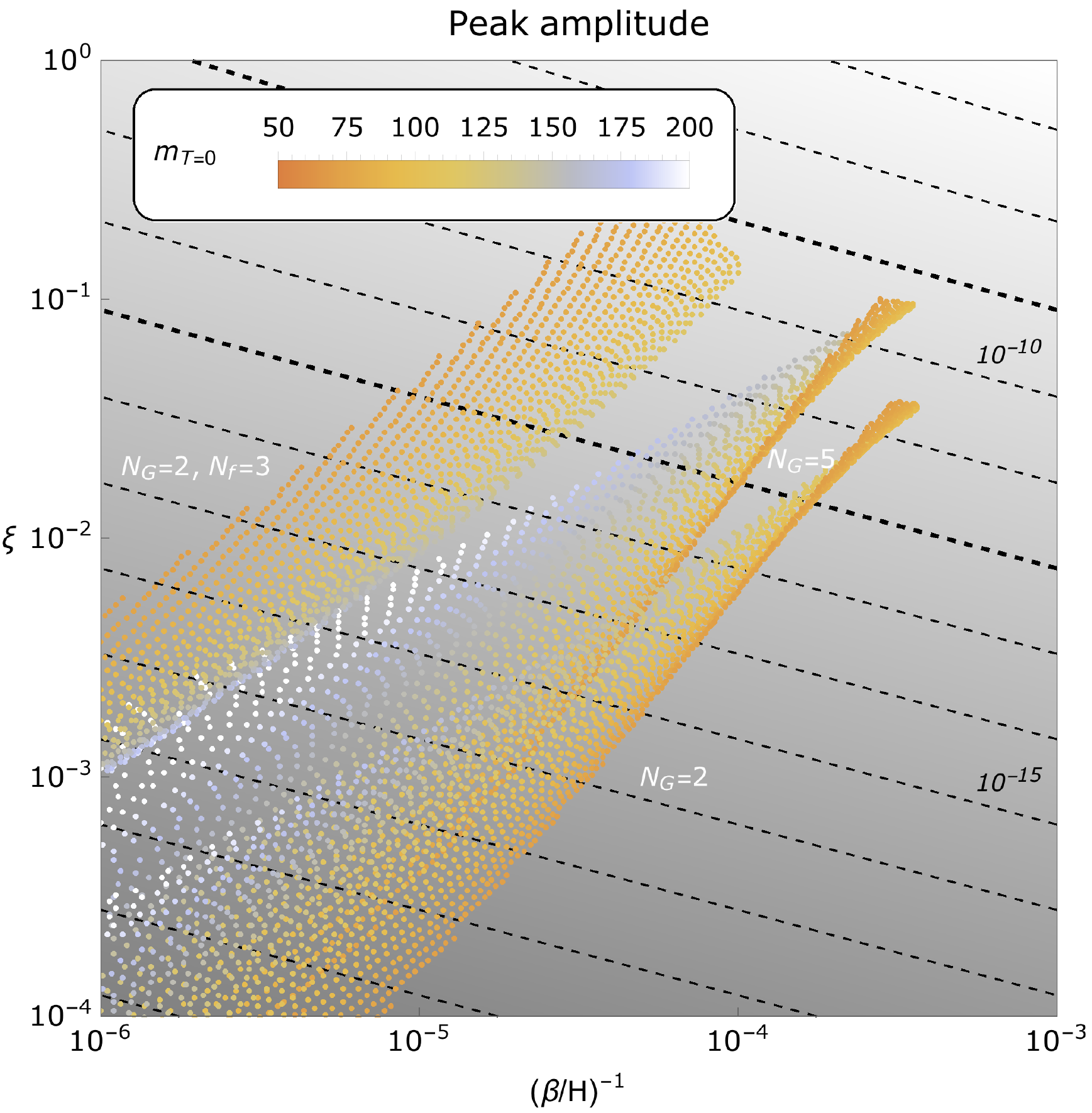}
     \includegraphics[width=0.48\textwidth]{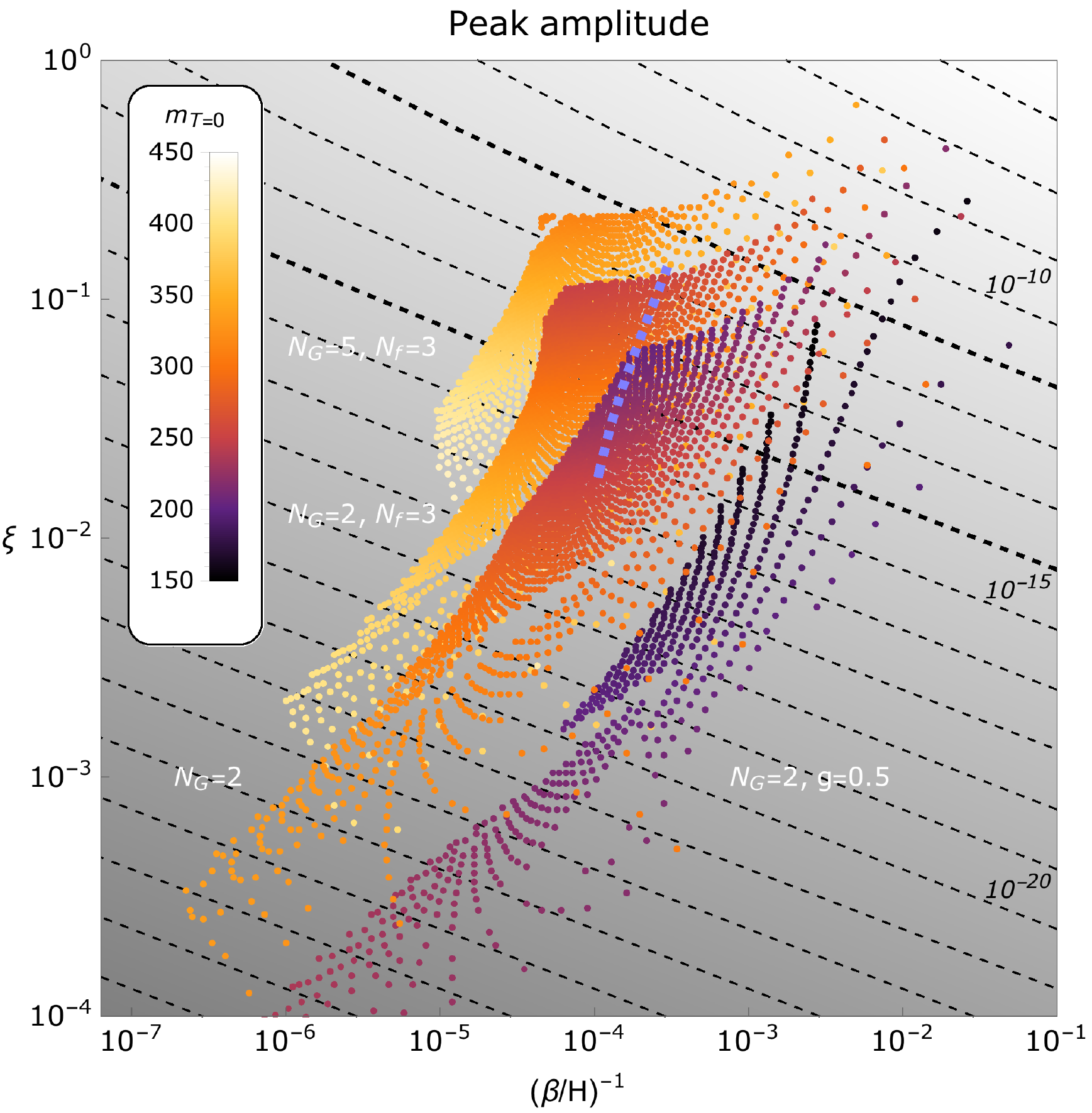}
    \caption{\emph{Thermal parameters from a renormalizable potential for a scalar field in the fundamental representation of $SU(N_G)$ which gets broken to $SU(N_G-1)$ (left). Right panel is the same aside from the inclusion of a non-renormalizable sextet term in the potential and the gauge coupling constant is fixed to unity. In the above $N_G$ denotes the order of the group, $N_f$ is the number of fermions in the fundamental representaion that are strongly coupled to the scalar field, in direct analogy with the SM save that the Yukawa couplings are set to unity. Note that $\xi$ is the ratio of latent heat to radiation energy density more commonly denoted $\alpha$. Contours of constant peak amplitude for the sound wave spectrum are shown with the darker line corresponding to LISA sensitivity range for a power law spectrum that has been integrated over frequency \cite{Thrane:2013oya}. Note in the above $v_w=0.5$ for the left plot and $v_w=1$ for the right plot using the efficiency terms in \cite{Espinosa:2010hh}. Figure taken from \cite{Croon:2018erz}.}}
    \label{fig:amps}
\end{figure}

Grand unification models can accommodate a rich scalar sector, which can result in a complicated cosmological history involving exotic phase transitions. Some GUT inspired possibilities are: a colour breaking phase transition where colour is broken and restored when leptoquarks acquire a vacuum expectation value in an intermediate transition, which can catalyse baryogenesis \cite{Patel:2013zla,Ramsey-Musolf:2017tgh}, $B-L$ and $L$ violating phase transitions \cite{Long:2017rdo,Schmitz:2012kaa,Buchmuller:2012wn,Buchmuller:2013lra}, hidden sector phase transitions \cite{Croon:2018erz,Schwaller:2015tja,Breitbach:2018ddu,Baldes:2018emh}, and a Pati-Salam transition \cite{Croon:2018kqn}. If any such phase transition occurs through bubble nucleation, an observable relic gravitational wave spectrum can be seen today, for a review see \cite{Mazumdar:2018dfl,Caprini:2018mtu,Weir:2017wfa}. Furthermore, GUT models often require the existence of extra singlets. For example, the $\mathrm{E}_6$SSM model studied in~\cite{Athron:2009bs} had $3$ generations of singlets and such singlets can, in principle, catalyse the electroweak phase transition to be strongly first order as well \cite{Chao:2014hya,Chiang:2017nmu,Profumo:2014opa}. \par 
The gravitational wave spectrum generated from a cosmic phase transition has three contributions: a contribution from the collision of scalar shells, and potentially long-lasting contributions from sound waves and turbulence in the plasma. The total spectrum can thus be written as,
\begin{equation}
    \Omega (f) h^2 = \Omega _{\rm coll}(f) h^2 +\Omega _{\rm sw} (f)h^2 +\Omega _{\rm turb}(f)h^2 \ .  
\end{equation}
Although much uncertainty remains about the precise form of these spectra, all three are controlled by four thermal parameters, which can be computed by first principles \cite{Caprini:2015zlo,Mazumdar:2018dfl}: the latent heat released during the transition (conventionally normalized by the radiation energy density), denoted $\alpha $, the nucleation rate (conventionally normalized to the Hubble parameter) $\beta/H$, the temperature at which the transition occurs $T_*$ and the velocity of the bubble wall $v_w$. The collision term is expected to be sub-dominant for transitions associated with a broken gauge group \cite{Bodeker:2017cim}. The sound wave contribution to the power spectrum is \cite{Caprini:2015zlo,Hindmarsh:2017gnf}
\begin{equation}
h^2\Omega _{\rm sw} = 8.5 \times 10^{-6} \left( \frac{100}{g_*} \right)^{-1/3} \Gamma ^2 \bar{U}_f^4  \left( \frac{\beta}{H} \right)^{-1}  v_w S_{\rm sw}(f) ,
\end{equation}
where $\bar{U}_f^2\sim (3/4) \kappa _f \alpha _T$ is the rms fluid velocity, $\Gamma \sim 4/3$ is the adiabatic index, $\kappa _f$ is the efficiency of converting the latent heat into gravitational waves and $g_*$ the number of relativistic degrees of freedom.
The frequency dependence is captured by the spectral state
\begin{equation}\notag
    S_{\rm sw} =  \left( \frac{f}{f_{\rm sw}} \right) ^3 \left( \frac{7}{4+3\left( \frac{f}{f_{\rm sw}}\right) ^2} \right)^{7/2},
\,\,\,\,\,\,\,\,
    f_{\rm sw} = 8.9 \times 10^{-7} {\rm Hz} \frac{1}{v_w} \left( \frac{\beta}{H} \right) \left( \frac{T_*}{{\rm Gev}} \right) \left( \frac{g_* }{100} \right)^{1/6} \ .
\end{equation}

The other notable, albeit sub-dominant, contribution is the contribution from magneto-hydrodynamic turbulence in the plasma. The power spectrum from this contribution is given by \cite{Caprini:2009yp},
\begin{eqnarray}
 h^2 \Omega _{\rm turb} = 3.354 \times 10^{-4} \left( \frac{\beta}{H} \right)^{-1} \left( \frac{\kappa \epsilon \alpha }{(1+\alpha } \right)^{3/2} \left( \frac{100}{g^*} \right) ^{1/3}v_w S_{\rm turb} (f),\label{eq:turb}
\end{eqnarray}
where $\epsilon$ is the fraction of the energy that contributes to turbulence, typically taken to be in the range $(0.05,0.1)$ \cite{Caprini:2015zlo}. In this case the spectral form is a function of two time scales,
\begin{equation}\notag
    S_{turb} = \frac{(f/f_{\rm turb})^3}{[1+(f/f_{\rm turb})]^{11/3}(1+\frac{8 \pi f}{h_*})} \ , 
    \,\,\,\,\,\,\,\,
     f_{\rm turb } = 27 \mu {\rm Hz} \frac{1}{v_w}  \left( \frac{T_N}{100 {\rm GeV}} \right) \frac{\beta }{H}  \left( \frac{g^*}{100} \right)^{1/6},
\end{equation}
where $h^*$ is the Hubble rate at the transition temperature.\footnote{Note that the existence of two time scales in the spectral form means that the peak amplitude for the turbulence contribution cannot be found simply be setting the frequency to either $h_*$ or $f_{\rm turb}$ in Eq. \ref{eq:turb}. } \par 
For a single scalar field transition, without a tree-level barrier between the true and the false vacuum, $\beta/H$ tends to be $O(10^3)$ or greater \cite{Camargo-Molina:2017kxd}. The transition temperature is the same order of magnitude as the mass of the scalar. Therefore transitions with scalar masses $O(10^5)$ GeV can be probed by ground-based interferometers such as the Einstein Telescope \cite{Punturo:2010zz}, Kagra \cite{Akutsu:2018axf} and cosmic explorer \cite{Evans:2016mbw}, whereas space-based LISA will probe transitions at the electroweak scale \cite{Caprini:2015zlo}. The former can be more directly related to studies of GUTs - we show benchmark examples for a Pati-Salam phase transition are shown in Fig.~\ref{fig:PSPT}.
The visibility of the spectrum tends to grow with the ratio $v/m$, the gauge coupling constant $g$, the rank of the (sub) group being broken and the number of other particles acquiring a mass during the transition \cite{Croon:2018erz}. Furthermore, it was found in \cite{Croon:2018erz} that some non-trivial model discrimination is possible if one observes a primordial power spectra due to the increase in visibility as well as moderate correlations between thermal parameters, shown in Fig.~\ref{fig:amps} for $SU(N)/SU(N-1)$ cosets. \par 
If multiple scalar fields are involved in a transition the barrier between the true and false vacuum can persist at zero temperature due to triscalar or non renormalizable operators \cite{Profumo:2014opa,Grojean:2004xa}. In such a case significantly more supercooling is possible and the transition temperature is no longer confined to be the same order of magnitude as the scalar mass. This implies that $\beta /H$ can be quite small and the latent heat can be large, increasing the visibility of the gravitational wave and reducing the peak frequency. A caveat to this is that recent work found that phase transitions that involve a large amount of supercooling may fail to complete due to the onset of inflation \cite{Ellis:2018mja}. Regardless, the thermal parameter space in the case of multifield phase transitions is broader, which minimizes model discrimination somewhat, though not completely \cite{Croon:2018erz}.

\subsection{Baryo-/leptogenesis}
\label{sec:baryonassymmetry}
The existence of a baryon asymmetry in the Universe (BAU) is one of the central problems of modern cosmology \cite{White:2016nbo,Morrissey:2012db}. At the same time, the concordance between different measurements of the primordial baryon asymmetry is a triumph of modern cosmology with BBN and CMB limits giving \cite{Aghanim:2018eyx,Riemer-Sorensen:2017vxj}
\begin{equation}
    \eta _B = \left\{ \begin{array}{cc}
(6.2 \pm 0.4)\times 10^{-10}         & {\rm BBN}  \\
(6.14 \pm 0.03)\times 10^{-10}         & {\rm CMB} 
    \end{array} \right.
\end{equation}
respectively.
Any explanation for the baryon asymmetry must satisfy the three Sakharov conditions \cite{Sakharov:1967dj}
\begin{itemize}
    \item Baryon number $B$ violation
    \item C and CP violation
    \item A departure from thermal equilibrium.
\end{itemize}
Early attempts at generating the BAU focused on $B$ violating decays (for a review see \cite{Riotto:1998bt}). Such decays typically violate $B+L$ while conserving $B-L$ (for an exception see \cite{Gu:2007bw}). For example, $SU(5)$ GUTs are invariant under changes to a global phase conjugate to $B-L$ number, whereas $SO(10)$ has a local $U(1)_{B-L}$ symmetry. However, any primordial $B+L$ asymmetry is washed out by $B+L$ violating electroweak sphalerons. Therefore, only a primordial $B- L $ asymmetry will be preserved unless sphalerons are quenched. \par 

Leptogenesis allows for a $B-L$ violating operator, $m \bar{\nu }_R ^c \nu _R$, that is also responsible for a light neutrino mass via type-I seesaw mechanism (see Section~\ref{sec:numasses}). CP violating decays of such sterile neutrinos ensure a net $B-L$ asymmetry which electroweak sphalerons convert to a baryon asymmetry. Electroweak baryogenesis by contrast uses the sphalerons themselves to generate a net $B+L$ asymmetry which cannot be washed out before the sphalerons are quenched \cite{White:2016nbo,Morrissey:2012db}. More specifically, if the electroweak phase transition is strongly first order, bubbles of electroweak broken phase populate a medium of symmetric phase with sphalerons quenched only inside such bubbles. Particles can have CP violating interactions with the bubble wall which biases the sphalerons to produce a net $B+L$ asymmetry. Some of this asymmetry is swept up in the expanding bubble wall where it is preserved.

GUTs are only relevant to electroweak baryogenesis if the GUT model motivates some light BSM states. Recent work on electroweak baryogenesis in  the $E_6$-SSM utilized three generations of singlet superfields to motivate a CPV source involving singlino-Higgsino interactions with the bubble wall \cite{Chao:2014hya}. \par 
A feature of leptogenesis during GUTs is the possibility of new CP violating decay channels due to the presence of leptoquarks \cite{Gehrlein:2015dxa,Falcone:2001im}. This allows a lower minimum mass for the lightest sterile neutrino than in the minimal scenario \cite{King:2008qb}.\footnote{This limit of course is for the non resonant regime. In case of resonant leptogenesis the masses of the sterile neutrinos can be very low \cite{Pilaftsis:2003gt}.} Much of the recent focus on baryogenesis within GUTs involves leptogenesis with some intriguing concordance achieved in the case of $SU(5)$~\cite{Blanchet:2008cj} and $SO(10)$ GUTs \cite{Ellis:2016ipm,SravanKumar:2018tgk,Chianese:2018rnq,DiBari:2017uka,DiBari:2015svd,DiBari:2014eya,DiBari:2013qja}. A generic feature of $SO(10)$ GUTs is normal ordering of neutrino masses and a negative Dirac phase \cite{Esteban:2018azc}, both of which are favoured by current observational limits \cite{DiBari:2013qja}. Furthermore, many GUTs, including $SO(10)$, predict a Dirac neutrino mass matrix that is not too different from the up quark mass matrix and $SO(10)$ leptogenesis also achieves agreement in the currently observed atmospheric mixing angle in the first octant. Realistic models with two right-handed neutrinos can emerge in $\Delta (27) \times SO(10)$ models and $A4 \times SU(5)$ supersymmetric models \cite{Bjorkeroth:2016lzs,Bjorkeroth:2015ora}. The third right-handed neutrino can either decouple because it is very heavy or because its Yukawa coupling is very small. The latter case implies a stable particle that can play the role of dark matter \cite{Anisimov:2008gg}.  

\section{Direct collider searches for GUTs}
\label{sec:collider}

\subsection{Searches for Supersymmetry}
\label{sec:SUSYsearches}

\begin{figure}[t]
 \centering
 \includegraphics[width=0.7\textwidth]{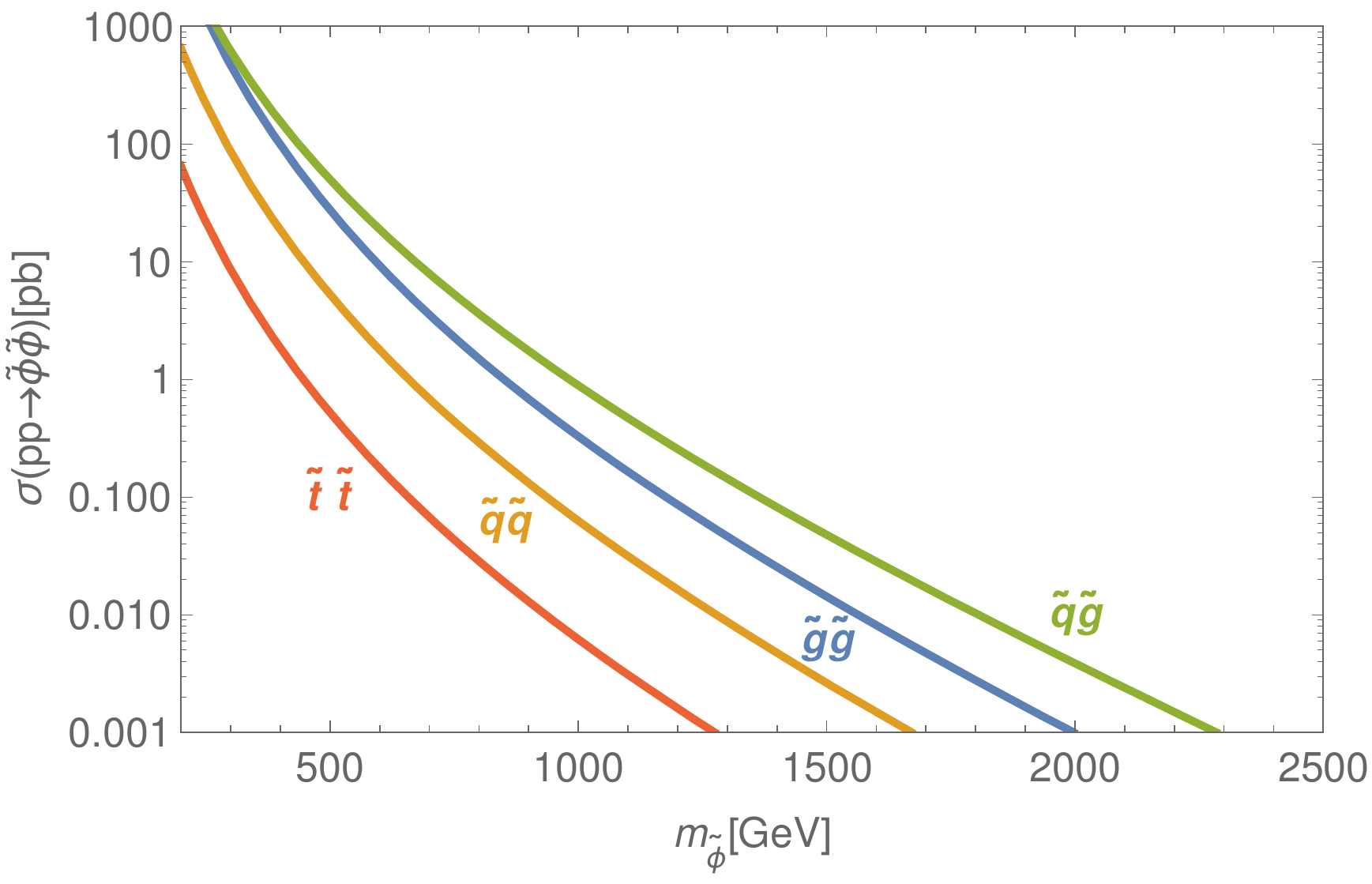}
 \caption{\emph{Pair production cross sections as a function of the mass of sparticle $\tilde{\phi}$ at centre of mass energy $\sqrt{s} = 13$ TeV.}}
 \label{fig:susycrossections}
\end{figure}

As we have seen in Section~\ref{sec:SUSYGUTs} supersymmetry plays a rather important role on many unified theories and motivates the unification of gauge couplings at large scales. However, both its solution to the hierarchy problem and gauge coupling unification often rely on a light sparticle spectrum, around or below the TeV scale. Thus searches for supersymmetric particles has been part of the research programme in collider physics for the last few decades, from searches at LEP and the Tevatron, to the recent results of the LHC, and it is still part of the proposed physics programme for future colliders, e.g. CLIC, ILC or VLHC~\cite{DeRoeck:2009id,Djouadi:2007ik,Baer:2013cma,Linssen:2012hp}.

In R-parity conserving SUSY the lightest supersymmetric particle (LSP) is stable. This has strong consequences for SUSY searches, for the LSP will escape the collider in the form of missing transverse energy (MET)\footnote{In cases where a charged next-to-lightest SUSY particle (NSLP) is stable at detector timescales, no clear MET signal is produced, since the NLSP will decay to the LSP outside the detector.}~\cite{Haber:1984rc}. In addition, R-parity requires that sparticles are pair-produced in colliders, hence the different searches for supersymmetry are classified according to the particle that is produced in pairs. The production cross sections of the different sparticle species are rather different and often determine the exclusion or detection power of a particular channel. For instance, the strongest exclusion limits at the LHC across the sparticle spectrum are on first and second generation squarks and gluinos which, as can be seen in Figure~\ref{fig:susycrossections}, have the largest production cross sections~\cite{Borschensky:2014cia,Beenakker:2016lwe}.

\begin{figure}[t]
 \centering
 \includegraphics[width=0.8\textwidth]{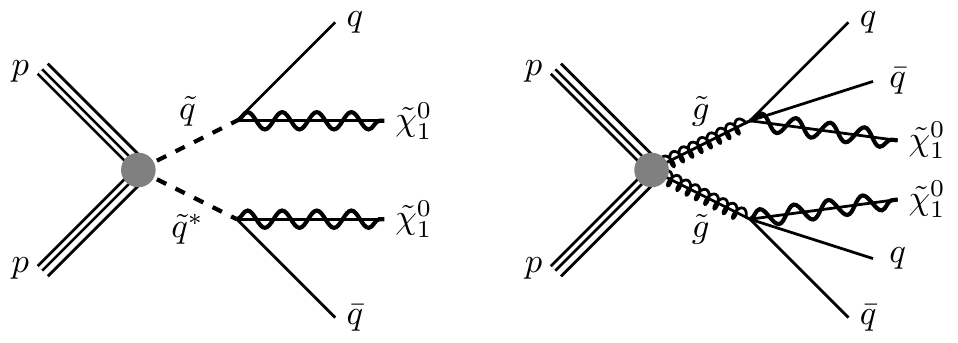}
 \caption{\emph{Most simple topologies for the production of squarks and gluinos at the LHC with decays to jets + MET.}}
 \label{fig:glusq}
\end{figure}

Squarks and gluinos are produced in pairs at the LHC in the combinations $\glu\glu$, $\tilde{q}\tilde{q}$ and $\tilde{q}\glu$ and their main decay channels are $\tilde{q} \to q \neut_1$ and $\glu \to q\bar{q}\neut_1$, with the neutralino LSP escaping the collider. Hence the typical signature for these processes has multiple jets and large missing energy. The decay topologies for these signatures are depicted in Figure~\ref{fig:glusq}. Other decay modes for squark and gluinos involve the production of charginos or heavier neutralinos, $\sq \to q \neut_2$, $\sq \to q' \cha_1$, $\glu \to q\bar{q}\neut_2$ and $\glu \to q\bar{q}\cha_1$, which then decay to $W$ and $Z$ bosons and $\neut_1$. The final state signatures depend on the decay modes of the gauge bosons, and can have (0-4) leptons, jets and MET. ATLAS and CMS have reported results from the last run of the LHC at 13 TeV and $36 ~\rm{fb}^{-1}$ for searches with jets and MET final states~\cite{Aaboud:2017vwy,Sirunyan:2018vjp,Sirunyan:2017cwe}, with one lepton, jets and MET~\cite{Aaboud:2017bac,Sirunyan:2017mrs,Sirunyan:2017fsj}, same and opposite-sign dilepton pairs, jets and MET~\cite{Aaboud:2018ujj,Sirunyan:2017leh}, two or three leptons, jets and MET~\cite{Aaboud:2017dmy,Sirunyan:2017hvp}, 3rd generation squarks (with and without Higgs reconstruction) and MET~\cite{Aaboud:2017hrg,Sirunyan:2017bsh,Sirunyan:2017pjw} and hadronic $\tau$ decays, jets and MET~\cite{Aaboud:2018mna}, among others. These searches set a lower limit for a range of simplified models on the mass of the gluino of $m_{\glu} \gtrsim 2.1$ TeV and the mass of the first and second generation squarks of $m_{\sq} \gtrsim 1.5$ TeV.

The next strongest production cross section is that of stop and sbottom pairs. The main decay channel for stops and sbottoms is $\stop \to t \neut_1$ and $\sbot \to b \neut_1$, respectively. This topology is similar to the decay of first and second generation squarks, with the added complexity that neither $t$ or $b$ produce a clean jet, but rather have many decay channels that can result in numerous jets, leptons and, of course, MET. Secondary decay channels for $\stop$ and $\sbot$ involve decays to a chargino, $\stop \to b \cha_1$ and $\sbot \to t \cha_1$, with subsequent decays involving $W$ bosons, or decays into a heavy neutralino, $\stop \to t \neut_2$ and $\sbot \to b \neut_2$, which in turn decays into a $Z$ or a Higgs boson and $\neut_1$. The latest searches of the LHC experiments for pair-produced stops and sbottoms target final states with jets and MET~\cite{Aaboud:2017ayj,Sirunyan:2017cwe,Sirunyan:2017xse}, b-jets and MET~\cite{Aaboud:2017wqg,Sirunyan:2017kiw}, one lepton, jets and MET~\cite{Aaboud:2017aeu,Sirunyan:2018omt}, two and three leptons, jets and MET~\cite{Aaboud:2017nfd,Aaboud:2017dmy,Sirunyan:2018lul,Sirunyan:2017hvp,Sirunyan:2017leh} and final states with a $h$ or a $Z$ boson and MET~\cite{Aaboud:2017ejf}, among others. These searches exclude masses of stops and sbottoms up to $m_{\stop} \sim 1$ TeV and $m_{\sbot} \sim 900$ GeV for some simplified models.

In the cases where the coloured sector of a supersymmetric model has large masses, the direct production of chargino, neutralino and slepton pairs dominate. A pair of directly produced sleptons decay typically like $\sl \to l \neut_1$. Neutralinos and charginos are produced in pairs in a number of different combinations, the most commonly studied of which are $\neut_2\cha_1$ and $\cha_1\cha_1$. The decays of heavy neutralinos and charginos produce $W$, $Z$ or Higgs bosons and the lightest neutralino. Further decay of $W$ and $Z$ sets the final states targeted by ATLAS and CMS searches, such as the final state with two leptons and MET~\cite{Aaboud:2017leg,Sirunyan:2018nwe,Sirunyan:2018ubx}, many leptons and MET~\cite{Aaboud:2018jiw,Aaboud:2018sua,Aaboud:2018zeb,Sirunyan:2018ubx}, leptons, jets and MET~\cite{Sirunyan:2018lul}, taus and MET~\cite{Aaboud:2017nhr,Sirunyan:2018vig}, and b-jets plus MET~\cite{Aaboud:2018htj}, among others. Due to their low production cross sections, the exclusion limits on slepton masses from direct production are quite weak and they only reach up to around $m_{\sl} \sim 500$ GeV. Stronger limits on slepton masses can be inferred from neutralino/chargino production with sleptons in the cascade, reaching up to $\sim 850$ GeV. The limits on electroweakinos (neutralinos and charginos) are very sensitive to the parameter choices for the simplified model analyses performed by the experiments, hence the exclusion limits on $\neut_2$ and $\cha_1$ vary from search to search and from signal region to signal region, roughly in the range $m_{\neut_2},m_{\cha_1} \in (500\text{ GeV}, 1.1\text{ TeV})$. Similarly the limits on the lightest neutralino varies in the range $m_{\neut_1} \in (200, 700)$ GeV. In addition, there is a hint of an excess in one of the two and three lepton final state analyses by the ATLAS collaboration in the low $m_{\neut_1}$ region, with a reported significance of 2 and 3$\sigma$ in the $2l$ and $3l$ channels respectively~\cite{Aaboud:2018sua}.

In addition to sparticle searches, SUSY can also be probed through searches for heavy and charged Higgs bosons. The MSSM predicts the existence of two CP-even scalars, $h$ and $H$, one CP-odd pseudoscalar, $A$, and a charged scalar $H^\pm$. The lightest CP-even scalar, $h$, is said to be ``SM-like'', as its mass and couplings are aligned with the Higgs boson discovered at the LHC~\cite{Aad:2012tfa,Chatrchyan:2012xdj}, the so called \textit{alignment limit}. Neutral heavy Higgses can be produced at the LHC in the same manner as the SM Higgs, that is by gluon fusion, vector boson fusion (VBF) and associated production, with a $t$ and/or $b$ quark. Thus the same mechanisms that lead to the discovery of the SM Higgs are employed to set exclusion limits on heavier neutral (pseudo)scalars, including signatures where $H$ is produced in resonance and decays into two light Higgs bosons $pp \to H \to hh$. The final states targeted by these exotic Higgs searches consist of 2-4 leptons, jets and MET from on- and off-shell $W$ and $Z$ bosons~\cite{Aaboud:2018ksn,Aaboud:2017rel}, two leptons and MET~\cite{Aaboud:2017gsl,Sirunyan:2018mgs}, final states with four b-jets~\cite{Aaboud:2018knk,Sirunyan:2018taj} or two b-jets and $WW$~\cite{Aaboud:2018zhh}, $\gamma\gamma$~\cite{Aaboud:2018ftw,Sirunyan:2018aui}, $\tau\tau$~\cite{Aaboud:2018sfw,Khachatryan:2017mnf}, $\mu\mu$~\cite{Sirunyan:2017uvf} or $t\bar{t}$~\cite{Aaboud:2018cwk} decays, ditau final states~\cite{Aaboud:2017sjh,Sirunyan:2018zut} and diphotons, with and without associated $W$ bosons~\cite{Aaboud:2017yyg,Aaboud:2018ewm,Aaboud:2018djx}. Charged Higgs bosons, $H^\pm$ can typically be produced with associated resonant and non-resonant top-quark production. Their main decay channels are $H^\pm \to W^\pm Z$~\cite{Aad:2015nfa,Sirunyan:2017sbn}, $H^\pm \to t(c)b$~\cite{Aaboud:2018cwk,Sirunyan:2018dvm} and $H^\pm \to \tau^\pm \nu$~\cite{Aaboud:2016dig,Aaboud:2018cwk}. Since no excess has been found for either heavy $H$ or $H^\pm$, the experiments set upper limits that strongly depend on the production cross section and, in turn, on $\tan\beta$. For $H$ the limits range from $m_H  <400$ GeV for $\tan\beta =2$ and production cross sections limits of 0.1 pb for larger masses. For $H^\pm$  with $m_{H^\pm} < 160$ GeV excluded for all values of $\tan\beta$ and $m_{H^{\pm}} < 1.1$ TeV excluded for $\tan\beta = 60$.

Many unified theories automatically preserve $R$-parity, such as left-right symmetric or Pati-Salam models, as well as supergroups of them, $SO(10)$ or $E_6$. This is because they contain a gauged $U(1)_{B-L}$ subgroup which effectively makes the LSP stable~\cite{Aulakh:1998nn}. Other models, such as $SU(5)$, may have $R$-parity violating (RPV) interactions, though in general they will be suppresed since they can lead to rapid proton decay. Since the LSP is no longer stable, collider signatures of RPV typically contain multiple leptons~\cite{Aaboud:2018zeb}, multiple jets~\cite{Aaboud:2018lpl,Aaboud:2017faq,Aaboud:2017nmi,Sirunyan:2017dhe} or both~\cite{Aaboud:2017opj,Sirunyan:2018hwm} in the final state. These searches often impose strong upper limits on sparticle masses that range from 150 GeV to a few TeV for $\stop$, depending on the channel, and from 1 TeV to 2 TeV, for $\glu$. 

If the LSP is metastable or the lightest chargino and neutralino are almost degenerate, they can live long enough to leave a displaced vertex or a disappearing track on the detector. Detailed searches have been performed by ATLAS~\cite{Aaboud:2018aqj,Aaboud:2018hdl,Aaboud:2017iio,Aaboud:2017mpt,Aaboud:2018jbr} and CMS~\cite{Sirunyan:2017jdo,Sirunyan:2018vlw,Sirunyan:2018pwn} to search for these long-lived particles, and they have reached exclusion limits comparable to those of the detailed searches above.

\begin{figure}[t]
 \centering
 \includegraphics[width=0.8\textwidth]{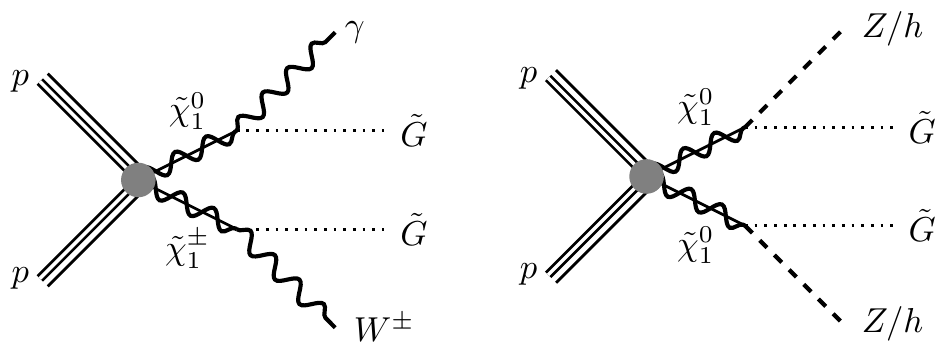}
 \caption{\emph{Topologies for SUSY searches with gravitino LSP and $\gamma$ and $h/Z$ final states.}}
 \label{fig:gravitino}
\end{figure}

Most of the searches described above assumed a neutralino LSP, which is typically the case in gravity mediated SUSY breaking. In gauge mediated SUSY breaking (GSMB) and general gauge mediation (GGM) the LSP is actually a nearly massless gravitino. In these cases new decay channels are open with photons~\cite{Sirunyan:2017nyt,Sirunyan:2017yse,ATLASCollaboration:2016wlb, Aaboud:2018doq}, Z's~\cite{Aaboud:2018htj,Aaboud:2018zeb}, Higgses~\cite{Aaboud:2018htj, Aaboud:2018zeb, Sirunyan:2017obz} and $\tau$s~\cite{Aaboud:2018mna,Aaboud:2018kya} in the final state (see Fig.~\ref{fig:gravitino}). 

Although the LHC results are the most recent and, for the most part, they supersede the results of previous collider experiments, such as those at the Tevatron, some experimental limits from LEP still remain relevant today. In particular for models with a significant production of neutralino/chargino or slepton pairs, the limits from ALEPH~\cite{Heister:2002mn,Heister:2002jca}, OPAL~\cite{Abbiendi:2003sc,Abbiendi:2003ji}, L3~\cite{Acciarri:1999km,Achard:2003ge} and DELPHI~\cite{Abdallah:2003xe} on sleptons and electroweakino masses are rather pertinent, as they are largely model independent. 

Many of the SUSY searches above are performed using simplified models, e.g ATLAS jets and MET search assumes a 50-50 split between the decay modes of gluinos~\cite{Aaboud:2017vwy}. Therefore, the mass and cross section limits obtained are weakened in more complicated models. In order to assess the relevance of many of these exclusion limits on several popular SUSY models, a full global fit of the parameter space of the model is required. Several of these fits have been performed for the CMSSM, NUMH1 and NUMH2~\cite{Roszkowski:2009sm,Bertone:2011nj, Bechtle:2012zk, Buchmueller:2013rsa, Buchmueller:2014yva, Han:2016gvr,Athron:2017qdc}, phenomenological MSSM models~\cite{Bagnaschi:2017tru,Athron:2017yua, Bertone:2015tza}, SUSY GUT models~\cite{Bagnaschi:2016afc} and electroweak-sector MSSM models~\cite{Athron:2018vxy}, by the Zfitter~\cite{Arbuzov:2005ma}\footnote{zfitter.desy.de/}, SuperBayes~\cite{deAustri:2006jwj,Trotta:2008bp}\footnote{www.ft.uam.es/personal/rruiz/superbayes}, Fittino~\cite{Bechtle:2004pc}\footnote{flcwiki.desy.de/Fittino}, MasterCode~\cite{Buchmueller:2009fn} \footnote{cern.ch/mastercode/} and GAMBIT~\cite{Athron:2017ard, Workgroup:2017bkh, Workgroup:2017lvb, Workgroup:2017htr, Balazs:2017moi, Workgroup:2017myk, Athron:2017kgt, Athron:2018hpc}\footnote{gambit.hepforge.org.} collaborations.

\subsection{Collider searches for leptoquarks}

Leptoquarks (LQs) are associated either with the vector (spin 1)
particles that correspond to the gauge bosons of the unified gauge
group or they can be scalars (spin 0) and belong to a Higgs sector of
a unified theory. Vector LQ mass is typically of the order of the
unification scale and can only be accessible directly at colliders if
the unification scale is low enough (e.g. Pati-Salam models). Scalar
representations can also contain light fields, most notably the SM Higgs,
but come at the cost of severe fine tuning, as discussed in the
Sec.~\ref{sec:basicsonguts} on the example of doublet-triplet splitting problem in
$5$-dim. representation of $SU(5)$. In non-supersymmetric unified models the presence of light colored scalars tends to aid unification~(see e.g.~\cite{Murayama:1991ah}). Another important difference is that the scalar LQ interactions can be analysed without specifying the concrete GUT completion in the ultraviolet. On the other hand, vector (gauge) LQs are sensitive to the mass generation
mechanism that is specified in the ultraviolet. Therefore, effective vector LQ models are not renormalizable~\cite{Biggio:2016wyy}. Furthermore, even the couplings of vector LQs to the SM gauge sector are not completely fixed by the gauge quantum numbers~\cite{Gabrielli:2015hua,Dorsner:2016wpm}.

Altogether there are six scalar and six vector leptoquarks, listed in Tab.~\ref{tab:LQs}, that couple to the SM matter at the renormalizable level~\cite{Buchmuller:1986zs, Davidson:1993qk, Dorsner:2016wpm}. The fermionic number $F \equiv 3 B + L$ of leptoquarks that do not couple to diquarks ($\phi q q$) and are potentially $B$ and $L$ conserving, must be $F = 0$, whereas LQs with $|F| = 2$ in general destabilize the proton via $B-L$ conserving decays.
\begin{table}[tbp]
\centering
\begin{tabular}{|c|ccc|}
\hline
LQ symbol & $(SU(3),SU(2),U(1))$ &  Spin &  $F$ \\
\hline 
  $S_3$ & $(\overline{\mathbf{3}},\mathbf{3},1/3)$ & 0  & $-2$\\ 
$\tilde{S}_1$ & $(\overline{\mathbf{3}},\mathbf{1},4/3)$ & 0  & $-2$\\
$S_1$ &$(\overline{\mathbf{3}},\mathbf{1},1/3)$ & 0  & $-2$\\
$\bar{S}_1$ & $(\overline{\mathbf{3}},\mathbf{1},-2/3)$ & 0  & $-2$\\
$R_2$ & $(\mathbf{3},\mathbf{2},7/6)$ & 0  & $0$ \\ 
$\tilde{R}_2$ & $(\mathbf{3},\mathbf{2},1/6)$ & 0  & $0$\\
\hline
\hline
$U_3$ & $(\mathbf{3},\mathbf{3},2/3)$ & 1  & $0$\\
$\tilde{U}_1$ & $(\mathbf{3},\mathbf{1},5/3)$ & 1  & $0$ \\
$U_1$ & $(\mathbf{3},\mathbf{1},2/3)$ & 1  & $0$\\
  $\bar{U}_1$ & $(\mathbf{3},\mathbf{1},-1/3)$ & 1  & $0$\\
  $V_2$ &$(\overline{\mathbf{3}},\mathbf{2},5/6)$ & 1  & $-2$\\
$\tilde{V}_2$ &$(\overline{\mathbf{3}},\mathbf{2},-1/6)$ & 1  & $-2$\\
\hline 
\end{tabular}
\caption{\emph{List of scalar and vector LQs.}}\label{tab:LQs} 
\end{table}

\begin{figure}[!ht]
  \centering
  \begin{tabular}{ccc}
    \includegraphics[scale=0.7]{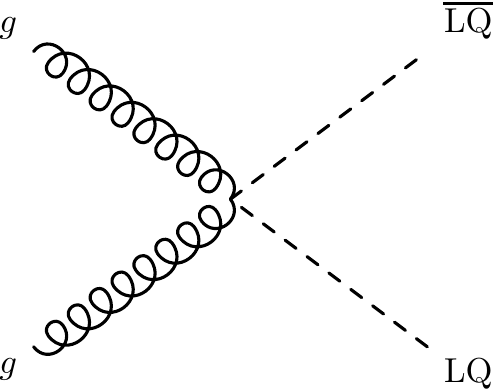}&&     \includegraphics[scale=0.7]{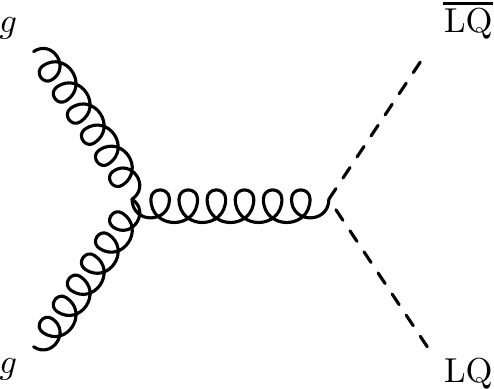}\\
        \includegraphics[scale=0.7]{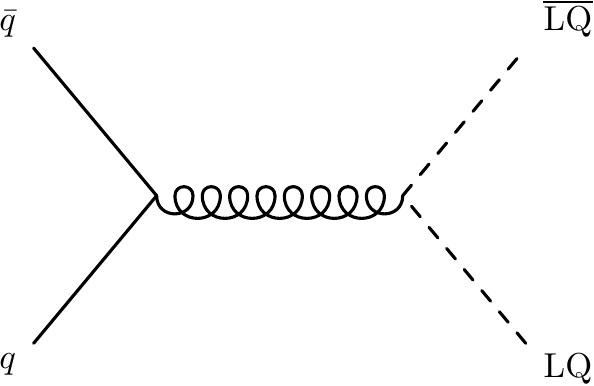}&&     \includegraphics[scale=0.7]{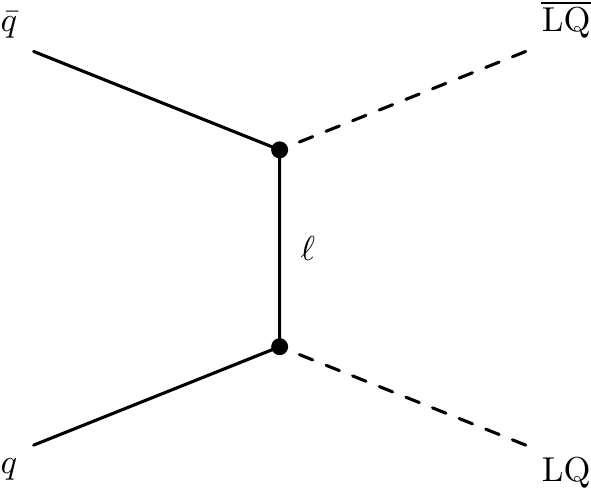}\\
  \end{tabular}
  \caption{\emph{Representative diagrams for leptoquark pair production at $pp$ colliders. Dots denote the $\mrm{LQ}$-$q$-$\ell$ coupling.}}
  \label{fig:LQpair}
\end{figure}

The most important phenomenological characteristic of light LQs (of mass of the order few TeV)
is their color triplet nature allowing them to be produced in pairs
via strong interactions in a largely model independent manner. In this
section we will focus on the on-shell production of LQs at $p p$
colliders, since the current mass constraints are dominated by
LHC searches. For specific signatures of LQ production in colliders
with alternative initial states, see~\cite{Dorsner:2016wpm}. Pair
production of leptoquarks is model independent for the
$g g \to \mrm{LQ} \overline{\mrm{ LQ}}$ partonic process, while the
parton level process $q\bar q \to \mrm{LQ} \overline{\mrm{LQ}}$ is affected also by the
$t$-channel lepton exchange diagram~(bottom right diagram in
Fig.~\ref{fig:LQpair}) that introduces some model dependence when the
leptoquark flavour couplings are non-negligible. The partonic cross sections for pair production at leading order are~\cite{Grifols:1981aq,Antoniadis:1983hz,Eichten:1984eu,Altarelli:1984ve,Dawson:1983fw,Blumlein:1996qp}:
\begin{align}
\label{eq:lqpair_gg}
\hat{\sigma}(gg \to \phi \bar\phi) &=   \frac{\alpha_3^2 \pi}{96 \hat{s}} \left[\beta (41-31 \beta^2)+(18 \beta^2-\beta^4-17)\log\frac{1+\beta}{1-\beta}\right],\\
\label{eq:lqpair_qq}
\hat{\sigma}(q\,\bar{q} \to \phi \bar \phi) &=   \frac{2 \alpha_3^2 \pi}{27 \hat{s}} \beta^3,
\end{align}
where $\hat s$ is the partonic center-of-mass energy squared, $\alpha_3$
is the strong coupling constant, and
$\beta = \sqrt{1-4m_\phi^2/\hat s}$.  A model independent study of weak
doublet scalar LQs at the LHC and the interplay with low energy flavour
processes was performed in~\cite{Dorsner:2014axa}, where it was also shown that searches for single LQ
production could be more sensitive in the regime of
large Yukawas and/or LQ masses. An analysis of pair and single
production, along with the corresponding UFO model files
\texttt{LQ\_NLO}, both for scalar and a vector leptoquark has been
presented in~\cite{Dorsner:2018ynv}.

\begin{figure}[!ht]
  \centering
  \begin{tabular}{ccc}
    \includegraphics[scale=0.7]{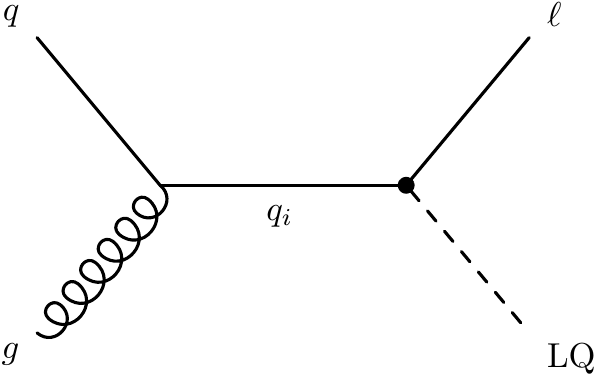}&&     \includegraphics[scale=0.7]{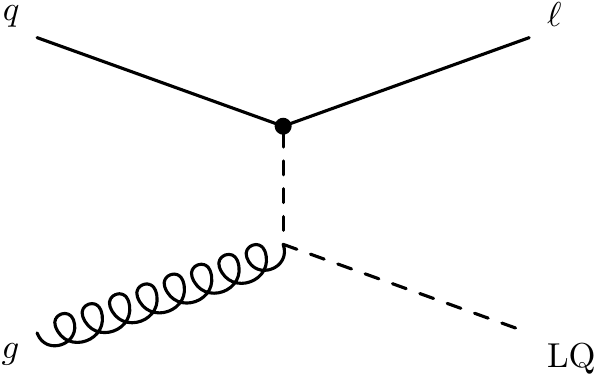}\\
  \end{tabular}
  \caption{\emph{Representative diagrams for single leptoquark production at $pp$ colliders. Dots denote the $\mrm{LQ}$-$q$-$\ell$ coupling.}}
  \label{fig:LQsingle}
\end{figure}

On the other hand, single leptoquark production at $pp$ colliders is always model dependent (Fig.~\ref{fig:LQsingle}). Single leptoquark searches are more effective at larger LQ masses~\cite{Dorsner:2014axa}.
On the decay side of the process, experimental searches for pair and single LQ production are targeting a resonance in the $j\ell$ channel. The decay width of a scalar leptoquark into a lepton-quark final state is given by~\cite{Plehn:1997az, Dorsner:2018ynv}:
\begin{equation}
  \label{eq:SLQWidth}
  \Gamma(\phi \to q \ell) = \frac{|y_{q\ell}|^2 m_\phi^2}{16\pi} \left[ 1+ \frac{\alpha_s}{\pi} \left(\frac{9}{2}-\frac{4\pi^2}{9}\right)\right].
\end{equation}
Current bounds from dedicated leptoquark pair production have been
commonly extracted in the framework that assumed LQ coupling only to
a single generation of SM fermions, whereas realistic LQ scenarios could posses richer flavour structure~\cite{Diaz:2017lit, Dorsner:2016wpm}. The experimental upper bounds are given for the product of cross section and the LQ branching fraction probability $\beta^2$, where $\beta$ is the probability for LQs to decay to a final state with charged leptons. There have been numerous analyses performed at the LHC for leptoquarks being either of  1st~\cite{Sirunyan:2018btu}, 2nd~\cite{Aaboud:2016qeg}, or 3rd~\cite{Aad:2015caa, Khachatryan:2016jqo, Sirunyan:2017yrk,Sirunyan:2018kzh,Sirunyan:2018nkj} generation. More recent studies, motivated by the observed lepton universality violation in $B$-meson decays, allow also for cross-generational couplings, e.g.~\cite{Sirunyan:2018ruf}.
Finally, also single LQ production channels are being studied \cite{Khachatryan:2015qda}. To conclude, the current lower bounds on leptoquark masses from direct searches at the LHC range from several $100\,$ GeV to above the TeV scale, where the exact bound depends on the size of the flavour couplings.

\subsection{Other exotic searches}

\begin{figure}[h]
 \centering
 \includegraphics[width=0.4\textwidth]{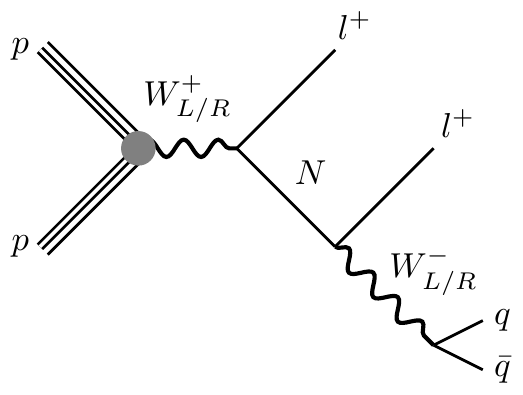}
 \caption{\emph{Diagram for direct searches of neutrinos via $W_{L/R}$ production.}}
 \label{fig:WRN}
\end{figure}

Many unified theories include heavy sterile neutrinos that contribute to the mass of the light neutrinos via type-I seesaw mechanism (see Sec.~\ref{sec:numasses}). These are often associated with the symmetry breaking of a left-right sector of the theory, and thus they are expected to be heavier than the EW scale. Direct searches for heavy neutrinos at colliders often target a decay channel where final state has two same-sign leptons, via $s$ or $t$-channel production of a gauge boson that can be left or right handed~\cite{Deppisch:2015qwa}. Figure~\ref{fig:WRN} shows the Feynman diagram for the golden channel for heavy neutrino searches, $pp \to W \to Nl \to Wll \to lljj$. The ATLAS and CMS experiments at the LHC have performed searches for heavy neutrinos in LR models with masses $M_N \approx (20, 1600)$ GeV and have imposed strong limits on the couplings between active and sterile neutrinos~\cite{Aaboud:2018spl,Sirunyan:2018pom,Sirunyan:2018mtv}.

Other searches for sterile neutrinos are performed in beam dump experiments~\cite{Drewes:2015iva}, where the neutrinos are produced in semileptonic decays of mesons, with masses below 2 GeV~\cite{Bergsma:1985is, Bernardi:1987ek, Vilain:1994vg, Vaitaitis:1999wq}. For intermediate masses below the $Z$ resonance, the strongest limits come from the decay $Z$ bosons by the LEP experiments DELPHI and L3~\cite{Adriani:1992pq,Abreu:1996pa}.

In addition to singlet fermions and coloured leptoquarks, the LHC looks for heavy colourless vector bosons as part of their exotic searches programme. Charged $W'$ and neutral $Z'$ vector bosons are predicted in a number of GUT frameworks and they can often live at low scales, which positions them within the reach of colliders. Clear examples of this are the left-right symmetric models described in Section~\ref{sec:lr}, that predict light $W_R$ and $Z_R$ bosons, or the light $Z_N$ appearing in $\mathrm{E}_6$SSM models .

These states are produced at $pp$ colliders through Drell-Yan processes $pp \to W'/Z'$ and subsequently decay into leptons or jets. One of the most targeted processes for $W'$ involve the decay into heavy neutrinos, as in Fig.~\ref{fig:WRN}, with two same or opposite sign leptons (depending on the Majorana or Dirac nature of the heavy neutrinos) and jets~\cite{Aaboud:2018spl,Sirunyan:2018pom}. These searches often use a simplified model where $g_R = g_L$ and $M_N = M_W'/2$ resulting in high exclusion limits with $M_{W'} \gtrsim 4.5$ TeV, but it has been shown that these limits weaken somewhat in more general models~\cite{Deppisch:2014qpa, Deppisch:2014zta, Deppisch:2015cua}. CMS also reported a search for $W'$ where the vector boson decays to $\tau \bar{\nu}_\tau$, the $\tau$ decaying hadronically~\cite{Sirunyan:2018lbg}, with slightly weaker limits.

Narrow resonance searches for $Z'$ have been performed by ATLAS and CMS, targeting final states with two opposite-sign leptons. These searches have yielded model-dependent exclusion limits on $M_{Z'}$. For $E_6$-inspired $Z'$, the limits vary around $M_{Z'} \gtrsim (3, 3.5)$ TeV, whereas for LR models they are moderately stronger $M_{Z'} \gtrsim 4$ TeV~\cite{Aaboud:2017buh,CMS:2016abv}.

Lastly, GUTs predict a plethora of different scalar states that can be observed at the LHC if they are light enough, e.g. $\Delta_{L,R}$ in LR symmetric models. Searches for neutral and singly charged scalar bosons are identical to the searches for supersymmetric Higgs bosons in Section~\ref{sec:SUSYsearches}, so we will not repeat them here. Doubly-charged scalars, such as the $\delta_L^{\pm\pm} \in \Delta_L$ in LR models, have been studied by ATLAS and CMS in multilepton final states~\cite{Aaboud:2017qph,CMS:2017pet}, diboson final states~\cite{Aaboud:2018qcu} and in long-lived particle studies~\cite{Aaboud:2018kbe} with model-dependent limits below 1 TeV.

\section{Precision tests of unification}
\label{sec:precision}

\subsection{Proton decay}

Unified theories may contain gauge or scalar bosons that mediate transitions between leptons and quarks. These transitions violate baryon $B$ and lepton $L$ number separately and hence can cause the rapid decay of nucleons~\cite{Abbott:1980zj, Marciano:1981un, Dimopoulos:1981dw, Nath:1985ub, Arnowitt:1985iy, Hisano:1992jj, Lucas:1996bc, Goto:1998qg, Murayama:2001ur, Dermisek:2000hr}.

In the language of Effective Field Theory, nucleon decay transitions are generated by higher dimensional operators suppressed by the mass of the heavy mediator. The most relevant contribution to nucleon decay comes from dimension-6 operators of the form $qqql$, mediated either by a gauge or scalar boson. In SUSY GUTs, however, dimension 4 and 5 operators can appear, involving $R$-parity violating interactions and mixing among sfermions, respectively. Assuming the conservation of $R$ parity and minimal flavour violation (MFV) in the sfermion sector, however, dimension 4 and 5 contributions can be made negligible~\cite{Bajc:2002bv, Bajc:2002pg, EmmanuelCosta:2003pu}. Therefore, dimension 6 operators dominate the contributions to nucleon decay. These are, in general, model dependent, so calculating the decay width requires knowledge of the flavour structure at the GUT scale and varies among the different decay channels~\cite{Dorsner:2004xa,FileviezPerez:2004hn}. However, with some simplifying assumptions one can approximate the decay width of the proton as~\cite{Nath:2006ut}
\begin{equation}
 \Gamma_p \approx \alpha_{GUT}^2 \frac{m_p^5}{M_{GUT}^4}.
\end{equation}

\begin{figure}[t]
 \centering
 \includegraphics[width=0.8\textwidth]{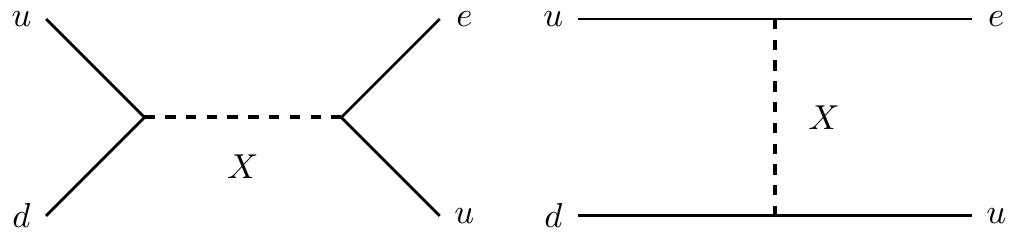}
 \caption{\emph{Parton level $s$ and $t$-channel diagrams for the proton decay channel $p \to e^+ \pi^0$ for a gauge or scalar boson mediator $X$.}}
 \label{fig:pdecay}
\end{figure}

There are several decay channels for the proton and neutron, each with a different experimental bound. The most stringent of them, known as the \textit{gold channel} for proton decay is $p \to e^+ \pi^0$, whose parton-level diagrams can be seen in Figure~\ref{fig:pdecay}, and with a lower limit on the half-life, set by the Super-Kamiokande, of $\tau > 1.6 \times 10^{34}$ years~\cite{Miura:2016krn}. Other processes with slighly lower bounds are $\tau(p \to \mu^+ \pi^0) > 7.7 \times 10^{33}$ years~\cite{Miura:2016krn}, $\tau(p \to \nu K^+) > 5.9 \times 10^{33}$ years~\cite{Abe:2014mwa} and $\tau(nn \to e^\pm\mu^\mp) > 4.4 \times 10^{33}$ years~\cite{Sussman:2018ylo}.

The next generation of experiments for nucleon decay has already been proposed. Hyper Kamiokande will take the place of Super-K and has a projected sensitivity in the golden channel $p \to e^+ \pi^0$ of $\sim 10^{35}$ years~\cite{Abe:2018uyc}. The Deep Underground Neutrino Experiment (DUNE)~\cite{Acciarri:2015uup} expects to improve the limit on $p \to \nu K^+$ to $\sim 3 \times 10^{34}$ years. These increased limits will probe unified teories at the highest scales and could be the smoking gun for them. In case of a positive signal from either of these experiments, precision calculations of proton decay processes with controlled uncertainties will become invaluable~\cite{Kolesova:2014mfa,Kolesova:2016ibq}.

\subsection{Flavour phenomenology of light leptoquarks}

Several flavour couplings of scalar leptoquarks to leptons and quarks, which are in general connected to the GUT contractions of scalar and fermionic representations, allow their virtual effects to be tested in low-energy flavour observables. Such are the decays of hadrons or leptons, precision observables at LEP, and static properties of particles such as dipole moments. On the high-$p_T$ front, the LHC is also becoming competitive as a flavour probe for virtual effects of particles that are too massive to be produced on-shell. 

The correlations between lepton-quark-LQ couplings are determined at low scales by the weak isospin and hypercharge. As an example, consider the weak doublet leptoquark $R_2 (3,2,7/6)$ (see Table~\ref{tab:LQs}), which can couple to two-types of lepton-quark bilinears:
\begin{equation}
  \label{eq:R2} 
\begin{split}
  \mathcal{L} &=  Y_{R}^{ij} \bar{Q}^\prime_i \ell^\prime_{Rj} R_2+ Y_L^{ij} \bar{u}^\prime_{Ri} \widetilde{R}_2^\dagger L^\prime_j \\
  &=(V Y_R)^{ij} \bar{u}_{Li}\ell_{Rj}R_2^{\frac{5}{3}} + Y_R^{ij} \bar{d}_{Li}\ell_{Rj} R_2^{\frac{2}{3}}\\
&\phantom{=}+Y_L^{ij} \bar{u}_{Ri} \nu_{Lj} R_2^{\frac{2}{3}}- Y_L^{ij} \bar{u}_{Ri}\ell_{Lj} R_2^{\frac{5}{3}}.
\end{split}
\end{equation}
Here $i,j$ are the flavour indices, primed fields are written in the interaction basis, unprimed fields are in the mass basis, except for the neutrinos which are aligned with charged leptons. There are three important features in the above Lagrangian. First, since $R_2$ is a weak doublet it must couple to left-handed quark doublets, which implies that CKM matrix $V$ relates the couplings of up-type and down-type quarks. Second, $R_2$ couples to both chiralities of quarks and leptons, which leads to scalar and/or tensor effective interactions and could lead to enhanced effects in meson mixing amplitudes, dipole moments, and radiative decays~\cite{Dorsner:2016wpm}. Third, as $F = 0$ for $R_2$ we cannot construct interaction term with diquark coupling, implying that proton cannot decay via $\Delta(B-L)=0$ process. Generalising to other LQ states, weak triplets only talk to the left-handed fermions ($2\otimes 2$), leading to strictly chiral interaction, whereas singlet LQs can talk to $2 \otimes 2$ and $1 \otimes 1$ fermion bilinears.

Among the flavour constraints, leptoquarks naturally (at tree-level) contribute to semileptonic effective operators at low scales, therefore the most relevant observables are (semi-)leptonic decays of mesons, baryons, or $\tau$ leptons. The most notable charged-current and flavour changing neutral current (FCNC) constraints, and the general framework to address them in leptoquark models, have been spelled out in~\cite{Leurer:1993em,Davidson:1993qk,Dorsner:2016wpm}. The most constraining are the FCNC observables, where the tree-level LQ contribution can easily stand out of the SM signal, which is 1-loop suppressed in the case of quark FCNC and absent in the case of lepton FCNC. Effective dimension-6 interactions for 4 lepton or 4 quark operators, which drive the $\ell \to \ell' \ell' \ell''$ (see Sec.~\ref{sec:lfv}) and meson mixing processes, occur at one-loop~\cite{Dorsner:2016wpm}. Therefore, meson mixing is in general not among the strictest constraints on LQs (for $B_s$ mixing see e.g.~\cite{DiLuzio:2017fdq}). 

\subsection{Lepton flavour universality}

Lepton flavour universality~(LFU) ratios, defined as ratios between rates for processes that differ only in lepton flavour, are very well suited to test the validity of the SM. The main advantage is that in the Standard Model LFU is respected by all gauge interactions, the only breaking comes from mass splitting among leptons, which leads to efficient cancellation of hadronic and parametric uncertainties in LFU ratios. Recently, two LFU ratios in $B$-meson decays have been observed
\begin{equation}
    R_{D^{(*)}} = \frac{ \mathcal{B}( B \to D^{(*)} \tau    \bar\nu_\tau)}{\mathcal{B}( B \to D^{(*)} l \bar\nu_l)}, \quad R_{K^{(\ast)}} = \frac{\mathcal{B} (B\to K^{(\ast)} \mu\mu)}{\mathcal{B} ( B\to K^{(\ast)} e e)},
\end{equation}
where $l = e,\mu$. Several experiments found that the ratios $R_{D^{(*)}}$ are larger than $R_{D^{(*)}}^\mathrm{SM}$. The measurements of $R_{D}$~\cite{Lees:2013uzd, Hirose:2016wfn, Aaij:2015yra} differ by $\sim 2\,\sigma$ with respect to the SM prediction~\cite{Lattice:2015rga} and by $\sim 3\,\sigma$ in the case of $R_{D^{\ast}}$~\cite{Fajfer:2012vx, Bernlochner:2017jka, Bigi:2017njr}. Combined significance reaches $4\sigma$ deviation from the SM~\cite{Amhis:2016xyh}.  
 The LHCb experiment has also measured $R_{K^{(\ast)}}$ LFU ratios, related to the neutral-current process $b\to s l l$, and found them to be lower than expected in the SM. While $R_K$ was measured in a single kinematical region, $q^2 \in [1.1,6]\,\mrm{GeV}^2$~\cite{Aaij:2014ora}, $R_{K^*}$ was measured also in the ultra-low region $q^2\in [0.045,1.1]\,\mrm{GeV}^2$~\cite{Aaij:2017vbb}. Each of the $R_{K^{(*)}}$ measurements is $\sim 2.5\,\sigma$ below the SM prediction level~\cite{Hiller:2003js,Bordone:2016gaq}, and furthermore, there are discrepancies in $b\to s\ell\ell$ driven decays that are coherent with the deviation in $R_K^{(*)}$ if there is $\sim 20\%$ reduction in the vector Wilson coefficient $C_9$~\cite{DAmico:2017mtc,Capdevila:2017bsm}.
 
Light leptoquarks are prime candidates to explain one or both of those puzzles. For the $R_{D^{(*)}}$ the effective Lagrangian contains four relevant operators:
\begin{equation}
\begin{split}
{\cal L}^{b \to c \tau \bar \nu_\tau}_{\mathrm{eff}} = -\frac{4 \, G_F}{\sqrt{2}} &V_{cb}\big[  (1+g_{V_L}) 
(\bar{c}_L \gamma_\mu b_L)(\bar{\tau}_L \gamma^\mu \nu_{\tau L})\\ 
&+ g_{S_L}(\mu)\, (\bar{c}_R  b_L)(\bar{\tau}_R \nu_{\tau L}) + g_{S_R}(\mu)\, (\bar{c}_L  b_R)(\bar{\tau}_R \nu_{\tau L})\\
&+ g_T(\mu)\, (\bar{c}_R  \sigma_{\mu \nu} b_L) (\bar{\tau}_{R} \sigma^{\mu \nu}\nu_{\tau L})
\big].
\end{split}
\end{equation}
Model independently it has been shown that $R_{D^{(*)}}$ can be explained either by rescaling the SM semileptonic operator ($g_{V_L}$), by turning on $g_T$, or by particular combinations of scalar and tensor operators, $g_{S_L} = \pm 4 g_T$ operators, that arise in presence of a non-chiral LQ~\cite{Bhattacharya:2014wla, Freytsis:2015qca,Buttazzo:2017ixm,Angelescu:2018tyl}. In order to address $R_{K^{(*)}}$ one has to modify the vector Wilson coefficient $C_9$ whereas the axial Wilson coefficient $C_{10}$ may also be present in the effective Lagrangian:
\begin{equation}
\mathcal{H}^{b\to s \mu\mu}_{\mathrm{eff}} = -\dfrac{\alpha G_F V_{tb} V_{ts}^* }{\sqrt{2}\pi}  \,  (\bar{s}_L\gamma_\mu  b_L) (\bar\mu\gamma^\mu (C_9 + C_{10}\gamma^5)\mu.
\end{equation}
Also purely left-handed scenarios with $C_9 = -C_{10} \approx -0.6$, which are characteristic of LQ weak-singlet or triplet exchange, are in good agreement with $R_{K^{(*)}}$ and the global fit of $b\to s \ell \ell$. Such left-handed leptoquark solutions have been put forward: triplet scalar $S_3$, singlet vector $U_1$, triplet vector $U_3$~\cite{Buttazzo:2017ixm,Angelescu:2018tyl}. For loop-level explanation of $R_{K^{(*)}}$ one can also invoke singlet $S_1$~\cite{Bauer:2015knc} or doublet $R_2$~\cite{Becirevic:2017jtw}, but at the price of large couplings. There are several proposals with scalar leptoquarks that address $R_{K^{(*)}}$ and/or $R_{D^{(*)}}$~\cite{Bauer:2015knc,Becirevic:2016yqi,Becirevic:2016oho,Chen:2017hir,deMedeirosVarzielas:2018bcy,Camargo-Molina:2018cwu,Crivellin:2017zlb}, some in the context of unified theories such as $SU(5)$~\cite{Cox:2016epl,Dorsner:2017ufx,Becirevic:2018afm}, left-right symmetry~\cite{Das:2016vkr}, Pati-Salam~\cite{Heeck:2018ntp,Aydemir:2018cbb}, $SO(10)$~\cite{Aydemir:2019ynb} and others~\cite{Faber:2018qon}. Recently it was realised that a singlet vector leptoquark $U^\mu_1(3,1,2/3)$ generates left-handed interactions and partially resolves both LFU puzzles, in many UV frameworks~\cite{Sahoo:2016pet,Crivellin:2018yvo,Sahoo:2018ffv}, including three-flavour extensions of the  Pati-Salam model~\cite{Assad:2017iib,Calibbi:2017qbu,DiLuzio:2017vat,Diluzio:2018zxy,Greljo:2018tuh,Bordone:2018nbg,Bordone:2017bld}.

Finally, moderately large leptoquark couplings dictated by the above LFU anomalies can be also observed in processes with virtual LQ exchange, typically in the $t$-channel,
resulting in a final state with at least one charged lepton. Inspired
by the abovementioned LFU anomalies, processes with final state leptons have
been studied, which can probe LQ scenarios for LFU
violation observed in $B$ meson
decays~\cite{Faroughy:2016osc,Greljo:2017vvb,Schmaltz:2018nls,Mandal:2018kau}. In
this case, LQ cannot be produced on-shell and the sensitivity does not
deteriorate abruptly with rising LQ mass. Instead there is a smooth
transition to the effective theory picture, where heavy LQ is
integrated out. Thus among the LQ induced processes the $t$-channel
has the best mass reach for LQs and it is thus complementary to pair and
single production~\cite{Dorsner:2014axa,Dorsner:2018ynv}.  Another recent set of
observables at the LHC, targeting the LQ scenarios that are well
suited to explain the LFU anomalies
$R_{D^{(*)}}$, are the searches with a single $\tau$ lepton in the
final state~\cite{Greljo:2018tzh}\footnote{See also~\cite{Bansal:2018eha} for constraints on couplings to light quarks.}.  Third generation leptoquarks could also be probed in
$t\bar t$ final states~\cite{Vignaroli:2018lpq}.

\subsection{Lepton flavour violation and dipole moments}
\label{sec:lfv}

In the SM with massive neutrinos, lepton flavour violation (LFV) can occur via the mixing in the neutrino sector. It is, however, heavily suppressed due to the GIM mechanism~\cite{Glashow:1970gm}, as the rate depends on the neutrino masses resulting in an unobservable prediction of order $10^{-55}$~\cite{Petcov:1976ff}. Extensions of the SM modify this prediction by introducing additional sources of lepton flavour violation~\cite{Cheng:1976uq,Bilenky:1977du,Kuno:1999jp}. New physics can then be probed by testing the deviations of certain lepton flavour violating processes with respect to the experimental limits. 

\begin{figure}[t]
 \includegraphics[width=0.3\textwidth]{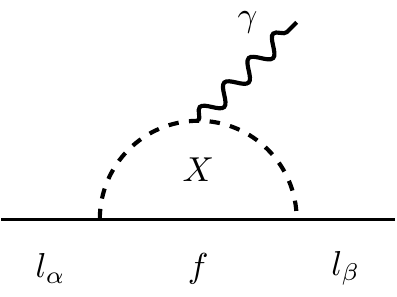} \quad
 \includegraphics[width=0.3\textwidth]{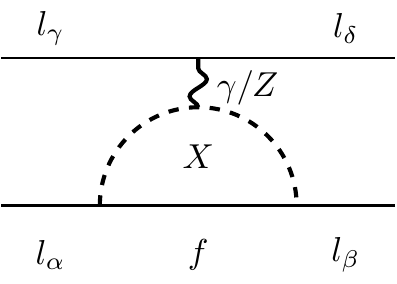} \quad
 \includegraphics[width=0.3\textwidth]{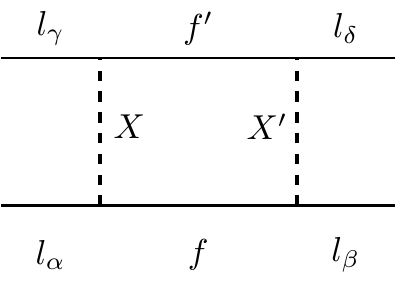}
 \caption{\emph{Diagrams contributing to LFV processes, $l^-_\alpha \to l^-_\beta \gamma$ (left) and $l^-_\alpha \to l^-_\beta l^-_\gamma l^+_\delta$ (centre and right) with $X$ and $X'$ scalar or vector mediators and $f$ and $f'$ fermions.}}
 \label{fig:lfv}
\end{figure}

Charged lepton flavour violating processes are typically of three types: $l^-_\alpha \to l^-_\beta \gamma$, $l^-_\alpha \to l^-_\beta l^-_\gamma l^+_\delta$ (with $\alpha \neq \beta$) and $\mu - e$ conversion in nuclei~\cite{Abada:2014kba}. One-loop contributions to the first two processes can occur through a dipole and box diagrams as depicted in Figure~\ref{fig:lfv}, with a scalar or vector mediator(s) $X(X')$ and a SM or exotic fermion(s) $f(f')$ running in the loop. Contributions to $\mu - e$ conversion follows from the penguin (centre) and box (right) diagrams with $l_\alpha = \mu$, $l_\beta = e$ and $l_{\gamma,\delta} = q$. 

Unified theories often contain a number of exotic states capable of fulfilling the role of $X$ and $f$ in Fig.~\ref{fig:lfv}, violating lepton flavour either by interactions between the leptons and mediators or by mixing in the leptonic sector. The latter case is realised in GUT models with heavy neutrinos (c.f. left-right models in Sec.~\ref{sec:lr}), where the mixing between active and sterile neutrinos enhances the LFV contribution~\cite{Bernabeu:1987gr, Ilakovac:1994kj, Deppisch:2005zm}. The contribution to the branching ratios of the most constraining LFV processes, $\mu \to e\gamma$, $\mu \to eee$ and $\mu -e$ conversion, in these models, with heavy neutrinos of mass $M_{N_I}$, active-sterile mixing $\Theta_{\alpha I}$, a right-handed gauge boson $W_R$ and left and right-handed scalar triplets $\delta_L$ and $\delta_R$, can be written as~\cite{Alonso:2012ji,Barry:2013xxa,Deppisch:2014zta}
\begin{align}
    BR(\mu \to e\gamma) &\sim 1.5\times 10^{-7} |\Theta_{e I}^*\Theta_{\mu I}|^2 \left(\frac{g_R}{g_L}\right)^4 \left(\frac{m_{N_I}}{m_{W_R}}\right)^4  \left(\frac{1\text{ TeV}}{M_{W_R}}\right)^4, \notag \\
    BR(\mu \to eee) &\sim \frac{1}{2} |\Theta_{e I}^*\Theta_{\mu I}|^2|\Theta_{e I}|^4 \left(\frac{g_R}{g_L}\right)^4 \left(\frac{m_{N_I}}{m_{W_R}}\right)^4  \left(\frac{M_{W_R}^4}{M_{\delta_R}^4} + \frac{M_{W_R}^4}{M_{\delta_L}^4}\right), \\
    R^N(\mu -e) &\sim0.73\times 10^{-9}  X_N |\Theta_{e I}^*\Theta_{\mu I}|^2
    \left(\frac{g_R}{g_L}\right)^4 \left(\frac{m_{N_I}}{m_{W_R}}\right)^4  \left(\frac{1\text{ TeV}}{M_{\delta_R}}\right)^4 \left(\log\frac{m_{\delta_R}^2}{m_\mu^2}\right)^2. \notag
\end{align}

In supersymmetric GUTs there are many possible sources of lepton flavour violation, parametrised by the mixing in the slepton sector of the MSSM, which has contributions to LFV processes of the type~\cite{Vicente:2015cka}
\begin{equation}
 BR(l_\alpha \to l_\beta \gamma) \approx \frac{48\pi^3\alpha_{\rm em}}{G_F^2} \frac{|(m_{\tilde{L}}^2)_{ij}|^2 + |(m_{\tilde{e}}^2)_{ij}|^2}{M_{SUSY}^8} BR(l_\alpha \to l_\beta \nu_\alpha \bar\nu_\beta).
\end{equation}
Off-diagonal entries in the slepton mass matrices can be the result of non-minimal flavour violating interactions or non-canonical Yukawa textures at the GUT scale, where the soft masses are supposed to unify~\cite{Grinstein:2006cg,Ciuchini:2007ha}. In addition, slepton mixing can be induced in minimal flavour violating (MFV) SUSY via seesaw mechanisms~\cite{Hisano:1995cp, Arganda:2005ji, Hirsch:2012ti, Rossi:2002zb, Deppisch:2004fa, Hirsch:2008dy, Esteves:2010ff} or, for Yukawa-unified theories (see Sec.~\ref{sec:yukawaunification}), it can depend on the CKM matrix at the GUT scale~\cite{Kuno:1999jp}. LFV contributions can also arise in SUSY models where $R$-parity is violated, explicitly or spontaneously, with interaction terms of the type $l_i l_j \tilde{\nu}_k$ that induce tree-level contributions to $l\to lll$ decays and $\mu \to e$ conversion, as well as new dipole contributions to $l\to l\gamma$~\cite{Dreiner:2006gu,deGouvea:2000cf,Arhrib:2012ax,Romao:1991tp,Hirsch:2009ee}.

\begin{table}[t]
 \centering
 \begin{tabular}{llc}
 \toprule
 \textbf{Process} & \textbf{Branch.~Frac.} & \textbf{Reference} \\
 \midrule
  $\mu^- \to e^- \gamma$ & $4.2\times 10^{-13}$ & MEG~\cite{TheMEG:2016wtm}\\
  $\tau^- \to e^- \gamma$ & $5.4\times 10^{-8}$ & BaBar~\cite{Aubert:2009ag},Belle~\cite{Hayasaka:2007vc}\\
  $\tau^- \to \mu^- \gamma$ & $5.0\times 10^{-8}$ & BaBar~\cite{Aubert:2009ag},Belle~\cite{Hayasaka:2007vc}\\
  \midrule
  $\mu^- \to e^-e^-e^+$ & $1.0\times 10^{-12}$ & SINDRUM~\cite{Bellgardt:1987du}\\
  $\tau^- \to e^-e^-e^+$ & $1.4\times 10^{-8}$ & BaBar~\cite{Lees:2010ez},Belle~\cite{Hayasaka:2010np}\\
  $\tau^- \to \mu^- \mu^- \mu^+$ & $1.2\times 10^{-8}$ & ATLAS~\cite{Aad:2016wce},BaBar~\cite{Lees:2010ez},Belle~\cite{Hayasaka:2010np},LHCb~\cite{Aaij:2014azz}\\
  $\tau^- \to \mu^-e^-e^+$ & $ 1.1\times 10^{-8}$ & BaBar~\cite{Lees:2010ez},Belle~\cite{Hayasaka:2010np}\\
  $\tau^- \to e^-e^-\mu^+$ & $0.84\times 10^{-8}$& BaBar~\cite{Lees:2010ez},Belle~\cite{Hayasaka:2010np}\\
  $\tau^- \to e^-\mu^-\mu^+$ & $1.6\times 10^{-8}$& BaBar~\cite{Lees:2010ez},Belle~\cite{Hayasaka:2010np}\\
  $\tau^- \to \mu^-\mu^-e^+$ & $0.98\times 10^{-8}$& BaBar~\cite{Lees:2010ez},Belle~\cite{Hayasaka:2010np}\\
  \midrule
  $\mu - e$ (Ti) & $1.7\times 10^{-12}$& SINDRUM II~\cite{Kaulard:1998rb}\\
  $\mu - e$ (Pb) & $4.6\times 10^{-11}$& SINDRUM II~\cite{Honecker:1996zf}\\
  $\mu - e$ (Au) & $8.0\times 10^{-13}$& SINDRUM II~\cite{Bertl:2006up}\\
  \midrule
  $d_e$ & $1.1 \times 10^{-29} ~e~cm$ & ACME II~\cite{Andreev:2018ayy} \\
  $d_\mu$ & $1.9 \times 10^{-19} ~e~cm$ & Muon g-2~\cite{Bennett:2008dy}\\
  $d_\tau$ & $4.5 \times 10^{-17} ~e~cm$ & Belle~\cite{Inami:2002ah}\\
  \midrule
  $a_e(10^{-13})$ & $11596521809.1 \pm 2.6$ & ~\cite{Hanneke:2008tm} \\
  $a_\mu(10^{-10})$ & $11659208.9 \pm 8.7$ & Muon g-2~\cite{Bennett:2006fi}\\
  \bottomrule
 \end{tabular}
 \caption{\emph{Upper bounds at $90\%$ C.L on LFV processes and EDMs, and measurements of AMMs, along with the experiments that provided them. The HFLAV average is quoted for limits by different experiments~\cite{Amhis:2016xyh}.}}
 \label{tab:lfvexp}
\end{table}

The anomalous electric, $d_i$, and magnetic, $a_i$, dipole moments of quarks and leptons follow from processes identical to the diagram on the left in Fig.~\ref{fig:lfv}, where $l_\alpha$ and $l_\beta$ have the same flavour. Hence contributions from heavy states running in the loops can have a strong effect that can be tested experimentally. As with LFV, the SM contribution to electric dipole moments (EDMs) is tiny, as it is proportional to the CP-violation phase in the CKM matrix~\cite{Raidal:2008jk, Abada:2015trh}. EDMs have not been observed experimentally, so deviations from the SM prediction due to CP-violation in other sectors is strongly constrained~\cite{Dimopoulos:1994gj}. Other sources of CP violation can appear in neutrino mixing~\cite{Abada:2015trh}, phases in fermion-sfermion couplings~\cite{Romanino:2001zf} or extended Higgs sectors~\cite{Jung:2013hka}. Anomalous magnetic moments (AMM), on the other hand, have been measured with extreme accuracy. In fact, the precision of both the experimental measurement and theoretical prediction for $a_\mu$ has shown a discrepancy of more than 3 standard deviations~\cite{Tanabashi:2018oca}. New physics contributions have been shown to resolve that tension, particularly in the context of supersymmetry~\cite{Stockinger:2006zn}.

In the presence of light leptoquarks anomalous dipole moments of leptons or quarks are one-loop processes~\cite{Dorsner:2016wpm}. A special feature of non-chiral leptoquarks, such as $R_2$ with couplings~\eqref{eq:R2}, is that both $l_{ij}$ and $r_{ij}$ are non-zero in the interaction Lagrangian
$\bar q^i \left[l_{ij} P_R + r_{ij} P_L \right] \ell^j \phi$
which then leads to the anomalous moment of the muon:
\begin{equation}
  \label{eq:amu}
  \begin{split}
   a_\mu = -\frac{3 m_\mu^2}{8\pi^2 m_\phi^2} \sum_q \Bigg[&
    \left(|l_{q\mu}|^2 + |r_{q\mu}|^2 \right) \left(Q_\phi/4 -
      1/6\right)
 -\frac{m_q}{m_\mu} \log\frac{m_q^2}{m_\phi^2} \re (r_{q\mu}^* l_{q\mu})  \left(Q_\phi -1\right)   \Bigg],
  \end{split}
 \end{equation}
 where $m_\phi$ and $Q_\phi$ are the charge and mass of the leptoquark and $q$ is the flavour of the quark in the loop (see the leftmost diagram in Fig.~\ref{fig:lfv}). Shown is the leading order contribution in $m_q$. The first term increases $a_\mu$ only when $Q_\phi > 2/3$ and it is present for all scalar LQ states that couple to a muon. The second term is relevant for non-chiral LQs and it is chirally enhanced by $m_q/m_\ell$, possibly leading to large effects with moderate couplings to $b$ or $t$ quark. Furthermore, the sign of the non-chiral term is adjustable. On the other hand, the same mechanism also enhances dipole LFV transitions, e.g. $\mu \to e \gamma$, $\tau \to \mu \gamma$~\cite{Becirevic:2017jtw}. Non-chiral LQs may also generate quark or lepton electric dipole moments at 1-loop~\cite{Dekens:2018bci,Fuyuto:2018scm}.

Whichever the mechanism, it is clear that LFV and anomalous dipole moments are predicted by many GUT models, with varying strengths for different processes. The full list of processes and their current experimental upper bounds and measurements can be seen in Table~\ref{tab:lfvexp}, where the experiments that have studied the processes are detailed. Furthermore, new experiments are being developed at this moment that will attempt to improve the limits on processes like $\mu \to eee$ (Mu3e~\cite{Blondel:2013ia}) and $\mu - e$ conversion (COMET~\cite{Kurup:2011zza}, Mu2e~\cite{Kutschke:2011ux}), with projected limits up to four orders of magnitude stronger than previous studies. Additionally, a new measurement of $a_\mu$ has been performed by the Muon g-2 experiment and it is expected to be released soon~\cite{Chapelain:2017syu}, which may confirm the deviation observed before, and thus further motivate the need of new physics.

\subsection{Neutrinoless Double Beta Decay}

This rare nuclear process corresponding to a simultaneous conversion of two neutrons to two protons and two electrons within the nucleus~\cite{Furry:1939qr} is of great interest for particle physics, as it clearly does not conserve lepton number, and thus violates the corresponding accidental Abelian global symmetry of the SM. Consequently, a strong experimental effort is being made to observe this unique process. Unfortunately, its observation will be very difficult, as $\ovbb$ decay is expected to be extremely rare. There is a number of experiments, some in operation, other being constructed or planned, attempting to measure the decay. An overview of the major collaborations is shown in Tab.~\ref{tab:0vbbExp}. The current experimental lower limits on its half-life are around $10^{26}$ years~\cite{Agostini:2017iyd,KamLAND-Zen:2016pfg} and the future searches should reach sensitivities by one or two orders of magnitude higher.

\begin{table}[t!]
\centering
\begin{tabular}{ccccccc} 
\toprule 
Experiment & Isotope & Status & $M$ [kg] & $T_{1/2}^{\ovbb}$ limit [y] \\ 
\midrule
CUORE & ${}^{130}$Te & running & 200 & ($3.5\times 10^{26}$) \\
EXO-200 & ${}^{136}$Xe & running & 110 & $1.1\times 10^{25}$ \\
nEXO & ${}^{136}$Xe & R\&D & 5000 & ($10^{27}\text{-}10^{28}$) \\
GERDA & ${}^{76}$Ge & running & 21.6 & $5.3\times 10^{25}$ \\
& & in progress & ~40 & ($\sim 10^{26}$) \\
KamLAND-Zen & ${}^{136}$Xe & running & 383 & $1.1\times 10^{26}$ \\
& & in progress & 600 & ($2\times 10^{26}$) \\
LEGEND & ${}^{76}$Ge & R\&D & 200 & ($\sim 10^{27}$) \\
& & R\&D & 1000 & ($\sim 10^{28}$) \\
Majorana Dem. & ${}^{76s}$Ge & running & 44.1 & $1.9\times 10^{25}$ \\
NEXT & ${}^{136}$Xe & in progress (demo) & 100 & ($5.9\times 10^{25}$) \\
SNO+ & ${}^{130}$Te & in progress & 1300 & ($2\times 10^{26}$) \\
SuperNEMO & ${}^{82}$Se (${}^{150}\text{Nd}$) & in progress (demo) & 100 & ($\sim 10^{26}$)\\
\bottomrule
\end{tabular}
\caption{\emph{An overview of both current and future major $\ovbb$ decay searches. For each experiment the following information is shown: used isotope, operational status, the deployed mass $M$ of the isotope in question and the measured or expected (for experiment in preparation these values are shown in parentheses) sensitivity $T_{1/2}^{\ovbb}$. For some experiments (GERDA, KamLAND-Zen, LEGEND) characteristics of more stages of development are given. In case of SuperNEMO, the primary isotope to be tested is ${}^{82}$Se and in the future the measurement will be repeated with a ${}^{150}\text{Nd}$ source.}}
\label{tab:0vbbExp}
\end{table}
\begin{figure}[h]
 \centering
 \includegraphics[width=0.3\textwidth]{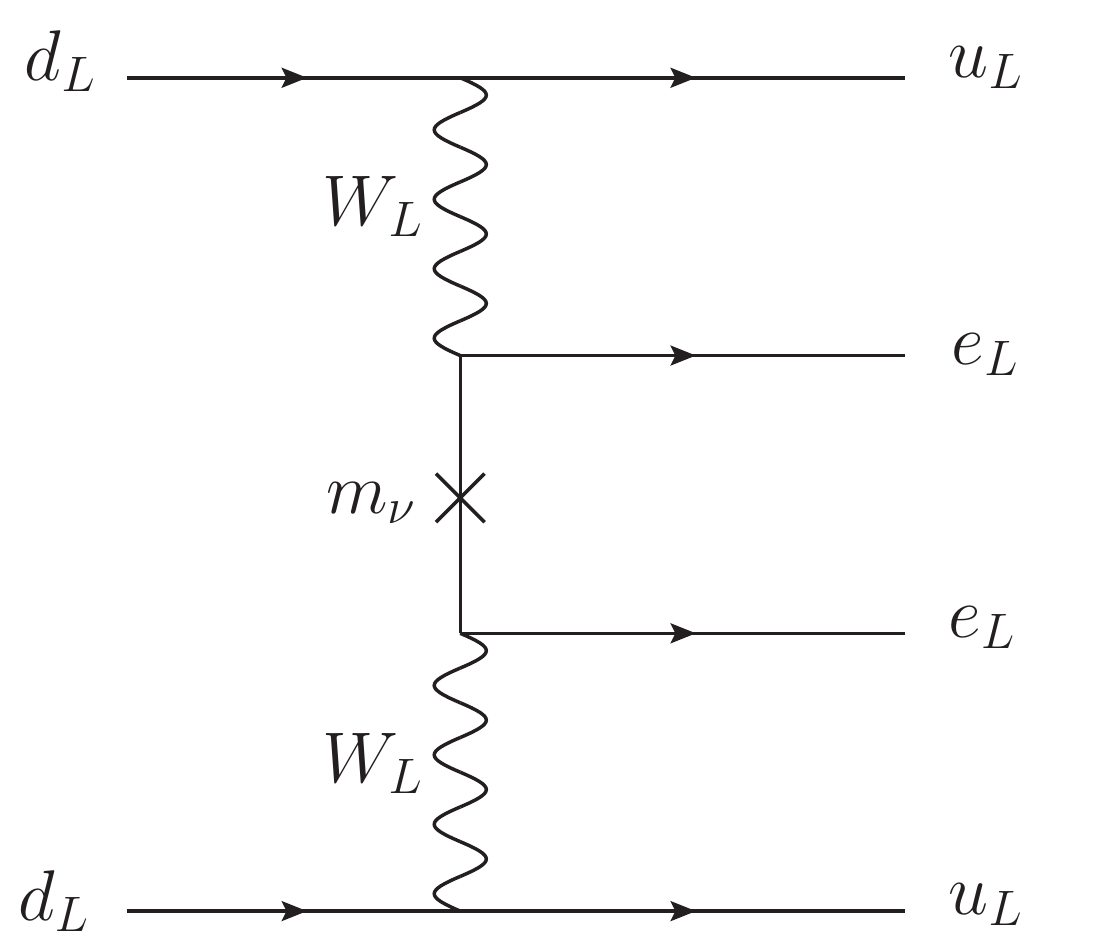}\quad
  \includegraphics[width=0.3\textwidth]{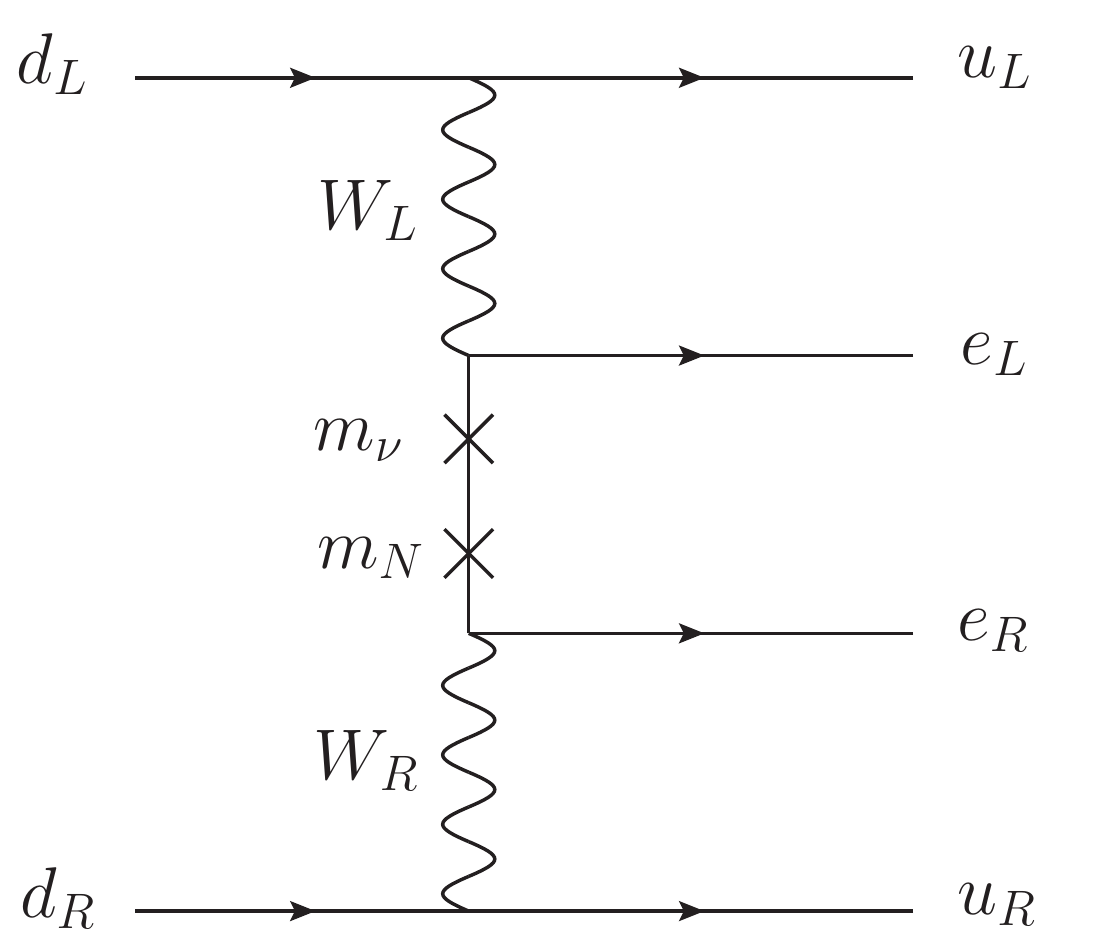}
 \caption{\emph{The standard mass mechanism of $\ovbb$ decay (left) and a non-standard contribution that can be constructed in the left-right symmetric models using vector currents of opposite chiralities (right).}}
\label{fig:lrovbb1}
\end{figure}

As can be shown, the existence of $\ovbb$ decay implies the Majorana nature of neutrinos (and vice versa)~\cite{Schechter:1981bd,Nieves:1984sn,Takasugi:1984xr} and as such it represents one of the best probes of this BSM hypothesis. Therefore, in GUT models allowing for Majorana neutrino mass generation $\ovbb$ decay can be in principle always triggered. This, however, does not say anything about the underlying mechanism and the resulting $\ovbb$ decay rate. The prominent \emph{standard (mass) mechanism} of $\ovbb$ decay assumes a light Majorana neutrino exchange between the two beta-decaying neutrons. In the SM with light massive neutrinos this process can be depicted as shown in Fig.~\ref{fig:lrovbb1} (left). Besides the standard scenario a number of \emph{non-standard mechanisms} triggering $\ovbb$ decay can be constructed. The effective treatment of these exotic mechanisms can be conveniently employed, see e.g. Refs.~\cite{Pas:1999fc,Pas:2000vn,Deppisch:2017ecm,Cirigliano:2018yza,Graf:2018ozy}.

As for the UV-complete $\ovbb$ decay mechanisms, a variety of interesting ones can be constructed within GUTs. For instance, in the left-right symmetric models (where, of course, the standard light neutrino exchange is available) one can think of several exotic mechanisms involving exchange of heavy neutrino as well as light and heavy $W$ vector bosons~\cite{Deppisch:2012nb}. In the simplest exotic case the light neutrino exchange is substituted by a heavy right-handed neutrino exchange, which means that the involved vector currents and emitted electrons must be also right-handed. Due to the large mass of the propagating neutrino, the interaction can be considered to be contact and we refer to this contribution as to short-range mechanism. Since the right-handed currents are present in left-right symmetric models, it is also possible to draw $\ovbb$ decay mechanisms, in which the neutrino exchange does not violate chirality. This means that the contribution is not proportional to the neutrino mass and the two outgoing electrons are of opposite chiralities. A possible mechanism of this type is depicted in Fig.~\ref{fig:lrovbb1} (right). As apparent, the diagram involves one right-handed and one left-handed vector current and since the light neutrino propagator is present, one refers to this contribution as to a long-range $\ovbb$ decay mechanism.

\begin{figure}[h]
 \centering
  \includegraphics[width=0.3\textwidth]{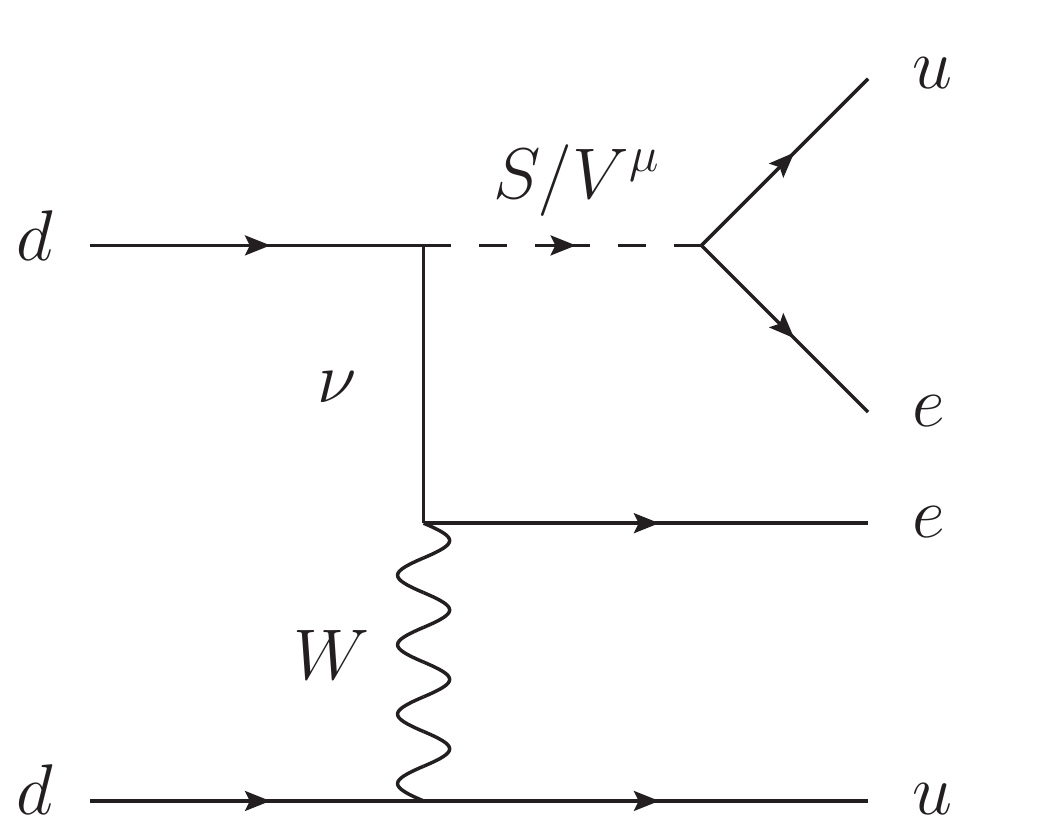}\quad
  \includegraphics[width=0.3\textwidth]{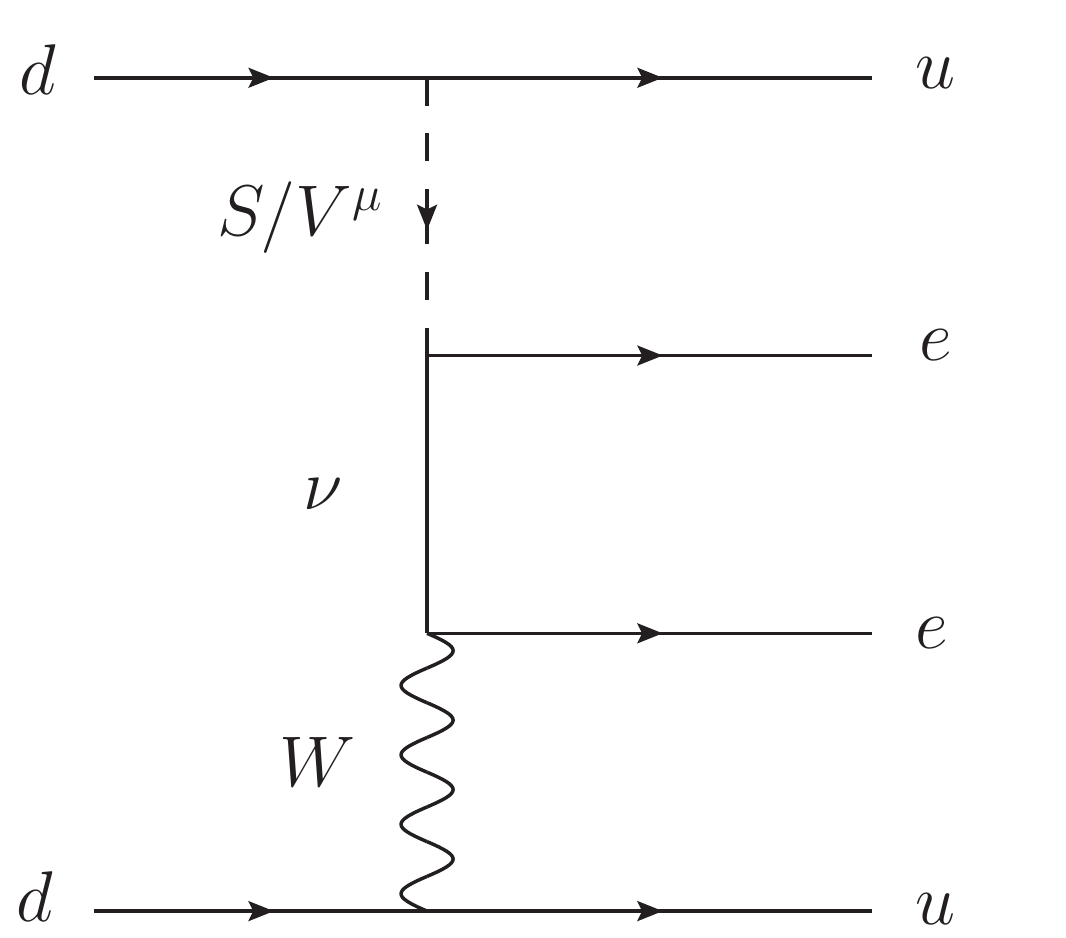}
 \caption{\emph{The exotic $\ovbb$ decay mechanisms involving scalar or vector leptoquarks $S$, $V^{\mu}$.}}
\label{fig:ovbbLQ}
\end{figure}

Leptoquarks, particles appearing prominently in GUTs, can also trigger non-standard $\ovbb$ decay contributions. It has been described that this is the case, when different leptoquark multiplets mix via a possible leptoquark-Higgs coupling violating lepton number~\cite{Hirsch:1996qy,Hirsch:1996ye}. Diagrams of this type of contributions to $\ovbb$ decay are shown in Fig.~\ref{fig:ovbbLQ}. The specific helicity structure of the effective four-fermion interaction leads to the fact that this contribution can dominate over the standard mass mechanism. The current lower limits on $\ovbb$ decay half-life then allow to derive the bounds on corresponding leptoquark parameters.

\begin{figure}[h]
 \centering
  \includegraphics[width=0.3\textwidth]{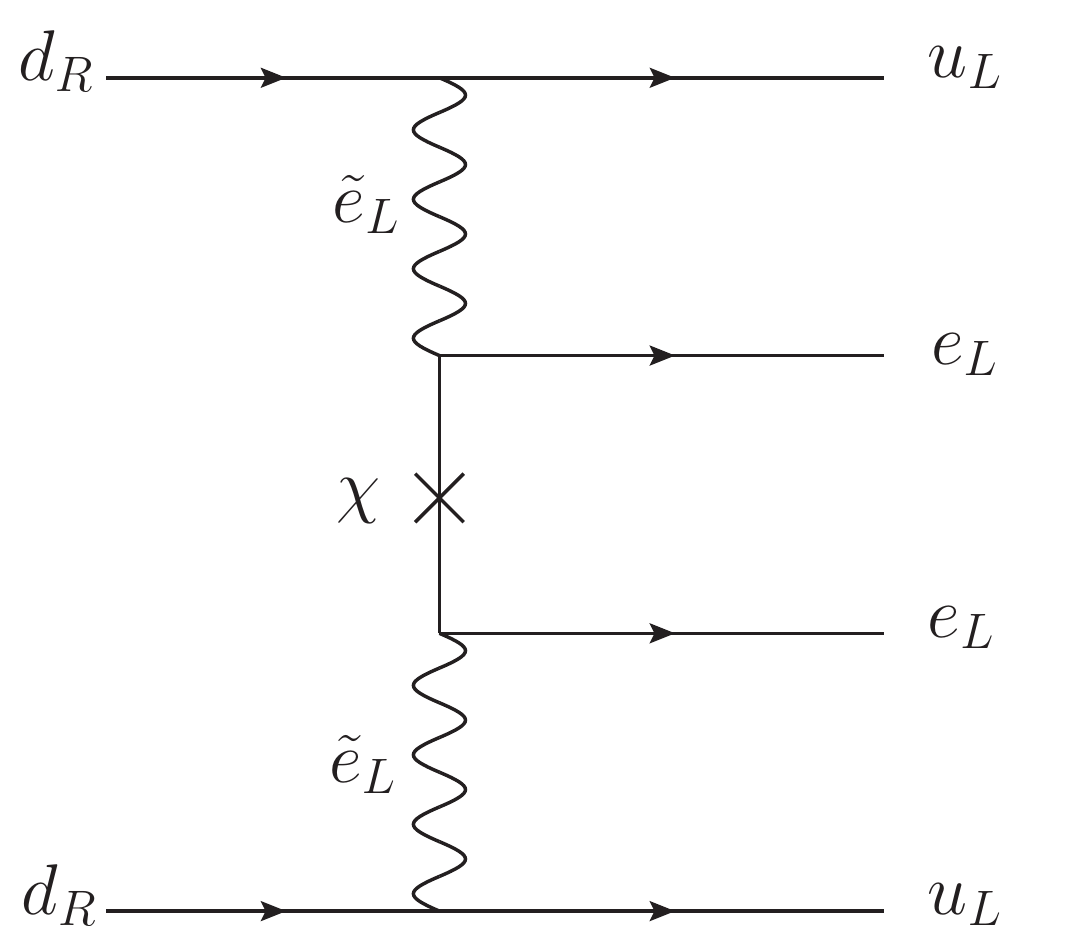}
 \caption{\emph{An example of a $\ovbb$ decay mechanism that can be triggered in R-parity-violating MSSM.}}
\label{fig:mssmovbb}
\end{figure}

Neutrinoless double beta decay can be triggered also in supersymmetric theories aspiring for grand unification. In the simplest case, if the MSSM with broken $R$ parity is considered, $\ovbb$ decay diagrams involving $R$-parity-violating couplings and supersymmetric mediators can be drawn~\cite{Mohapatra:1986su,Hirsch:1995zi,Hirsch:1995ek,Hirsch:1995cg,Pas:1998nn}. An example of a supersymmetric $\ovbb$ decay mechanism is depicted in Fig.~\ref{fig:mssmovbb}. Again, from non-observation of $\ovbb$ decay it is possible to derive limits on the unknown model parameters.

%

\section{Outlook and future prospects for GUTs}
\label{sec:outlook}

Among the vast landscape of theories beyond the Standard Model, Grand Unified Theories stand out as appealing candidates. As we have seen, GUTs are a collection of ideas from group theory, supersymmetry, neutrino physics, flavour physics and more, which positions them as some of the most complete and attractive theories in the literature. Indeed they are among the few BSM theories capable of simultaneously affecting the highest energy scales, influencing the cosmology of the early Universe, and the low energies, within reach of colliders and terrestrial experiments.

Throughout this review article we have provided a rough sketch of the status of GUTs and some of the associated research in the field. We have described the basic principles behind them and their first appearance in the world of particle physics in the late 70s. A lot of effort was spent in the years after and many of the greatest models were designed at that time. Research in GUTs has continued since, focusing either on specific topics within and alongside them, or on particular models that compiled a few advances in the field. We have thus summarized a non-exhaustive selection of topics and models that are at the forefront of research in unified theories, aiming to provide an overview of the current state of the art.

We are fortunate enough to live in a time where experimental searches are abundant and they cover a rather vast range of fronts. The most cutting-edge technologies have been and are being developed to push the boundaries of our current understanding of particle physics and cosmology. Grand unified theories are and will be put under the microscope by many of these experimental advances, which will confirm, constrain or outright exclude some of the existing models.	

The recent observation of gravitational wave signatures opens a new window into the history of the Universe, where events and phenomena that ocurred in the early Universe can be observed with gravitational wave detectors. Cosmic phase transitions associated with patterns of symmetry breaking in unified theories are such events, as they can be the source of stochastic gravitational waves that can be observed today. Transition temperatures above the EW scale, typically associated with the breaking of some intermediate step in a GUT model, can be studied by future gravitational wave experiments such as a LISA, the Einstein telescope, Kagra, the Cosmic Explorer, BBO and DEIGO. Also in the cosmological frontier, GUTs can have a serious impact on the inflationary epoch of the Universe, testable in measurements of the CMB, and can contribute to the baryon asymmetry of the Universe, via baryo and leptogenesis.

At the time of writing we have reached the end of the second run of the LHC, with an outstanding recorded integrated luminosity of about 150 ${\rm fb}^{-1}$. Analyses of the accumulated data, however, are still under way and they will probably spill well into the start run 3 in 2021. Many of the analyses already published have strong consequences for the predictions of unified theories, as are direct searches for supersymmetry, leptoquarks or other exotics at ATLAS and CMS. Upcoming results from ongoing and future analyses of the results from the LHC experiments may strengthen the bounds on light states as predicted by SUSY GUTs and other models, or they might show hints of the existence of new particles, whose relevance for GUTs would need to be determined. Upgraded versions of the LHC (HL-LHC, VLHC or FCC) or other future colliders (ILC, CLIC) will certainly boost this programme  with increased accuracy and higher energies, which will further probe the low-hanging states predicted by GUTs.

Where colliders search for the low scale predictions of GUTs, precision experiments can explore the intermediate and high scales associated with unification. Nucleon decay limits are often among the strongest probes of fully unified theories and future experiments such as Hyper-Kamiokande and DUNE may set even stronger exclusion limits or perhaps measure signs of proton decay, which would be a smoking gun for GUTs. Furthermore, GUTs can provide contributions to a number of flavour and precision observables, such as LFU, LFV, EDM, AMM or $\ovbb$, some of which are in tension with the SM. Confirmation of these flavour anomalies with more collected data by LHCb and other experiments would be undeniable evidence of the need for new physics models and GUTs are very well suited for that purpose.

To conclude, Grand Unified Theories are still at the vanguard of research in BSM models. They can explain many of the issues of the SM and can accommodate the recent results from the cosmological, precision and collider frontiers with relative ease. Contrary to ``simplified'' models, GUTs are complete theories that can simultaneously make a large number of testable predictions on the different fronts. Fortunately, these predictions can be explored by upcoming analyses and future experiments, which can set strong exclusion limits in a subset of GUT models. On a more optimistic note, any observation in, for instance, SUSY searches at colliders, gravitational waves signatures or proton decay will stack the odds in favour of some GUT models and will significantly shape the future of the research in particle physics.

\section*{Acknowledgements}

The authors would like to thank P.~Athron, F.~Deppisch, A.~Kvellestad and Y.~Zhang for helpful discussions. T.E.G. was partly funded by the Research Council of Norway under FRIPRO project number 230546/F20 and partly supported by the ARC Centre of Excellence for Particle Physics at the Tera-scale, grant CE110001004. N.K.\ was supported by the Slovenian Research Agency under the research core funding grant P1-0035 and No.  J1-8137. TRIUMF receives federal funding via a contribution agreement with the National Research Council of Canada and the Natural Science and Engineering Research Council of Canada.

\appendix


\bibliographystyle{h-physrev4}
\bibliography{gut_review}

\end{document}